\documentclass[twocolumn,superscriptaddress,showpacs,nofootinbib]{revtex4}
\usepackage{bm,color}

\usepackage{graphicx}
\usepackage{amsmath}
\usepackage{amssymb}
\usepackage{amsfonts}
\usepackage{bm}
\usepackage{color}

\def\be{\begin{equation}}
\def\ee{\end{equation}}
\def\bea{\begin{eqnarray}}
\def\eea{\end{eqnarray}}

\begin{document}

\title{Hydrodynamic representation of the Klein-Gordon-Einstein equations\\
 in
the weak field limit: I. General formalism and perturbations analysis}

% First author block:
\author{Abril Su\'arez}
%\email{suarez@irsamc.ups-tlse.fr}
% \homepage{An author's web page; optional}
\affiliation{Laboratoire de Physique Th\'eorique, Universit\'e Paul
Sabatier, 118 route de Narbonne 31062 Toulouse, France}
% You may list several affiliation, using separate commands for each:
%\affiliation{The third affiliation is shared by both co-authors}
\author{Pierre-Henri Chavanis}
%\email{chavanis@irsamc.ups-tlse.fr}
\affiliation{Laboratoire de Physique Th\'eorique, Universit\'e Paul
Sabatier, 118 route de Narbonne 31062 Toulouse, France}
% For other authors please repeat the author block as needed
%\author{Second Author}
% Note how REVTeX 4 deals with identical affiliations
%\affiliation{The third affiliation is shared by both co-authors}

\begin{abstract}
Using a generalization of the Madelung transformation, we derive the
hydrodynamic representation of the Klein-Gordon-Einstein equations in the weak
field limit. We consider a complex self-interacting scalar field with a
$\lambda|\varphi|^4$ potential. We study the evolution of the
spatially homogeneous
background in the fluid representation and derive the linearized equations
describing the evolution of small perturbations in a static and in an expanding
universe. We compare the results with simplified models in which the
gravitational potential is introduced by hand in the Klein-Gordon equation, and
assumed to satisfy a (generalized) Poisson equation. Nonrelativistic
hydrodynamic equations based on the
Schr\"odinger-Poisson equations or on the Gross-Pitaevskii-Poisson equations are
recovered in the limit $c\rightarrow +\infty$. We study the evolution
of the perturbations in the matter era using the nonrelativistic limit of our
formalism. Perturbations whose wavelength is below the Jeans length
oscillate in time while perturbations whose wavelength is above the Jeans
length grow linearly with the scale factor as in the cold dark matter
model. The growth of perturbations in the scalar field model is substantially
faster than in the cold dark matter model. When the wavelength of the
perturbations approaches the cosmological horizon (Hubble length), a
relativistic
treatment is
mandatory. In that case, we find that relativistic effects attenuate or even
prevent the growth of perturbations. This paper exposes the general
formalism and provides illustrations in simple cases. Other applications of our
formalism will be considered in companion papers.
\end{abstract}

% Insert suggested PACS numbers (up to 4) in braces.
% The PACS (Physics and Astronomy Classification Scheme)
% can be accessed on the web at http://www.aip.org/pacs/
\pacs{95.35.+d, 98.80.-k, 98.80.Jk, 04.40.-b, 95.30.Sf}

% Insert keywords (up to about 4) in braces; optional.
% \keywords{Up to four keywords}

\maketitle

% Here the text of your article begins
%---------------------------------------------------------------------------------------------------------------------------------------------------

\section{\label{intro}Introduction}
% References should be done using the \cite, \ref, and \label commands.
% Put \label in argument of \section for cross-referencing like this:
%\section{\label{}}

Scalar fields (SF) play an important role in particle physics,
astrophysics, and cosmology \cite{kolb,zee,dodelson}. Their evolution
is usually
described by the Klein-Gordon (KG) equation \cite{klein1,gordon,klein2} which
can be
viewed as a relativistic extension of the Schr\"odinger equation
\cite{schrodinger1,schrodinger2}. The KG equation was one of the first attempts
to
unify the ideas of quantum mechanics and Einstein's theory of special
relativity.\footnote{The KG equation was actually discovered by Schr\"odinger
before he found the equation that now bears his name \cite{zee}.}  It describes
the evolution of a SF 
$\varphi(\vec x,t)$ whose excitations are bosonic particles of
spin zero. These particles are
neutral for a real SF and charged for a complex SF.  The KG
equation usually involves a potential  $V(\varphi)$ that takes self-interaction
into
account. For example, $\pi$-mesons,
Higgs bosons and  axions are described by the KG equation. SF are also
present in the theory of superstrings as well as in a large number of
Kaluza-Klein and supergravity models.

The coupling between the KG equation and gravity through
the
Einstein equations,  leading to the
Klein-Gordon-Einstein (KGE) equations, 
was first considered in the context of boson
stars
\cite{kaup,rb,thirring,breit,takasugi,colpi,bij,gleiserseul,
ferrellgleiser,gleiser,seidel90,
kusmartsev,kusmartsevbis,leepang,jetzer,seidel94,balakrishna,schunckliddle,
mielkeschunck,torres2000,wang,mielke,guzmanbh,chavharko}.\footnote{Boson stars
are described by complex SFs. Self-gravitating
systems described by real SFs are not static but, instead, are periodic with
both the spacetime geometry and the matter field oscillating in time. For that
reason they are called oscillatons \cite{oscillatons1,oscillatons2}.}
  Initially, the study
of boson stars
was motivated by the
axion field, a pseudo-Nambu-Goldstone boson of the Peccei-Quinn phase
transition, that was proposed as a possible solution to the strong CP
problem in QCD. In the early works of Kaup \cite{kaup} and Ruffini and Bonazzola
\cite{rb}, it was assumed that the bosons have no self-interaction. For a boson
mass $m\sim
1\, {\rm GeV}/c^2$, the
maximum mass of boson stars is very small, of the order of
$M_{\rm max}\sim 10^{-19}M_{\odot}$.  This corresponds to mini
boson stars like axion black holes. The
mass of these mini boson stars may be
too small to be astrophysically relevant. They could play a
role,
however, if they exist in the universe in abundance or if the axion
mass is extraordinarily small (less than $m\sim 10^{-10}{\rm eV}/c^2$) leading
to
macroscopic objects with a mass $M_{\rm max}$ comparable to the mass of
the sun (or even larger) \cite{mielke}. It has also been proposed that stable
boson stars with
a boson mass $m\sim 10^{-17}{\rm eV}/c^2$ could mimic supermassive black holes
($M\sim 10^6\, M_{\odot}$, $R\sim 10^7\, {\rm km}$) that reside at the center of
galaxies \cite{torres2000,guzmanbh}. On the other hand, Colpi {\it et al.}
\cite{colpi} considered the case where  the bosons have a repulsive
self-interaction described by a $\lambda|\varphi|^4$ potential and showed that, 
for $m\sim 1{\rm GeV}/c^2$, the  maximum mass of boson stars can be of the order
of the solar mass $M_{\odot}$, similar to the mass of neutron
stars. Therefore, a
self-interaction can significantly change the physical dimensions of boson
stars, making them much more astrophysically interesting.

It has also been proposed that dark matter (DM) halos may
be made of a SF
described by the KGE equations (see,
e.g., \cite{revueabril,revueshapiro,bookspringer} for recent reviews). Actually,
at the galactic scale, the Newtonian limit is valid so DM halos can be
described by the Schr\"odinger-Poisson (SP) equations or by the
Gross-Pitaevskii-Poisson (GPP) equations. In that case, the wavefunction
$\psi(\vec
x,t)$ describes a Bose-Einstein condensate (BEC) at $T=0$, and the
self-interaction of the bosons is measured by their scattering length $a_s$.
Therefore, DM halos could be gigantic quantum objects made of BECs. The wave
properties of bosonic DM may
stabilize the system against gravitational collapse, providing halo cores
and sharply suppressing small-scale linear power. This may
solve the
problems of the cold dark matter (CDM) model such as the cusp problem and the
missing
satellite
problem. The  scalar
field dark matter (SFDM) model and the BEC dark matter (BECDM) model, also
called $\Psi$DM models,
have received much attention in the last years. In the non-interacting case
\cite{baldeschi,membrado,sin,jisin,schunckpreprint,matosguzman,guzmanmatos,
hu,mu,
arbey1,silverman1,matosall,silverman,bmn,sikivie,mvm,lee09,ch1,lee,prd1,prd2,
mhh,ch2,ch3},  the mass of the bosons must
be extremely
small, of the order of $m\sim 2.57\times 10^{-20}\, {\rm eV}/c^2$,  in order to
account for the mass
and size of dwarf DM halos that are completely
condensed. Ultralight scalar
fields like axions may
have such small
masses. This
corresponds to ``fuzzy cold dark matter'' \cite{hu}. On the other hand, when a
repulsive self-interaction is taken into account
\cite{leekoh,peebles,goodman,arbey,lesgourgues,bohmer,prd1,prd2,briscese,harko,
pires,rmbec,rindler,lora,lensing,glgr1}, the characteristics of dwarf DM halos
can
be reproduced with a much larger boson mass $m=1.69\times 10^{-2}\,
{\rm eV}/c^2$  (satisfying the limit $m<1.87\, {\rm eV}/c^2$
obtained from cosmological
considerations \cite{limjap}) and a scattering length $a_s=1.73\times 10^{-5}\,
{\rm fm}$ (satisfying the constraint $4\pi a_s^2/m<1.25\, {\rm cm}^2/{\rm g}$
set by the
Bullet Cluster \cite{bullet}). Dwarf halos are
purely condensed. Large dark
matter halos have a
core-halo structure resulting from gravitational cooling \cite{seidel94}. They
are made of a solitonic core (BEC) surrounded by a halo of scalar
radiation.

Scalar fields have also been introduced in cosmology. The phase of inflation in
the
early universe is usually described by some
hypothetical SF, called the inflaton, with its origin in the quantum
fluctuations of the vacuum \cite{guth0,guth1,guth2,guth3,linde}. A variety of
inflationary
models that include SFs have been proposed \cite{olive}, and SF are expected to
play
an important role in determining the dynamics of the early universe.  Inflation
is generally considered to be a reasonable
solution to many of the fundamental problems within the standard cosmological
model.  Inspired
by the analogy
with the inflation, some authors have represented the dark energy responsible
for the present acceleration of the Universe by a SF
called quintessence
\cite{quintessence1,quintessence2,quintessence3,quintessence4,quintessence5,
quintessence6,quintessence7,quintessence8,quintessence9,quintessence10,
quintessence11,quintessence12,quintessence13}. This treatment seems to fit
current
observations, and unlike the cosmological constant $\Lambda$, the SF
evolves
dynamically, leaving distinctive imprints in the Cosmic
Microwave Background (CMB) and matter power spectrum.

Since DM may be a SF, it is of
considerable interest to study the cosmological implications of this scenario.
The cosmological evolution of a spatially
homogeneous non-interacting real SF described by the KGE equations competing
with baryonic matter, radiation and dark energy, was considered by
Matos {\it et al.} \cite{mvm}. They found that real SFs display fast
oscillations but that, on the mean, they reproduce the cosmological
predictions of the standard
$\Lambda$CDM model. The study of
perturbations was considered by Su\'arez and
Matos
\cite{abrilMNRAS} for a self-interacting real SF
described by the Klein-Gordon-Poisson (KGP) equations and by
Maga\~na {\it et al.} \cite{abrilJCAP} for a non-interacting
real SF described by the KGE equations. These studies show that the
perturbations can grow in the linear regime, leading, in the nonlinear regime,
to the formation of structures corresponding to DM halos. This is in
agreement with the early work of Khlopov {\it et al.} \cite{khlopov} who studied
the
Jeans instability of a relativistic SF in a static
background.

The case of a complex self-interacting SF representing BECDM was considered by
Chavanis
\cite{prd1,chavaniscosmo} in the context of
Newtonian cosmology. His study is based on the GPP equations. The
Jeans instability of a homogeneous self-gravitating BEC in a static background
is studied in \cite{prd1}. The evolution of
perturbations of BECDM in an expanding Einstein-deSitter (EdS) universe is
considered in
\cite{chavaniscosmo}. It is
found that the perturbations grow faster in BECDM as compared to
$\Lambda$CDM. Harko \cite{harkocosmo} and Chavanis \cite{chavaniscosmo}
independently developed a relativistic BEC cosmology by assuming that the
equation of state of BECDM is given by $P=2\pi
a_s\hbar^2\epsilon^2/m^3c^4$, which corresponds to the classical equation of
state of BECs \cite{revuebec} where $\rho$ is replaced by $\epsilon/c^2$.
However, the
identification of the energy density $\epsilon$ with the rest-mass density
$\rho$ is only valid in a weakly relativistic regime, so the extrapolation of
their results to early times is not correct (see the discussion in
\cite{becstiff}). Recently, Li {\it et al.} \cite{shapiro} developed an
exact relativistic
cosmology
for a complex self-interacting SF/BEC based on the KGE equations. They studied 
the
evolution of the homogeneous background and showed that the Universe
undergoes three successive phases: a stiff matter era, followed by a radiation
era due to the SF (that exists only for self-interacting SFs), and finally a
matter era similar to CDM.  They compared their theoretical results with
observations in order to constrain the parameters $(m,\lambda)$ of the SF.

Instead of working directly in terms of a SF, we can adopt a
fluid approach and work with hydrodynamic equations. In the case of the
Schr\"odinger equation, this hydrodynamic approach was introduced by
Madelung  \cite{madelung}. 
He showed that the Schr\"odinger equation is equivalent to the Euler equations
for an irrotational fluid with an additional quantum
potential\footnote{The  results of Madelung \cite{madelung}  were rediscovered
by Bohm \cite{bohm} so the quantum potential is sometimes called the Bohm
potential. The formulation of
classical and relativistic quantum mechanics in terms of 
hydrodynamic equations  was
also developed by Takabayasi \cite{takabayasi1,takabayasi2}.}
arising from the finite value of $\hbar$. This hydrodynamic formulation was
criticized, or disregarded, by many authors (e.g., Pauli)
in the early years
of quantum mechanics because it lacks a clear physical interpretation in the
case
where the Schr\"odinger equation describes just one particle. However, it takes
more sense when the Schr\"odinger equation (or the GP equation) describes a BEC
made of many particles in the same quantum state \cite{revuebec}. In that case,
the BEC can be
interpreted as a real fluid described by quantum Euler
equations.\footnote{One interesting aspects of BECs is related to
their superfluid properties. Their velocity field is irrotational
 but there may exist vortical motion due to
singular point vortices with quantized circulation $h/m$
\cite{onsager,feynman}.}  This
hydrodynamic representation has been used by B\"ohmer and Harko \cite{bohmer}
and by Chavanis \cite{prd1,prd2,chavaniscosmo} among others in the case of BECDM
and in
the case of BEC stars
\cite{chavharko}. This hydrodynamic approach has been generalized by Su\'arez
and Matos
\cite{abrilMNRAS}
in the context of the KG equation. They used it to
study the formation of structures in the Universe, assuming that DM is
in the form of a fundamental SF with a $\lambda \varphi^4$ potential
\cite{abrilMNRAS}. They also studied the phase transition of a real SF due to a
$Z_2$ symmetry of its potential \cite{smz}.

In the works \cite{abrilMNRAS,smz}, the SF is taken to be real and the
gravitational
potential is
introduced by hand in the KG equation, and assumed to be determined by the
classical 
Poisson equation where the source is the rest-mass density $\rho$. This leads to
the
KGP equations. However, this treatment is not self-consistent since it combines
relativistic and nonrelativistic equations. In this article,
we derive the hydrodynamic
representation of a complex SF coupled to gravity through
the Einstein
equations in the weak field approximation. In this way, we develop a
self-consistent relativistic treatment. Throughout this work, we use the
Newtonian gauge which takes into account metric perturbations up to first order.
We  consider only scalar perturbations. This is sufficient if we are interested
in calculating
observational consequences of the SF dynamics in the linear regime.
After having derived the hydrodynamic equations, we study the evolution of the
homogeneous background and the evolution of small perturbations in a static and
in an
expanding universe using the hydrodynamic
representation. We compare our results with those
obtained from the heuristic KGP equations. We argue that this simplified model
is not sufficient to study the evolution of the perturbations in the linear
relativistic regime. 

The main purpose of the present
paper is to develop a general
formalism. Applications of this formalism will be considered in following
papers. However, for illustration, we already present a few applications of our
formalism in simple cases.

The paper is organized as follows. In Sec. \ref{sec_tb}, we present the
theoretical
background and introduce the main equations of the the paper that are the KG
equation and the Einstein field equations. This allows us to set the
notations. In Sec. \ref{sec_kge}, we consider the KGE equations in the weak
field limit. We first write the KG equation and
the Einstein equations in the Newtonian
gauge. Then, we transform the KGE equations into the Gross-Pitaevskii-Einstein
(GPE) equations to facilitate
the connection with the nonrelativistic limit $c\rightarrow +\infty$. Finally,
we write the GPE equations in
the form of hydrodynamic equations. In Sec. \ref{sec_jeans}, we consider the
case of a
static background and study the evolution of small perturbations in an
infinite homogeneous Universe (Jeans problem). We determine the energy of the
homogeneous SF, the dispersion relation of the perturbations, and the
Jeans length. In Sec. \ref{sec_b}, we consider the evolution of the Universe
induced by
a homogeneous SF. We recover from the fluid equations the three
phases previously obtained by Li {\it et al.} \cite{shapiro}: a stiff matter
phase, a
radiation phase, and
a matter phase. In Sec. \ref{sec_lp}, we consider the evolution of the
perturbations in an expanding homogeneous Universe. We provide the exact set of
linearized
equations, then propose a closed approximate equation for the density contrast
that has the correct Jeans length. In Sec. \ref{sec_evolution}, 
we study the evolution of the perturbations in the matter era. We show that the
nonrelativistic limit can be used when the wavelength of the
perturbations is much smaller than the cosmological horizon (Hubble length).
Perturbations whose wavelength is below the Jeans length
oscillate in time while perturbations whose wavelength is above the Jeans
length grow linearly with the scale factor as in the CDM model. The growth
of perturbations in the scalar field model is substantially
faster than in the CDM model. When the wavelength of the
perturbations approaches the cosmological horizon, general relativity
attenuates or even
prevents the growth of perturbations. The
Appendices
present useful complements.  In Appendix
\ref{sec_cons}, we
determine the constant needed to transform the KG equation into the
GP equation.  In Appendix \ref{sec_inh}, we derive the hydrostatic 
equations describing spatially inhomogeneous
relativistic SF/BEC clusters. In Appendix \ref{sec_nr}, we recover from
our
formalism the nonrelativistic equations obtained in the framework of the GPP
equations.  In Appendix \ref{sec_kgp}, we introduce the  generalized
KGP equations in which the
gravitational potential is introduced by hand in the KG equation in a flat
Friedmann-Lema\^itre-Robertson-Walker (FLRW) background
space-time, and assumed to be given by a generalized Poisson equation where the
source is the energy density $\epsilon$. We show that this model corresponds to
the limit $\Phi/c^2\rightarrow +\infty$ of the KGE equations. In Appendices
\ref{sec_typ}-\ref{sec_smallness}, we regroup all the elements necessary to
make the numerical applications needed in the paper. A summary
of our results can be found in the Proceedings of the X Mexican School on
Gravitation and Mathematical Physics \cite{proceedings}.

\section{Theoretical Background}
\label{sec_tb}

\subsection{The Lagrangian of the scalar field}

We assume that DM can be described by a  complex
SF\footnote{There are several reasons for considering a complex
rather than a real SF \cite{shapiro}. Complex SFs have been invoked in many
different sectors of
elementary particle physics in relation to the Higgs mechanism responsible for
mass generation. On the other hand, the $U(1)$ symmetry implies the dark
matter particle number (charge) conservation \cite{zee}. Finally, complex scalar
fields can form singular vortices leading to a rich dynamics of the halos when
they rotate \cite{rindler}.} which is a continuous
function of space and time defined at each point by
$\varphi(x^\mu)=\varphi(x,y,z,t)$. The action of the relativistic SF
is
\begin{equation}
S_\varphi=\int d^4x\sqrt{-g}\mathcal{L}_\varphi,
\label{tb1}
\end{equation}
where $\mathcal{L}_\varphi=\mathcal{L}_\varphi(\varphi,
\varphi^*,\partial_\mu\varphi,\partial_\mu\varphi^*)$
is the Lagrangian density and $g={\rm det}(g_{\mu\nu})$ is the determinant of
the metric tensor. We adopt the following generic Lagrangian density 
\begin{eqnarray}
{\cal L}_\varphi=\frac{1}{2}g^{\mu\nu}\partial_{\mu}\varphi^*
\partial_{\nu}\varphi-V(|\varphi|^2),
\label{tb2}
\end{eqnarray}
which is written for a metric signature $(+,-,-,-)$.
Specifically, we consider a SF
potential of the form
\begin{equation}
V(|\varphi|^2)=\frac{m^2c^2}{2\hbar^2}|\varphi|^2+\frac{m^2}{2\hbar^4}\lambda
|\varphi|^4,
\label{tb3}
\end{equation}
where the quadratic term is the rest-mass term and the quartic term
is a self-interaction term. In view of future applications, we shall assume
that the SF describes a BEC at $T=0$ in which all the particles are
in the same ground state. In that case, the self-interacting
constant $\lambda$ can be expressed in terms of the scattering length of the
bosons $a_s$ and of their mass $m$ by $\lambda=4\pi a_s\hbar^2/m$. The
potential
of the SF can be rewritten as
\begin{equation}
V(|\varphi|^2)=\frac{m^2c^2}{2\hbar^2}|\varphi|^2+\frac{2\pi a_s
m}{\hbar^2}|\varphi|^4.
\label{tb4}
\end{equation}
The self-interaction is repulsive when $a_s>0$ and attractive when $a_s<0$
\cite{revuebec}. The case of a general potential of the form
$V(|\varphi|^2)$ is treated in \cite{proceedings}.

\subsection{The Klein-Gordon equation}

The equation of motion for the SF can be obtained from the principle
of least action. Imposing  $\delta S_{\varphi}=0$ for arbitrary variations 
$\delta\varphi$
and $\delta\varphi^*$, we obtain the Euler-Lagrange
equation  
\begin{eqnarray}
D_{\mu}\left\lbrack\frac{\partial {\cal L}_\varphi}{\partial
(\partial_{\mu}\varphi)^*}\right\rbrack-\frac{\partial {\cal
L}_\varphi}{\partial\varphi^*}=0,
\label{tb5}
\end{eqnarray}
where $D$ is the covariant derivative. For the generic Lagrangian (\ref{tb2}),
 this leads to the KG
equation
\begin{equation}
\Box\varphi+2V(|\varphi|^2),_{\varphi^*}=0,
\label{tb6}
\end{equation}
where $\Box$ is the d'Alembertian operator
\begin{equation}
\Box\equiv D_{\mu}(g^{\mu\nu}\partial_{\nu})=\frac{1}{\sqrt{-g}}
\partial_\mu(\sqrt{-g}\, g^{\mu\nu}\partial_\nu)
\label{tb7}
\end{equation}
and
\begin{equation}
V(|\varphi|^2),_{\varphi^*}=\frac{dV}{d|\varphi|^2}\varphi.
\label{tb7b}
\end{equation}
For the specific SF potential (\ref{tb3}), the KG equation
takes the form
\begin{equation}
\square \varphi+\frac{m^2c^2}{\hbar^2}\varphi+\frac{8\pi a_s
m}{\hbar^2}|\varphi|^2\varphi=0.
\label{tb8}
\end{equation}
The d'Alembertian operator can be written in terms of the covariant
derivative and of
the Christoffel symbols as 
\begin{eqnarray}
\square
\varphi=D_{\mu}(g^{\mu\nu}\partial_{\nu}\varphi)=g^{\mu\nu}D_{\mu}(\partial_{\nu
} \varphi)\nonumber\\
=g^{\mu\nu}\partial_{\mu}\partial_{\nu}\varphi-g^{\mu\nu}\Gamma_{
\mu\nu } ^ {
\sigma}\partial_{\sigma}\varphi.
\label{tb9}
\end{eqnarray}

\subsection{The energy-momentum tensor}

Taking the variation of the SF action (\ref{tb1}) with respect to $g^{\mu\nu}$,
we
get
\begin{equation}
\delta S_\varphi=\frac{1}{2}\int d^4x \sqrt{-g}\, T_{\mu\nu}\delta g^{\mu\nu},
\label{tb10}
\end{equation}
where
\begin{eqnarray}
T_{\mu\nu}=2\frac{\partial\mathcal{L}_\varphi}{\partial
g^{\mu\nu}}-g_{\mu\nu}\mathcal{L}_{\varphi}
\label{tb11}
\end{eqnarray}
is the energy-momentum tensor of the SF. For the generic Lagrangian (\ref{tb2}),
it takes
the form
\begin{eqnarray}
T_{\mu\nu}=\frac{1}{2}(\partial_{\mu}\varphi^*
\partial_{\nu}\varphi+\partial_{\nu}\varphi^* \partial_{\mu}\varphi)\nonumber\\
-g_{\mu\nu}\left
\lbrack\frac{1}{2}g^{\rho\sigma}\partial_{\rho}\varphi^*\partial_{\sigma}
\varphi-V(|\varphi|^2)\right \rbrack.
\label{tb12}
\end{eqnarray}
By analogy with the energy-momentum tensor of a
perfect fluid, the energy
density and the pressure tensor of the SF are defined by $\epsilon=T_0^0$
and $P_{i}^{j}=-T_i^j$. The conservation of the energy-momentum
tensor, which
results from the  Noether theorem, writes $D_{\nu}T^{\mu\nu}=0$.

\subsection{The Einstein equations}

The Einstein-Hilbert action in general relativity is defined by
\begin{equation}
S_g=\frac{c^4}{16\pi G}\int d^4x\sqrt{-g}R,
\label{kge7}
\end{equation}
where  $R$ is the Ricci scalar and $G$ is Newton's
gravitational
constant.  Its variation with respect to $g^{\mu\nu}$
is given by \cite{weinberg}:
\begin{equation}
\delta S_g=-\frac{c^4}{16\pi
G}\int d^4x\,\sqrt{-g} \left(R_{\mu\nu}-\frac{1}{2}g_{\mu\nu}R\right)\delta
g^{\mu\nu},
\label{kge8}
\end{equation}
where $R_{\mu\nu}$ is the Ricci tensor. The total action (SF $+$
gravity) is $S=S_{\varphi}+S_g$. The field equations can be obtained from the
principle of least action. Imposing 
$\delta S=0$ for arbitrary variations in $g^{\mu\nu}$, and using Eqs.
(\ref{tb10}) and
(\ref{kge8}), we get the Einstein equations
\begin{equation}
R_{\mu\nu}-\frac{1}{2}g_{\mu\nu}R=\frac{8\pi G}{c^4}T_{\mu\nu}.
\label{kge9}
\end{equation}
These are a set of $10$ equations that describe the fundamental
interaction
between gravity and matter as a result of the curvature of space-time. The
energy-momentum tensor  $T^{\mu\nu}$ is the source of the
gravitational field in the Einstein field equations of general relativity
in the sense that it determines the metric $g^{\mu\nu}$. The conservation of
the energy-momentum tensor is automatically included in the Einstein equations.

\section{The Klein-Gordon-Einstein equations in the weak field limit}
\label{sec_kge}

In this paper, we study the  KGE
equations in the weak field limit $\Phi/c^2\ll 1$. The equations that we
derive are valid at the order $O(\Phi/c^2)$. For $\Phi/c^2\rightarrow 0$, we
obtain the generalized KGP equations (see Appendix \ref{sec_kgp}). Of course,
the limit $\Phi/c^2\rightarrow 0$ is
different from the nonrelativistic limit $c\rightarrow +\infty$ leading to the
GPP equations (see Appendix \ref{sec_nr}).

\subsection{The conformal Newtonian Gauge}

We work with the conformal Newtonian gauge \cite{ma}. In general relativity,
this
gauge is a perturbed form of the FLRW line element.
We assume that the Universe is flat in
agreement with the observations of the cosmic microwave background (CMB). The
general perturbed FLRW metric in the comoving frame has the form
\begin{eqnarray}
ds^2=\left(1+2\frac{\Psi}{c^2}\right)c^2
dt^2-2a(t)w_i c dt dx^i\nonumber\\
-a(t)^2\left\lbrack \left(1-2\frac{\Phi}{c^2}
\right)\delta_{ij}+H_{ij}\right \rbrack dx^idx^j,
\label{kge1}
\end{eqnarray}
where the perturbed quantities $\Psi/c^2$, $\Phi/c^2$, $w_i$, and $H_{ij}$ are
all $\ll 1$. In this metric, $\Phi$ represents the gravitational potential of
classical Newtonian gravity while $\Psi$ is the lapse function \cite{ma}. 

We consider the simplest form of the Newtonian gauge, only taking
into account scalar perturbations which are the ones that contribute to
the formation of structures in cosmology. Vector (which are supposed to be
always small) vanish during cosmic inflation and tensor contributions (which
account for gravitational waves) are neglected \cite{mfb}. We also
assume $\Psi=\Phi$, supposing a universe without anisotropic stress. Therefore,
our line element is given by
\begin{equation}
ds^2=c^2\left(1+2\frac{\Phi}{c^2}\right)dt^2-a(t)^2\left(1-2\frac{\Phi}{c^2}
\right)\delta_{ij}dx^idx^j,
\label{kge2}
\end{equation}
where we recall that ${\Phi}/{c^2}\ll 1$. Working with this metric enables us
to obtain exact equations at the
order  $O(\Phi/c^2)$ taking into account both relativistic and gravitational
contributions inside an expanding Universe without having to introduce
the gravitational potential $\Phi$ by hand.

\subsection{The Klein-Gordon equation}

Computing the d'Alembertian (\ref{tb7}) with the Newtonian gauge (\ref{kge2}),
we obtain the
KG equation
\begin{eqnarray}
\frac{1}{c^2}\frac{\partial^2\varphi}{\partial
t^2}&+&\frac{3H}{c^2}\frac{\partial\varphi}{\partial
t}-\frac{1}{a^2}\left(1+\frac{4\Phi}{c^2}\right)\Delta\varphi\nonumber\\
&+&2\left(1+2\frac{\Phi}{c^2}\right)V,_{\varphi^*}
-\frac{4}{c^4}\frac{\partial\Phi}{\partial t}\frac{\partial\varphi}{\partial
t}=0,\label{kge3}
\end{eqnarray}
where $H=\dot a/a$ is the Hubble constant and
$\Delta=\vec\nabla^2$ is the usual Laplacian ($\vec\nabla$ is the usual nabla
operator). For the specific SF potential (\ref{tb4}), the KG equation takes the
form
\begin{eqnarray}
\frac{1}{c^2}\frac{\partial^2\varphi}{\partial
t^2}+\frac{3H}{c^2}\frac{\partial\varphi}{\partial
t}-\frac{1}{a^2}\left(1+\frac{4\Phi}{c^2}\right)\Delta\varphi\nonumber\\
-\frac{4}{c^4}\frac{\partial\Phi}{
\partial t}\frac{\partial\varphi}{\partial
t}+\left (1+\frac{2\Phi}{c^2}\right) \frac{m^2
c^2}{\hbar^2}\varphi\nonumber\\
+\left (1+\frac{2\Phi}{c^2}\right)
\frac{8\pi a_s m}{\hbar^2}|\varphi|^2\varphi =0.\label{kge4}
\end{eqnarray}
Using the expression (\ref{tb12}) of the energy-momentum
tensor, the energy
density and the pressure are given by 
\begin{eqnarray}
\epsilon=T_0^0=\frac{1}{2c^2}\left (1-\frac{2\Phi}{c^2}\right
)\left
|\frac{\partial\varphi}{\partial t}\right |^2 \nonumber\\
+\frac{1}{2a^2}\left (1+\frac{2\Phi}{c^2}\right
)|\vec\nabla\varphi|^2+V(|\varphi|^2),
\label{kge5}
\end{eqnarray}
\begin{eqnarray}
P=-\frac{1}{3}(T_1^1+T_2^2+T_3^3)=\frac{1}{2c^2}\left
(1-\frac{2\Phi}{c^2}\right
)\left
|\frac{\partial\varphi}{\partial t}\right |^2 \nonumber\\
-\frac{1}{6a^2}\left (1+\frac{2\Phi}{c^2}\right
)|\vec\nabla\varphi|^2-V(|\varphi|^2).
\label{kge6}
\end{eqnarray}

\subsection{The Einstein equations}

The time-time component of the Einstein equations is 
\begin{eqnarray}
R_0^0-\frac{1}{2}R=\frac{8\pi G}{c^4}T_0^0.
\label{kge10}
\end{eqnarray}
With the Newtonian conformal gauge, the left hand side of Eq. (\ref{kge10}) is
given by
\begin{eqnarray}
R_0^0-\frac{1}{2}R=\frac{3H^2}{c^2}+\frac{2}{a^2c^2}\Delta\Phi-\frac{6}{c^4}
H\left
(\frac{\partial\Phi}{\partial t}+H\Phi\right ).
\label{kge11}
\end{eqnarray}
Therefore, the time-time component of the Einstein
equations can be rewritten as
\begin{eqnarray}
\frac{\Delta\Phi}{4\pi Ga^2}=\frac{\epsilon}{c^2}-\frac{3H^2}{8\pi
G}+\frac{3H}{4\pi G
c^2}\left (\frac{\partial\Phi}{\partial t}+H\Phi\right ).
\label{kge11b}
\end{eqnarray}
Using the time-time component of the energy-momentum
tensor, which represents the energy density $\epsilon$ given by Eq.
(\ref{kge5}), we get
\begin{eqnarray}
\frac{\Delta\Phi}{4\pi G a^2}=\frac{1}{2c^4}\left (1-\frac{2\Phi}{c^2}\right
)\left |\frac{\partial\varphi}{\partial t}\right |^2 \nonumber\\
+\frac{1}{2a^2c^2}\left (1+\frac{2\Phi}{c^2}\right
)|\vec\nabla\varphi|^2+\frac{m^2}{2\hbar^2}|\varphi|^2\nonumber\\
+\frac{2\pi a_s m}{\hbar^2c^2}|\varphi|^4-\frac{3H^2}{8\pi G}+\frac{3H}{4\pi G
c^2}\left (\frac{\partial\Phi}{\partial t}+H\Phi\right ).
\label{kge12}
\end{eqnarray}
Eqs. (\ref{kge4}) and (\ref{kge12}) form the KGE equations. This system of
equations  is closed because we consider a universe without anisotropic stress
($\Psi=\Phi$).
Otherwise, we need to write the other components of the Einstein equations.

\subsection{Spatially homogeneous scalar field}

For a spatially homogeneous SF with $\varphi_b(\vec
x,t)=\varphi_b(t)$ and $\Phi_b(\vec
x,t)=0$, the KG equation (\ref{kge3}) reduces to
\begin{eqnarray}
\frac{1}{c^2}\frac{d^2\varphi_b}{dt^2}+\frac{3H}{c^2}\frac{d\varphi_b}{dt}
+2V(|\varphi_b|^2),_{\varphi_b^*}=0.
\label{kgp6bz}
\end{eqnarray}
For the specific SF potential (\ref{tb4}), we explicitly have
\begin{eqnarray}
\frac{1}{c^2}\frac{d^2\varphi_b}{dt^2}+\frac{3H}{c^2}\frac{d\varphi_b}{dt}+\frac
{m^2
c^2}{\hbar^2}\varphi_b
+\frac{8\pi a_s m}{\hbar^2}|\varphi_b|^2\varphi_b=0.\nonumber\\
\label{kgp6b}
\end{eqnarray}
For a homogeneous SF, the energy-momentum tensor is diagonal and isotropic,
$T^{\mu}_{\nu}={\rm diag} (\epsilon_b,-P_b,-P_b,-P_b)$. The energy density
$\epsilon_b(t)$ and the pressure $P_b(t)$ are given by
\begin{equation}
\epsilon_b=\frac{1}{2c^2}\left |\frac{d\varphi_b}{d
t}\right|^2+V(|\varphi_b|^2),
\label{kgp7b}
\end{equation}
\begin{equation}
P_b=\frac{1}{2c^2}\left |\frac{d\varphi_b}{d
t}\right|^2-V(|\varphi_b|^2).
\label{kgp8b}
\end{equation}
From these equations, we obtain the continuity equation
\begin{equation}
\frac{d\epsilon_b}{dt}+3H(\epsilon_b+P_b)=0
\label{frid1}
\end{equation}
which is one of the Friedmann equations \cite{weinberg}. The other Friedmann
equation is
obtained from the Einstein equation (\ref{kge11b}) that reduces, when
$\Phi_b=0$, to
\begin{eqnarray}
H^2=\frac{8\pi G}{3c^2}\epsilon_b.
\label{kge12b}
\end{eqnarray}
This relation shows that the term $-3H^2/8\pi G=-\epsilon_b/c^2$ in the Einstein
equation (\ref{kge11b}) with $\Phi(\vec{x},t)\neq 0$ plays the role of a
neutralizing background.
From Eqs. (\ref{frid1}) and (\ref{kge12b}), we easily obtain
\begin{eqnarray}
\frac{\ddot a}{a}=-\frac{4\pi G}{3c^2}(\epsilon_b+3P_b).
\label{kge12bw}
\end{eqnarray}

\subsection{The Gross-Pitaevskii-Einstein equations}

The KG equation without self-interaction can be viewed as a
relativistic generalization of the Schr\"odinger equation. Similarly, the
KG equation with a self-interaction can be viewed as a
relativistic generalization of the GP equation. In
order to recover the  Schr\"odinger and GP equations in the
nonrelativistic limit $c\rightarrow +\infty$, we make the transformation
\cite{zee}:
\begin{eqnarray}
\varphi(\vec{x},t)=\frac{\hbar}{m}e^{-i m c^2 t/\hbar}\psi(\vec{x},t).
\label{kgp12}
\end{eqnarray}
The prefactor $\hbar/m$ is
justified in
Appendix \ref{sec_cons}. Mathematically, we can
always make this change of variables. 
However, we emphasize that it is only in the nonrelativistic limit $c\rightarrow
+\infty$ that $\psi$ has the interpretation of a wave function, and that
$|\psi|^2=\rho$ has the interpretation of a rest-mass density. In the
relativistic regime, $\psi$ and $\rho=|\psi|^2$ do not have a clear
physical
interpretation. We will call them ``pseudo wave function'' and ``pseudo
rest-mass
density''. Nevertheless, it is perfectly legitimate to work with these
variables and, as we shall see,  the equations written in terms of these
quantities take relatively simple forms that generalize naturally the
nonrelativistic
ones.

Substituting Eq.
(\ref{kgp12}) in Eqs. (\ref{kge4}) and
(\ref{kge12}), we obtain
\begin{eqnarray}
i\hbar\frac{\partial\psi}{\partial t}-\frac{\hbar^2}{2m
c^2}\frac{\partial^2\psi}{\partial t^2}-\frac{3}{2}H\frac{\hbar^2}{m
c^2}\frac{\partial\psi}{\partial t}\nonumber\\
+\frac{\hbar^2}{2 m a^2}\left
(1+\frac{4\Phi}{c^2}\right )\Delta\psi-m\Phi \psi\nonumber\\
-\frac{4\pi a_s \hbar^2}{m^2}\left
(1+\frac{2\Phi}{c^2}\right )|\psi|^2\psi+\frac{3}{2}i\hbar
H\psi\nonumber\\
+\frac{2\hbar^2}{m c^4}\frac{\partial\Phi}{\partial t}\left
(\frac{\partial \psi}{\partial t}-\frac{i m c^2}{\hbar}\psi\right )=0,
\label{kge13}
\end{eqnarray}
\begin{eqnarray}
\frac{\Delta\Phi}{4\pi G a^2}=\left (1-\frac{\Phi}{c^2}\right
)|\psi|^2\nonumber\\
+\frac{\hbar^2}{2m^2c^4}\left (1-\frac{2\Phi}{c^2}\right )\left
|\frac{\partial\psi}{\partial t}\right |^2+\frac{\hbar^2}{2a^2m^2c^2}\left
(1+\frac{2\Phi}{c^2}\right )|\vec\nabla\psi|^2\nonumber\\
+\frac{2\pi a_s\hbar^2}{m^3c^2}|\psi|^4-\frac{\hbar}{m c^2}\left
(1-\frac{2\Phi}{c^2}\right ){\rm Im} \left (\frac{\partial\psi}{\partial
t}\psi^*\right )  \nonumber\\ -\frac{3H^2}{8\pi G}+\frac{3H}{4\pi G c^2}\left
(\frac{\partial\Phi}{\partial t}+H\Phi\right ).\nonumber\\
\label{kge14}
\end{eqnarray}
Eq. (\ref{kge13})
can be interpreted as a generalized
Schr\"odinger equation (in the absence of self-interaction $a_s=0$) or
as a generalized GP equation (in the presence of self-interaction
$a_s\neq 0$). It is coupled to the Einstein equation (\ref{kge14}).

The energy density and the pressure can be written as
\begin{eqnarray}
\epsilon=\frac{\hbar^2}{2m^2c^2}\left (1-\frac{2\Phi}{c^2}\right )\left
|\frac{\partial\psi}{\partial t}\right |^2\nonumber\\
-\frac{\hbar}{m}\left
(1-\frac{2\Phi}{c^2}\right ){\rm Im} \left
(\frac{\partial\psi}{\partial
t}\psi^*\right )\nonumber\\
+\frac{\hbar^2}{2a^2m^2}\left (1+\frac{2\Phi}{c^2}\right
)|\vec\nabla\psi|^2\nonumber\\
+\frac{1}{2}c^2\left (1-\frac{2\Phi}{c^2}\right )|\psi|^2+V(|\psi|^2),
\label{kgp14b}
\end{eqnarray}
\begin{eqnarray}
P=\frac{\hbar^2}{2m^2c^2}\left (1-\frac{2\Phi}{c^2}\right )\left
|\frac{\partial\psi}{\partial t}\right |^2\nonumber\\
-\frac{\hbar}{m}\left
(1-\frac{2\Phi}{c^2}\right ){\rm Im} \left
(\frac{\partial\psi}{\partial
t}\psi^*\right )\nonumber\\
-\frac{\hbar^2}{6a^2m^2}\left (1+\frac{2\Phi}{c^2}\right
)|\vec\nabla\psi|^2\nonumber\\
+\frac{1}{2}c^2\left (1-\frac{2\Phi}{c^2}\right )|\psi|^2-V(|\psi|^2).
\label{kgp14c}
\end{eqnarray}

Eqs. (\ref{kge13}) and (\ref{kge14}) form the  GPE equations.
In the nonrelativistic limit $c\rightarrow +\infty$, we recover
the GPP equations (\ref{nr1})-(\ref{nr2}) of Appendix
\ref{sec_nr}.

\subsection{The hydrodynamic representation}

Important characteristics of the system are revealed  by reformulating the
KGE equations in the form of hydrodynamic equations. This can
be done
at the level of the GPE equations
(\ref{kge13})-(\ref{kge14}) via the Madelung  transformation \cite{madelung}. To
that purpose, we write the pseudo wavefunction $\psi$ as
\begin{eqnarray}
\psi(\vec{x},t)=\sqrt{\rho(\vec{x},t)} e^{iS(\vec{x},t)/\hbar},\label{kgp15}
\end{eqnarray}
where $\rho=\psi \psi^*=|\psi|^2$ plays the role of a pseudo rest-mass density
and $S=(1/2)i\hbar\ln(\psi^*/\psi)$ plays
the role of a pseudo action. Following
Madelung, we also define a pseudo velocity field as
\begin{eqnarray}
\vec{v}(\vec{x},t)=\frac{\vec\nabla S}{ma},
\label{kgp16}
\end{eqnarray}
where the scale factor $a$ has been introduced in order to
take into account the expansion of the Universe \cite{abrilMNRAS}. We
note that this velocity field is irrotational.

Substituting Eqs. (\ref{kgp15})-(\ref{kgp16}) in the 
GPE equations (\ref{kge13})-(\ref{kge14}), and 
separating real and imaginary
parts, we obtain the system of hydrodynamic equations
\begin{eqnarray}
\frac{\partial\rho}{\partial t}+3H\rho+\frac{1}{a}\vec\nabla\cdot (\rho
{\vec v})=\frac{1}{mc^2}\frac{\partial }{\partial t}\left (\rho \frac{\partial
S}{\partial t}\right )\nonumber\\
+\frac{3H\rho}{mc^2}\frac{\partial S}{\partial t}+
\frac{4\rho}{mc^4}\frac{\partial\Phi}{\partial t}\left (mc^2-\frac{\partial
S}{\partial t}\right )-\frac{4\Phi}{ac^2}\vec\nabla\cdot (\rho {\vec
v}),\nonumber\\
\label{kge15}
\end{eqnarray}
\begin{eqnarray}
\frac{\partial S}{\partial t}+\frac{({\vec\nabla} S)^2}{2 m
a^2}=-\frac{\hbar^2}{2 m
c^2}\frac{\frac{\partial^2\sqrt{\rho}}{\partial
t^2}}{\sqrt{\rho}}\nonumber\\
+\left (1+\frac{4\Phi}{c^2}\right )\frac{\hbar^2}{2
m a^2}\frac{\Delta\sqrt{\rho}}{\sqrt{\rho}}-\frac{2\Phi}{m c^2
a^2}({\vec\nabla} S)^2\nonumber\\
-m\Phi-\frac{4\pi a_s \hbar^2\rho}{m^2}\left
(1+\frac{2\Phi}{c^2}\right )\nonumber\\
+\frac{1}{2 mc^2}\left (\frac{\partial
S}{\partial t}\right )^2-\left (3H-\frac{4}{c^2}\frac{\partial\Phi}{\partial
t}\right )\frac{\hbar^2}{4 m c^2 \rho}\frac{\partial\rho}{\partial t},
\label{kge18}
\end{eqnarray}
\begin{eqnarray}
\frac{\partial {\vec v}}{\partial t}+H{\vec v}+\frac{1}{a}({\vec v}\cdot
\vec\nabla){\vec v}=-\frac{\hbar^2}{2am^2c^2}\vec\nabla \left
(\frac{\frac{\partial^2\sqrt{\rho}}{\partial t^2}}{\sqrt{\rho}}\right
)\nonumber\\
+\frac{\hbar^2}{2m^2a^3}\vec\nabla\left\lbrack \left (1+\frac{4\Phi}{c^2}\right
)\frac{\Delta\sqrt{\rho}}{\sqrt{\rho}}\right\rbrack-\frac{1}{a}
\vec\nabla\Phi-\frac{
1}{\rho a}\vec\nabla p\nonumber\\
-\frac{8\pi a_s\hbar^2}{a m^3c^2}\vec\nabla (\rho\Phi)-\frac{2}{a c^2}\vec\nabla
(\Phi
v^2)+\frac{1}{2am^2c^2}\vec\nabla \left\lbrack\left (\frac{\partial S}{\partial
t}\right )^2\right \rbrack\nonumber\\
-\frac{3\hbar^2}{4 a m^2 c^2} H\vec\nabla \left
(\frac{1}{\rho}\frac{\partial\rho}{\partial t}\right )+\frac{\hbar^2}{a m^2
c^4}\vec\nabla \left (\frac{\partial\Phi}{\partial
t}\frac{1}{\rho}\frac{\partial\rho}{\partial t}\right ),\nonumber\\
\label{kge16}
\end{eqnarray}
\begin{eqnarray}
\frac{\Delta\Phi}{4\pi G a^2}=\left (1-\frac{\Phi}{c^2}\right )\rho\nonumber\\
+\frac{\hbar^2}{2m^2c^4}\left (1-\frac{2\Phi}{c^2}\right )\left\lbrack
\frac{1}{4\rho}\left (\frac{\partial\rho}{\partial t}\right
)^2+\frac{\rho}{\hbar^2}\left (\frac{\partial S}{\partial t}\right
)^2\right\rbrack\nonumber\\
+\frac{\hbar^2}{2a^2m^2c^2}\left (1+\frac{2\Phi}{c^2}\right )\left\lbrack
\frac{1}{4\rho}(\vec\nabla\rho)^2+\frac{\rho}{\hbar^2}(\vec\nabla
S)^2\right\rbrack
\nonumber\\
+\frac{2\pi a_s \hbar^2}{m^3c^2}\rho^2-\frac{1}{m c^2}\left
(1-\frac{2\Phi}{c^2}\right )\rho\frac{\partial S}{\partial t}\nonumber\\
-\frac{3H^2}{8\pi G}+\frac{3H}{4\pi G c^2}\left
(\frac{\partial\Phi}{\partial
t}+H\Phi\right ),\nonumber\\
\label{kge17}
\end{eqnarray}
where $p$ is a pseudo pressure
given by the
polytropic (quadratic) equation of
state\footnote{The equation of state $p(\rho)$ associated
with a general potential of interaction of the form $V(|\varphi|^2)$ is given
in \cite{proceedings}.}
\begin{eqnarray}
p=\frac{2\pi a_s\hbar^2}{m^3}\rho^2.
\label{kgp20}
\end{eqnarray}
We note that this equation of state coincides with the equation
of state of a nonrelativistic BEC with a self-interaction \cite{revuebec}.
This coincidence is not obvious because Eqs. (\ref{kge15})-(\ref{kge17}) are
valid in the relativistic regime. The
interpretation of this equation of state is, however, not direct because $\rho$
and $p$ are a pseudo density and a pseudo pressure that coincide with the real
density
and the real pressure of a BEC only in the nonrelativistic limit $c\rightarrow
+\infty$.

The energy density and the
pressure can be written in terms of hydrodynamic
variables as
\begin{eqnarray}
\epsilon=\frac{\hbar^2}{2m^2c^2}\left (1-\frac{2\Phi}{c^2}\right )\left\lbrack
\frac{1}{4\rho}\left (\frac{\partial\rho}{\partial t}\right
)^2+\frac{\rho}{\hbar^2}\left (\frac{\partial S}{\partial t}\right
)^2\right\rbrack\nonumber\\
+\frac{\hbar^2}{2a^2m^2}\left (1+\frac{2\Phi}{c^2}\right )\left\lbrack
\frac{1}{4\rho}(\vec\nabla\rho)^2+\frac{\rho}{\hbar^2}(\vec\nabla
S)^2\right\rbrack\nonumber\\
-\left (1-\frac{2\Phi}{c^2}\right )\frac{\rho}{m}\frac{\partial S}{\partial
t}+\frac{1}{2}\left (1-\frac{2\Phi}{c^2}\right )\rho c^2+V(\rho),
\label{kgp22}
\end{eqnarray}
\begin{eqnarray}
P=\frac{\hbar^2}{2m^2c^2}\left (1-\frac{2\Phi}{c^2}\right )\left\lbrack
\frac{1}{4\rho}\left (\frac{\partial\rho}{\partial t}\right
)^2+\frac{\rho}{\hbar^2}\left (\frac{\partial S}{\partial t}\right
)^2\right\rbrack\nonumber\\
-\frac{\hbar^2}{6a^2m^2}\left (1+\frac{2\Phi}{c^2}\right )\left\lbrack
\frac{1}{4\rho}(\vec\nabla\rho)^2+\frac{\rho}{\hbar^2}(\vec\nabla
S)^2\right\rbrack\nonumber\\
-\left (1-\frac{2\Phi}{c^2}\right )\frac{\rho}{m}\frac{\partial S}{\partial
t}+\frac{1}{2}\left (1-\frac{2\Phi}{c^2}\right )\rho c^2-V(\rho).
\label{kgp23}
\end{eqnarray}
We note that, in general, the pressure $P$ defined by Eq. (\ref{kgp23}) is
different from the
pressure $p$ defined by Eq. (\ref{kgp20}). However, we shall
find that
they coincide for a spatially homogeneous SF.

The hydrodynamic equations  (\ref{kge15})-(\ref{kge17}) have a clear physical
interpretation. Eq. (\ref{kge15}), corresponding to the imaginary
part of the GPE equations, is the continuity equation. We note that
$\int \rho\, d^3x$ is not conserved in the relativistic regime. However, we
will see in Sec. \ref{sec_b} that Eq. (\ref{kge15}) is consistent with the
conservation of the charge of a spatially homogeneous SF.  Eq. (\ref{kge18}),
corresponding to the real part of the GPE equations,
is the Bernoulli or Hamilton-Jacobi equation. Eq. (\ref{kge16}), obtained by
taking the gradient of  Eq. (\ref{kge18}), is the momentum equation. Eq.
(\ref{kge17}) is the Einstein equation. We stress
that the hydrodynamic equations  (\ref{kge15})-(\ref{kge17}) are equivalent to
the GPE equations (\ref{kge13})-(\ref{kge14}) which are themselves equivalent
to the 
KGE equations (\ref{kge4}) and (\ref{kge12}). In the nonrelativistic limit
$c\rightarrow +\infty$, we recover
the quantum Euler-Poisson equations (\ref{nr3})-(\ref{nr5}) of
Appendix \ref{sec_nr}.

\section{The case of a static Universe: Jeans-type instability}
\label{sec_jeans}

We first consider the case of a static Universe. This amounts
to taking $H=0$ and $a=1$ in the previous equations.

\subsection{The fluid equations}

In a static Universe, the hydrodynamic equations  (\ref{kge15})-
(\ref{kge17}) reduce to
\begin{eqnarray}
\frac{\partial\rho}{\partial t}+\vec\nabla\cdot (\rho
{\vec v})=\frac{1}{mc^2}\frac{\partial }{\partial t}\left (\rho \frac{\partial
S}{\partial t}\right )\nonumber\\
+\frac{4\rho}{mc^4}\frac{\partial\Phi}{\partial t}\left (mc^2-\frac{\partial
S}{\partial t}\right )-\frac{4\Phi}{c^2}\vec\nabla\cdot (\rho {\vec
v}),
\label{kge15s}
\end{eqnarray}
\begin{eqnarray}
\frac{\partial S}{\partial t}+\frac{({\vec\nabla} S)^2}{2 m}=-\frac{\hbar^2}{2 m
c^2}\frac{\frac{\partial^2\sqrt{\rho}}{\partial
t^2}}{\sqrt{\rho}}\nonumber\\
+\left (1+\frac{4\Phi}{c^2}\right )\frac{\hbar^2}{2
m}\frac{\Delta\sqrt{\rho}}{\sqrt{\rho}}-\frac{2\Phi}{m c^2}({\vec \nabla}
S)^2\nonumber\\
-m\Phi-\frac{4\pi a_s \hbar^2\rho}{m^2}\left
(1+\frac{2\Phi}{c^2}\right )\nonumber\\
+\frac{1}{2 mc^2}\left (\frac{\partial
S}{\partial t}\right )^2+\frac{\hbar^2}{m c^4 \rho}\frac{\partial\rho}{\partial
t}\frac{\partial\Phi}{\partial
t},
\label{kge18s}
\end{eqnarray}
\begin{eqnarray}
\frac{\partial {\vec v}}{\partial t}+({\vec v}\cdot
\vec\nabla){\vec v}=-\frac{\hbar^2}{2m^2c^2}\vec\nabla \left
(\frac{\frac{\partial^2\sqrt{\rho}}{\partial t^2}}{\sqrt{\rho}}\right
)\nonumber\\
+\frac{\hbar^2}{2m^2}\vec\nabla\left\lbrack \left (1+\frac{4\Phi}{c^2}\right
)\frac{\Delta\sqrt{\rho}}{\sqrt{\rho}}\right\rbrack-
\vec\nabla\Phi-\frac{
1}{\rho}\vec\nabla p\nonumber\\
-\frac{8\pi a_s\hbar^2}{m^3c^2}\vec\nabla (\rho\Phi)-\frac{2}{c^2}\vec\nabla
(\Phi
v^2)+\frac{1}{2m^2c^2}\vec\nabla \left\lbrack\left (\frac{\partial S}{\partial
t}\right )^2\right \rbrack\nonumber\\
+\frac{\hbar^2}{m^2
c^4}\vec\nabla \left (\frac{\partial\Phi}{\partial
t}\frac{1}{\rho}\frac{\partial\rho}{\partial t}\right ),\nonumber\\
\label{kge16s}
\end{eqnarray}
\begin{eqnarray}
\frac{\Delta\Phi}{4\pi G}=\left (1-\frac{\Phi}{c^2}\right )\rho\nonumber\\
+\frac{\hbar^2}{2m^2c^4}\left (1-\frac{2\Phi}{c^2}\right )\left\lbrack
\frac{1}{4\rho}\left (\frac{\partial\rho}{\partial t}\right
)^2+\frac{\rho}{\hbar^2}\left (\frac{\partial S}{\partial t}\right
)^2\right\rbrack\nonumber\\
+\frac{\hbar^2}{2m^2c^2}\left (1+\frac{2\Phi}{c^2}\right )\left\lbrack
\frac{1}{4\rho}(\vec\nabla\rho)^2+\frac{\rho}{\hbar^2}(\vec\nabla
S)^2\right\rbrack
\nonumber\\
+\frac{2\pi a_s \hbar^2}{m^3c^2}\rho^2-\frac{1}{m c^2}\left
(1-\frac{2\Phi}{c^2}\right )\rho\frac{\partial S}{\partial t}.\nonumber\\
\label{kge17s}
\end{eqnarray}

\subsection{Infinite homogeneous background}

We consider  a homogeneous SF at rest extending in an infinite space. We
have $\rho(\vec x,t)=\rho_b$, $\vec
v_b(\vec x,t)=\vec 0$, and $\Phi_b(\vec x,t)=0$. We must be careful, however,
that the phase can depend on time: $S_b(\vec x,t)=S_b(t)$.
Actually, the equation of continuity (\ref{kge15s}) implies that $dS_b/dt$ is a
constant that
we write $-E$ as it represents the energy of the homogeneous SF
(with the opposite sign). Indeed, for a stationary state, the pseudo
wavefunction of the SF writes $\psi_b(\vec x,t)=\sqrt{\rho_b}e^{-i
Et/\hbar}$, so
that $S_b(t)=-Et$. The energy of the SF is given by
the Hamilton-Jacobi equation (\ref{kge18s}) which reduces to
\begin{eqnarray}
\frac{E^2}{2 m c^2}+E-\frac{4\pi
a_s \hbar^2\rho_b}{m^2}=0,
\label{j5}
\end{eqnarray}
yielding
\begin{eqnarray}
E=mc^2\left\lbrack -1+\sqrt{1+\frac{8\pi
a_s\hbar^2\rho_b}{m^3c^2}}\right\rbrack.
\label{j6}
\end{eqnarray}

From Eqs. (\ref{kgp22}) and (\ref{kgp23}), the energy density and the pressure
of the SF are
\begin{eqnarray}
\epsilon_b=\rho_b c^2
+\frac{\rho_b}{m}\left (\frac{E^2}{2 m c^2}+E\right )+\frac{2\pi a_s
\hbar^2}{m^3}\rho_b^2,
\label{j8w}
\end{eqnarray}
\begin{eqnarray}
P_b=\frac{\rho_b}{m}\left (\frac{E^2}{2 m c^2}+E\right )-\frac{2\pi
a_s
\hbar^2}{m^3}\rho_b^2.
\label{j8b}
\end{eqnarray}
Combining Eq. (\ref{j5}) with Eqs. (\ref{j8w}) and (\ref{j8b}), we obtain
\begin{eqnarray}
\epsilon_b=\rho_b c^2+\frac{6\pi a_s
\hbar^2}{m^3}\rho_b^2,\qquad P_b=\frac{2\pi a_s
\hbar^2}{m^3}\rho_b^2.
\label{j9}
\end{eqnarray}
We note that the pressure $P_b$ of a spatially homogeneous SF
coincides with the pseudo pressure $p$ given by Eq. (\ref{kgp20}).
We also note the identities
\begin{eqnarray}
\left (\frac{E}{mc^2}+1\right
)^2=1+\frac{2c_s^2}{c^2},
\label{j10}
\end{eqnarray}
\begin{eqnarray}
\left (\frac{E}{2mc^2}+1\right
)\frac{E}{mc^2}=\frac{c_s^2}{c^2},
\label{j11}
\end{eqnarray}
where
\begin{eqnarray}
c_s^2=P'(\rho_b)=\frac{4\pi a_s\hbar^2\rho_b}{m^3}
\label{j11b}
\end{eqnarray}
is the square of the speed
of sound in the homogeneous background.  In terms of the speed of sound, the
energy
of the SF can be written
as
\begin{eqnarray}
\frac{E}{mc^2}=-1+\sqrt{1+\frac{2c_s^2}{c^2}}.
\label{j12}
\end{eqnarray}

{\it Remark:} The expression (\ref{j6}) of the energy $E$ of the SF can be
directly obtained from the GP equation (\ref{kge13}) with $H=0$, $a=1$, and
$\Phi_b=0$ by looking for a stationary solution of the form $\psi_b(\vec
x,t)=\sqrt{\rho_b}e^{-i Et/\hbar}$.  Then, using Eq.
(\ref{kgp12}), implying
$\varphi_b(\vec x,t)=\frac{\hbar}{m}\sqrt{\rho_b}e^{-i(mc^2+E)t/\hbar}$,
we find that the total energy of the SF, including its rest mass, is
$E_{\rm tot}=E+mc^2$. Using Eq. (\ref{j6}), we obtain
\begin{eqnarray}
E_{tot}=mc^2\sqrt{1+\frac{8\pi
a_s\hbar^2\rho_b}{m^3c^2}}.
\label{j7}
\end{eqnarray}

\subsection{Linear wave equations}

We now slightly perturb the homogeneous SF and consider the evolution
of the perturbations in the linear regime. This is the relativistic SF
generalization of the classical Jeans problem \cite{jeans}. As usual, we make
the Jeans swindle \cite{bt}. We write $\rho(\vec
x,t)=\rho_b+\delta\rho(\vec
x,t)$ and $S(\vec
x,t)=-Et+\delta S(\vec
x,t)$ with $\delta\rho(\vec
x,t)\ll 1$
and $\partial\delta S/\partial t(\vec
x,t)\ll 1$. We also recall that $|\vec v(\vec
x,t)|\ll 1$ and $\Phi(\vec
x,t)\ll 1$.
Introducing the notations  
\begin{eqnarray}
\delta=\frac{\delta\rho}{\rho_b},\qquad \sigma=\frac{1}{m}\frac{\partial
\delta S}{\partial t},
\label{j8}
\end{eqnarray}
we can write the linearized  equations for the perturbations as
\begin{eqnarray}
\left (1+\frac{E}{mc^2}\right )\frac{\partial\delta}{\partial
t}+\vec\nabla\cdot\vec{v}=\frac{1}{c^2}\frac{\partial\sigma}{\partial
t}+\frac{4}{c^2}\frac{\partial\Phi}{\partial
t}\left (1+\frac{E}{mc^2}\right ),\nonumber\\
\label{kge19}
\end{eqnarray}
\begin{eqnarray}
\left (1+\frac{E}{mc^2}\right
)\sigma=-\frac{\hbar^2}{4m^2c^2}\frac{\partial^2\delta}{\partial
t^2}+\frac{\hbar^2} {4m^2}\Delta\delta\nonumber\\
-c_s^2\delta-\left
(1+2\frac{c_s^2}{c^2}\right )\Phi,
\label{kge22}
\end{eqnarray}
\begin{eqnarray}
\left (1+\frac{E}{mc^2}\right )\frac{\partial\vec{v}}{\partial
t}=-\frac{\hbar^2}{4m^2c^2}\vec\nabla\left
(\frac{\partial^2\delta}{\partial t^2}\right
)+\frac{\hbar^2} { 4m^2}\vec\nabla(\Delta\delta)\nonumber\\
-c_s^2\vec\nabla\delta-\left
(1+2\frac{c_s^2}{c^2}\right ) \vec\nabla\Phi,\qquad
\label{kge20}
\end{eqnarray}
\begin{eqnarray}
\frac{\Delta\Phi}{4\pi
G\rho_b}=\left(1+\frac{c_s^2}{c^2}\right)\delta-\frac{1}{c^2}\left
(1+\frac{E}{mc^2}\right )\sigma-\frac{\Phi}{c^2}\nonumber\\
+\frac{E}{mc^2}\left
(1+\frac{E}{2mc^2}\right )\delta-\frac{2E}{mc^2}\frac{\Phi}{c^2}\left
(1+\frac{E}{2mc^2}\right ).
\label{kge21}
\end{eqnarray}

Taking the time derivative of Eq. (\ref{kge19}) and using
$\frac{\partial}{\partial t}\nabla\cdot \vec v=\Delta\sigma$,
we get
\begin{eqnarray}
\frac{1}{c^2}\frac{\partial^2\sigma}{\partial
t^2}-\Delta\sigma=\left (1+\frac{E}{mc^2}\right
)\frac{\partial^2\delta}{\partial
t^2}-\frac{4}{c^2}\frac{\partial^2\Phi}{\partial t^2}\left
(1+\frac{E}{mc^2}\right
).\nonumber\\
\label{kge23}
\end{eqnarray}
Taking the Laplacian of Eq. (\ref{kge22}), we obtain
\begin{eqnarray}
\left (1+\frac{E}{mc^2}\right
)\Delta\sigma&=&\frac{\hbar^2}{4m^2}\Delta\left(\Delta\delta-\frac{1}{c^2}\frac{
\partial^2\delta}{\partial t^2}\right)-c_s^2\Delta\delta\nonumber\\
&-&\left(1+\frac{2c_s^2}{c^2}\right)\Delta\Phi.
\label{kge24}
\end{eqnarray}
Eqs. (\ref{kge23}) and (\ref{kge24}), together with Eq. (\ref{kge21}),  form a
system of three coupled wave equations for the evolution of the
 perturbations ($\delta$, $\sigma$, $\Phi$) in the linear regime.

\subsection{Dispersion relation and Jeans length}
\label{sec_di}

Decomposing the perturbations in Fourier modes of the form
$\delta(\vec x,t)=\delta_k\exp{i(\vec{k}\cdot\vec{r}-\omega t)}$, $\sigma(\vec
x,t)=\sigma_k\exp{i(\vec{k} \cdot\vec{r}-\omega t)}$, and $\Phi(\vec
x,t)=\Phi_k\exp{i(\vec{k}\cdot\vec{r}-\omega t)}$, we obtain after
straightforward but lenghty
calculations the dispersion relation
\begin{eqnarray}
\frac{\hbar^2}{4m^2c^4}\omega^4-\left\lbrack
\frac{1+\gamma}{3\gamma+1}\frac{\hbar^2k^2}{2
m^2c^2}+1+\frac{3c_s^2}{c^2}\right\rbrack\omega^2\nonumber\\
+\frac{1}{1+3\gamma}\left\lbrack
(1-\gamma)\frac{\hbar^2k^4}{4m^2}-\gamma k^2 c^2+(1-3\gamma) k^2 c_s^2
\right\rbrack=0,\nonumber\\
\label{kge25}
\end{eqnarray}
where we have introduced the abbreviation
\begin{eqnarray}
\gamma=\frac{4\pi G\rho_b}{k^2c^2}\left (1+\frac{2c_s^2}{c^2}\right ).
\label{kge26}
\end{eqnarray}
In the nonrelativistic limit $c\rightarrow +\infty$, the dispersion relation
reduces to \cite{prd1}:
\begin{eqnarray}
\omega^2=\frac{\hbar^2k^4}{4m^2}+c_s^2k^2-4\pi
G\rho_b.
\label{nr13}
\end{eqnarray}

\begin{figure}[!ht]
\includegraphics[width=0.98\linewidth]{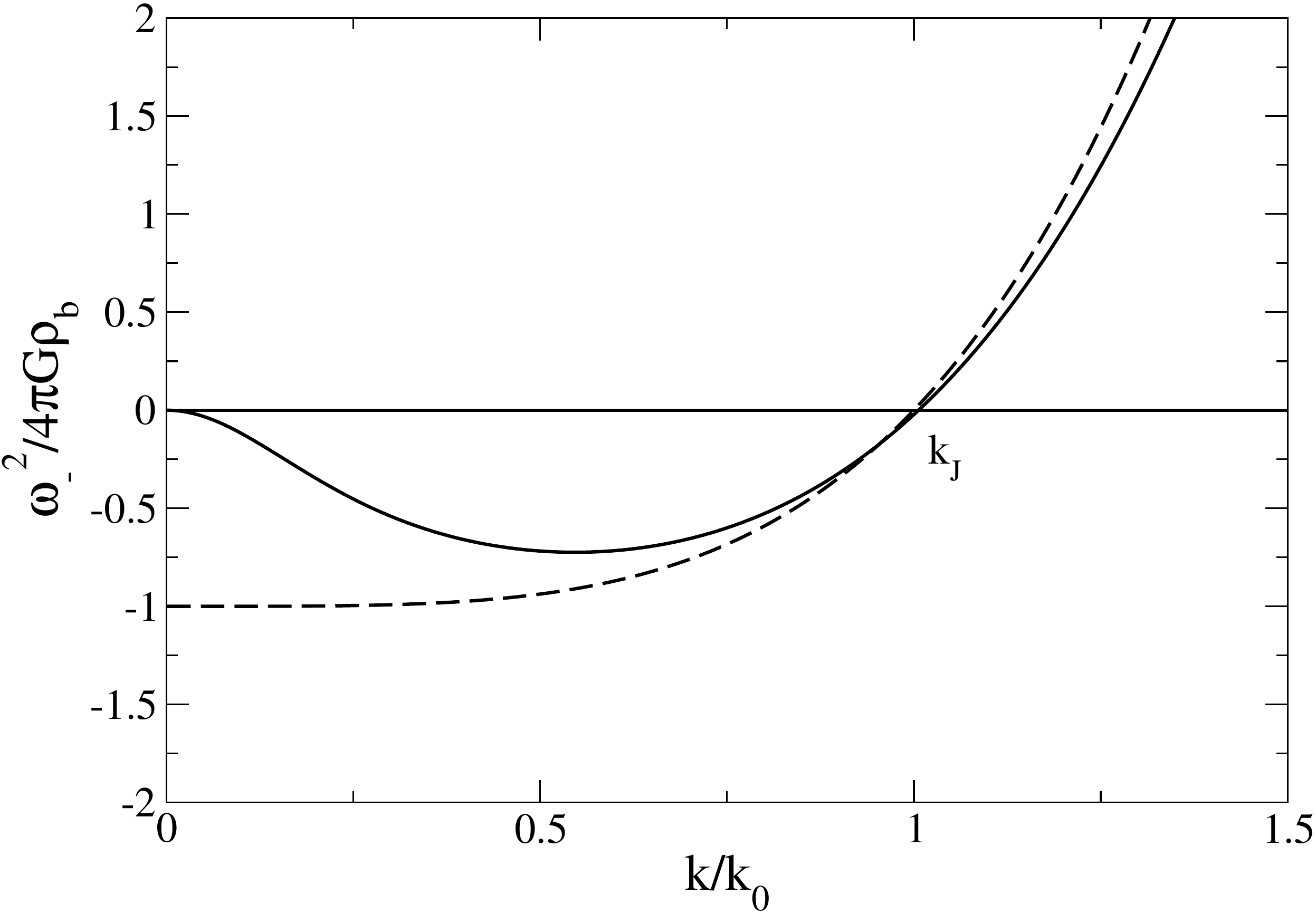}
\caption{Relativistic (solid line) and nonrelativistic (dashed line)  dispersion
relation. Here $k_0=(16\pi Gm^2\rho_b/\hbar^2)^{1/4}$. We have
taken $a_s=0$
and $\chi=(4\pi G \hbar^2\rho_b/m^2c^4)^{1/2}=0.05$.
\label{dispersionCHI0p05}}
\end{figure}

Taking $\omega=0$ in Eq. (\ref{kge25}), we find that the Jeans length $k_J$ is
determined by the equation
\begin{eqnarray}
\frac{\hbar^2k_J^4}{4m^2}+\left\lbrack c_s^2-\frac{\pi \hbar^2
G\rho_b}{m^2c^2}\left (1+\frac{2c_s^2}{c^2}\right )\right\rbrack
k_J^2\nonumber\\
-4\pi G\rho_b \left (1+\frac{2c_s^2}{c^2}\right ) \left
(1+\frac{3c_s^2}{c^2}\right )=0.
\label{kge27}
\end{eqnarray}
This is a second degree equation in $k_J^2$. Only the solution with the sign $+$
has $k_J^2>0$. In the nonrelativistic limit  $c\rightarrow +\infty$,  Eq.
(\ref{kge27})
reduces to \cite{prd1}:
\begin{eqnarray}
\frac{\hbar^2k_J^4}{4m^2}+c_s^2k_J^2-4\pi
G\rho_b=0.
\label{nr14}
\end{eqnarray}

Eq. (\ref{kge25}) is a second degree equation in $\omega^2$. We can show that
the discriminant is positive so the relativistic dispersion relation has two
branches $\omega_{\pm}^2(k)$. For $k\rightarrow 0$,
\begin{eqnarray}
\omega_+^2(k)\simeq \frac{4m^2c^4}{\hbar^2}\left
(1+\frac{3c_s^2}{c^2}\right )+k^2c^2+...,
\label{dis1}
\end{eqnarray}
\begin{eqnarray}
\omega_-^2(k)\sim -\frac{1}{3}k^2c^2.
\label{dis2}
\end{eqnarray}
For $k\rightarrow +\infty$, 
\begin{eqnarray}
\omega_{\pm}^2(k)\sim k^2c^2.
\label{dis3}
\end{eqnarray}
The branch $\omega_+^2(k)$ is always positive, corresponding to stable
(oscillating) modes. The branch $\omega_-^2(k)$ starts from $0$, decreases,
reaches a minimum, increases, vanishes at $k=k_J$, and finally tends to
$+\infty$ (see Fig. \ref{dispersionCHI0p05}). The modes are stable (oscillating)
for $k>k_J$  and unstable (growing) for $k<k_J$. However, in the relativistic
case, the system is stabilized at very large scales since
$\omega_-^2(k)\rightarrow 0$ for $k\rightarrow 0$ while in the nonrelativistic
case $\omega^2(0)=-4\pi G\rho_b<0$. This shows that the limits $c\rightarrow
+\infty$ and $k\rightarrow 0$ do not commute. We shall encounter this
large-scale stabilization again in Secs. \ref{sec_appdelta} and 
\ref{sec_relat}.

\section{Cosmological evolution of a spatially homogeneous scalar field}
\label{sec_b}

We consider the evolution of a universe induced solely by a spatially
homogeneous
SF. In the comoving frame, we have $\rho(\vec x,t)=\rho_b(t)$,
$\vec v_b(\vec x,t)=\vec 0$, $\Phi_b(\vec x,t)=0$, and
$S_b(\vec x,t)=S_b(t)$. We introduce the notation $E(t)=-dS_b/dt$ which can
be considered as the time-dependent energy of the spatially homogeneous SF in
the comoving
frame. The wavefunction of the SF is $\psi_b(\vec
x,t)=\sqrt{\rho_b(t)}e^{-(i/\hbar)\int E(t)\, dt}$. Using Eq. (\ref{kgp12}),
we have $\varphi_b(\vec x,t)=\frac{\hbar}{m}\sqrt{\rho_b(t)}e^{-(i/\hbar)(m c^2
t + \int E(t)\, dt)}$ so the total energy of the SF, including its
rest mass, is $E_{\rm tot}(t)=E(t)+mc^2$.

\subsection{Hydrodynamic equations for a homogeneous scalar field}

For a spatially homogeneous SF, the hydrodynamic equations
(\ref{kge15})-(\ref{kge17})  reduce to
\begin{eqnarray}
\frac{d\rho_b}{dt}+3H\rho_b=-\frac{1}{mc^2}\frac{d}{dt}\left
(\rho_b E\right )-\frac{3H\rho_b} {
mc^2}E,
\label{b1}
\end{eqnarray}
\begin{eqnarray}
\left (\frac{E}{2 mc^2}+1\right )E=\frac{\hbar^2}{2 m
c^2}\frac{\frac{d^2\sqrt{\rho_b}}{d
t^2}}{\sqrt{\rho_b}}
+\frac{4\pi a_s \hbar^2\rho_b}{m^2}\nonumber\\
+\frac{3H\hbar^2}{4 m c^2
\rho_b}\frac{d\rho_b}{dt},
\label{b2}
\end{eqnarray}
\begin{eqnarray}
\frac{3H^2}{8\pi G}=\rho_b
+\frac{\hbar^2}{2m^2c^4}\left\lbrack
\frac{1}{4\rho_b}\left (\frac{d\rho_b}{d t}\right
)^2+\frac{\rho_b}{\hbar^2}E^2\right\rbrack\nonumber\\
+\frac{2\pi a_s \hbar^2}{m^3 c^2}\rho_b^2
+\frac{E}{m
c^2}\rho_b.
\label{b3}
\end{eqnarray}

In terms of the total energy $E_{\rm tot}(t)=E(t)+mc^2$, the equation of
continuity (\ref{b1}) becomes
\begin{eqnarray}
\frac{1}{\rho_b}\frac{d\rho_b}{dt}+\frac{3}{a}\frac{da}{dt}+\frac{1}{E_{\rm
tot}} \frac{dE_{\rm tot}}{dt}=0.
\label{b4}
\end{eqnarray}
It can be rewritten as a conservation law:
\begin{eqnarray}
\frac{d}{dt}(E_{\rm tot}\, \rho_b
a^3)=0. 
\label{b5}
\end{eqnarray}
Therefore, the total energy is exactly given by
\begin{eqnarray}
\frac{E_{\rm tot}}{mc^2}=\frac{Qm}{\rho_b a^3},
\label{b6}
\end{eqnarray}
where $Q$ is a constant. This conservation law was found by Gu and Hwang
\cite{gh} directly from the KG equation (see also Appendix B of
\cite{shapiro}). It can be shown that $Q=\int J^0\, d^3x$ represents the
conserved charge
density of the complex SF.  

The energy density and the
pressure of a homogeneous SF are
\begin{eqnarray}
\epsilon_b=\frac{\hbar^2}{8m^2c^2}\frac{1}{\rho_b}\left
(\frac{d\rho_b}{dt}\right
)^2+\frac{\rho_b}{m}E\left (1+\frac{E}{2mc^2}\right
)\nonumber\\
+\rho_b c^2+\frac{2\pi a_s\hbar^2}{m^3}\rho_b^2,
\label{kgp22h}
\end{eqnarray}
\begin{eqnarray}
P=\frac{\hbar^2}{8m^2c^2}\frac{1}{\rho_b}\left
(\frac{d\rho_b}{dt}\right
)^2+\frac{\rho_b}{m}E\left (1+\frac{E}{2mc^2}\right
)\nonumber\\
-\frac{2\pi a_s\hbar^2}{m^3}\rho_b^2.
\label{kgp23h}
\end{eqnarray}

Equations (\ref{b2}), (\ref{b3}) and (\ref{b6}) determine the complete
evolution of a
universe induced by a spatially homogeneous SF.\footnote{A remark may be in
order. A SF
$\varphi_b(t)$ oscillates in time because it has a phase
$S_b(t)$. The kinetic and potential energies $\dot\varphi_b^2$ and
$V(\varphi_b^2)$ of a real SF also present oscillations because they keep track
of the phase $S_b(t)$ of the SF
\cite{mvm,abrilMNRAS,abrilJCAP}. By contrast, the kinetic and potential energies
 $|\dot\varphi_b|^2$ and $V(|\varphi_b|^2)$ of a complex SF do not oscillate in
time. It is not clear how one
can measure the oscillations of a complex SF because we do not have
directly access to field variables such as  $\varphi_b(t)$ but rather to
hydrodynamic variables such as $\epsilon_b(t)$, $\rho_b(t)$, and $P_b(t)$. In
this sense, the hydrodynamic representation of the SF may be more physical than
the KG equation itself.} Working directly on
the homogeneous KG equation, Li {\it et al.}  \cite{shapiro} have shown that
a universe filled with a relativistic complex SF first undergoes an intrinsic
stiff matter era, followed by a radiation era due to its self-interaction,
before finally entering in the matter era. The stiff matter era occurs when the 
SF oscillations are slower than the Hubble expansion while the radiation and
matter eras occur when the SF oscillations are faster than the Hubble expansion.
Since we propose in this paper to use
hydrodynamic equations instead of the KG equation, it is important to show,
for self-consistency, that these different regimes can be directly obtained from
the hydrodynamic equations (\ref{b2}), (\ref{b3}) and (\ref{b6}).
One
advantage of the hydrodynamic
representation is that we do not have to make
averagings over the oscillations of the SF as in \cite{shapiro}.

\subsection{Stiff
matter era}
\label{sec_stiff}

At early times ($a\rightarrow 0$, $t\rightarrow 0$),  we can
approximate Eqs. (\ref{b2}) and (\ref{b3}) by\footnote{We can check {\it a
posteriori} that the terms that have been neglected are indeed negligible when
$a\rightarrow 0$ and $t\rightarrow 0$.}
\begin{eqnarray}
\frac{\hbar^2}{m^2
c^4}\frac{\frac{d^2\sqrt{\rho_b}}{d
t^2}}{\sqrt{\rho_b}}
+\frac{3H\hbar^2}{2 m^2 c^4
\rho_b}\frac{d\rho_b}{dt}=0,
\label{b10}
\end{eqnarray}
\begin{eqnarray}
\frac{3H^2}{8\pi G}=\frac{\hbar^2}{8m^2c^4\rho_b}\left
(\frac{d\rho_b}{dt}\right )^2.
\label{b11}
\end{eqnarray}
Recalling that $H=\dot a/a$, Eq. (\ref{b11}) can be integrated into 
\begin{eqnarray}
\rho_b\sim \frac{3m^2c^4}{4\pi G\hbar^2}(-\ln a)^2.
\label{b12}
\end{eqnarray}
Substituting Eq. (\ref{b12}) in Eq. (\ref{b10}), we obtain 
\begin{eqnarray}
\frac{dH}{dt}+3H^2=0
\label{b13}
\end{eqnarray}
which yields
\begin{eqnarray}
H=\frac{1}{3t},\qquad a\propto t^{1/3}.
\label{b14}
\end{eqnarray}
Substituting Eq. (\ref{b14}) in Eq. (\ref{b12}) we find that the pseudo
rest-mass density behaves as
\begin{eqnarray}
\rho_b\sim \frac{m^2c^4}{12\pi G\hbar^2}(-\ln t)^2.
\label{b15}
\end{eqnarray}
According to Eq. (\ref{b6}) the energy of the
homogeneous SF behaves as
\begin{eqnarray}
\frac{E}{mc^2}\sim \frac{4\pi
QG\hbar^2}{3m c^4}\frac{1}{a^3(-\ln a)^2}\propto \frac{1}{(-\ln t)^2 t}.
\label{b16}
\end{eqnarray}
Finally, the energy density and the pressure of the homogeneous SF behave as
\begin{eqnarray}
P_b=\epsilon_b=\frac{3H^2c^2}{8\pi G}=\frac{c^2}{24\pi
Gt^2}\propto \frac{1}{a^6}.
\label{b18}
\end{eqnarray}

These results can be straightforwardly obtained from the KG
equation in the regime where the SF oscillations are slower than
the Hubble expansion \cite{shapiro}. Indeed, neglecting the potential energy in
Eqs. (\ref{kgp7b}) and (\ref{kgp8b}), we get $P=\epsilon$ which is the equation
of state of a stiff fluid for which the speed of sound
$c_s=\sqrt{P'(\epsilon)}c$ is equal to the speed of light ($c_s=c$).
Substituting this equation of state in the Friedmann equations (\ref{frid1})
and (\ref{kge12b}), we directly
obtain Eq. (\ref{b18}). Therefore, when
$t\rightarrow 0$, a relativistic SF behaves as a stiff fluid \cite{shapiro}. 
This stiff matter era is independent on the scattering length
$a_s$ of the bosons. Therefore, it exists for both self-interacting and
non-interacting
SFs.

\subsection{Radiation and matter eras}
\label{sec_radmatt}

At later times, following the stiff matter era, we can
approximate Eqs. (\ref{b2}) and (\ref{b3}) by\footnote{These equations
correspond to the terms that
have been neglected in Sec. \ref{sec_stiff} and that now become important.} 
\begin{eqnarray}
\left (\frac{E}{2 mc^2}+1\right )E=\frac{4\pi a_s
\hbar^2\rho_b}{m^2},
\label{b2b}
\end{eqnarray}
\begin{eqnarray}
\frac{3H^2}{8\pi G}=\rho_b
+\frac{2\pi a_s \hbar^2}{m^3 c^2}\rho_b^2
+\frac{E}{m
c^2}\left (\frac{E}{2 mc^2}+1\right )\rho_b.
\label{b3b}
\end{eqnarray}
Substituting Eq. (\ref{b2b}) in Eq. (\ref{b3b}), we obtain
\begin{eqnarray}
\frac{3H^2}{8\pi G}=\rho_b+\frac{6\pi a_s
\hbar^2}{m^3c^2}\rho_b^2.
\label{b21}
\end{eqnarray}
Eq. (\ref{b2b}) is the same as Eq.
(\ref{j5}) in the static
case, so we still have the identity (\ref{j6}),
where now $E$ and $\rho_b$ depend on time. Combining Eqs. (\ref{j6})
and (\ref{b6}) we
get
\begin{eqnarray}
\rho_b \left (1+\frac{8\pi a_s \hbar^2\rho_b}{m^3 c^2}\right
)^{1/2}=\frac{Qm}{a^3}.
\label{b22}
\end{eqnarray}
Equations (\ref{b21}) and (\ref{b22}) determine the evolution of the Universe
induced by the homogeneous SF in the regime where its oscillations
are faster than the Hubble expansion. 

The energy density and the pressure are given by
\begin{eqnarray}
\epsilon_b=\rho_bc^2+\frac{6\pi a_s
\hbar^2}{m^3}\rho_b^2, \qquad P_b=\frac{2\pi a_s
\hbar^2}{m^3}\rho_b^2.
\label{b21b}
\end{eqnarray}
We note that the pressure $P_b(t)$ of a spatially homogeneous scalar
field
coincides with the pseudo pressure $p(t)$ given by Eq.
(\ref{kgp20}). From Eq.
(\ref{b21b}) we get
\begin{eqnarray}
P_b=\frac{m^3c^4}{72\pi a_s\hbar^2}\left (\sqrt{1+\frac{24\pi
a_s\hbar^2}{m^3c^4}\epsilon_b}-1\right )^2.
\label{b21bc}
\end{eqnarray}
This is the equation of state obtained by Colpi {\it et al.} \cite{colpi} in
the context of boson stars (see also \cite{mu,chavharko,shapiro}).

We can now determine the asymptotic behavior of the homogeneous SF in
the regime where its oscillations are faster than the Hubble expansion, i.e. in
the regime that follows the stiff matter era.

(i) At late times $t\rightarrow +\infty$ ($\rho_b\rightarrow 0$, $a\rightarrow
+\infty$),
we get
\begin{eqnarray}
\rho_b\sim \frac{Qm}{a^3}, \qquad  \frac{E}{mc^2}\sim \frac{4\pi a_s \hbar^2
Q}{m^2c^2a^3}.
\label{b23}
\end{eqnarray}
\begin{eqnarray}
\frac{3H^2}{8\pi G}\sim \rho_b \sim \frac{Qm}{a^3},\qquad a\sim  (6\pi
GQmt^2)^{1/3},
\label{b24}
\end{eqnarray}
\begin{eqnarray}
\epsilon_b\sim \rho_bc^2, \qquad P_b\sim \frac{2\pi a_s
\hbar^2}{m^3c^4}\epsilon_b^2\simeq 0.
\label{b21c}
\end{eqnarray}
The SF/BEC universe behaves as a pressureless matter
fluid.

(ii) At early times $t\rightarrow 0$ ($\rho_b\rightarrow +\infty$, 
$a\rightarrow 0$), we
get
\begin{eqnarray}
\rho_b\sim \left (\frac{Q^2m^5c^2}{8\pi a_s\hbar^2}\right
)^{1/3}\frac{1}{a^2},\quad  \frac{E}{mc^2}\sim
\left
(\frac{8\pi a_s\hbar^2 Q}{m^2 c^2}\right
)^{1/3}\frac{1}{a},\nonumber\\
\label{b25}
\end{eqnarray}
\begin{eqnarray}
\frac{3H^2}{8\pi G}\sim \frac{6\pi a_s \hbar^2}{m^3
c^2}\rho_b^2
\sim \frac{3}{2}\left (\frac{Q^4 \pi ma_s\hbar^2}{c^2}\right
)^{1/3}\frac{1}{a^4},
\label{b26}
\end{eqnarray}
\begin{eqnarray}
a\sim 2\left (\frac{\pi^4 G^3 Q^4 m a_s \hbar^2}{c^2}\right )^{1/12}t^{1/2},
\label{b26b}
\end{eqnarray}
\begin{eqnarray}
\epsilon_b\sim \frac{6\pi a_s
\hbar^2}{m^3}\rho_b^2, \qquad P_b\sim \frac{1}{3}\epsilon_b.
\label{b21w}
\end{eqnarray}
The SF/BEC universe  behaves like radiation. As emphasized by Li
{\it et al.} \cite{shapiro}, the radiation era is due to the self-interaction
of the SF ($a_s\neq 0$). There is no such phase for a
non-interacting SF
($a_s=0$). 

{\it Remark:} the case of a SF with an attractive self-interaction ($a_s<0$) is
theoretically interesting because it leads to a model of universe
without
initial Big Bang singularity. At
$t=0$, the universe has a finite scale factor $a_*=(12\sqrt{3}\pi
Q|a_s|\hbar^2/m^2c^2)^{1/3}$ and a finite density
$\rho_*=m^3c^2/12\pi|a_s|\hbar^2$. Two evolutions are possible: (i) a normal
evolution in which the density decreases as the scale factor increases until the
system enters in the matter era, and (ii) a phantom-like evolution in which the
density increases as the scale factor increases until the universe enters in a
de Sitter era with a constant density $\rho_{max}=m^3c^2/8\pi|a_s|\hbar^2$. The
choice of the evolution between these two behaviors  cannot be decided {\it a
priori}. This interesting dynamical
system will be studied in a specific paper (in preparation).

\section{Evolution of the perturbations in the expanding Universe}
\label{sec_lp}

We now slightly perturb the homogeneous SF and consider the evolution
of the perturbations in the linear regime. 

\subsection{Linearized equations}

The linearized hydrodynamic equations obtained from Eqs. 
(\ref{kge15})-(\ref{kge17}) describing the evolution of small
inhomogeneities in
the expanding Universe are
\begin{eqnarray}
\left (1+\frac{E}{mc^2}\right )\frac{\partial\delta}{\partial
t}+\frac{1}{a}\vec\nabla\cdot
{\vec v}=\frac{1}{c^2}\frac{\partial\sigma}{\partial
t}\nonumber\\
+\frac{1}{c^2}\left
(\frac{1}{\rho_b}\frac{d\rho_b}{dt}+3H\right )\sigma
+\frac{4}{c^2}\frac{\partial\Phi}{\partial t}\left
(1+\frac{E}{mc^2}\right ),
\label{lp1}
\end{eqnarray}
\begin{eqnarray}
\left (1+\frac{E}{mc^2}\right )\sigma=\frac{\hbar^2}{4
m^2 a^2}\Delta\delta-c_s^2\delta\nonumber\\
-\frac{\hbar^2}{4 m^2
c^2}\frac{1}{\sqrt{\rho_b}}\left\lbrack\frac{\partial^2}{\partial
t^2}(\sqrt{\rho_b}\delta)-\delta\frac{d^2}{
dt^2}(\sqrt{\rho_b})\right\rbrack\nonumber\\
-\frac{3H\hbar^2}{4m^2c^2}\frac{\partial\delta}{\partial t}
-\left (1+\frac{2c_s^2}{c^2}\right
)\Phi+\frac{\hbar^2}{m^2c^4}\frac{1}{\rho_b}\frac{d\rho_b}{dt}\frac{\partial\Phi
}{\partial t},
\label{lp4}
\end{eqnarray}
\begin{eqnarray}
\frac{\partial {\vec v}}{\partial t}+H{\vec v}=
\frac{\hbar^2}{4m^2a^3}\vec\nabla(\Delta\delta)-\frac{1}{a}c_s^2 \vec\nabla
\delta\nonumber\\
-\frac{\hbar^2}{4 a m^2
c^2}\frac{1}{\sqrt{\rho_b}}\nabla\cdot\left\lbrack\frac{\partial^2}{\partial
t^2}(\sqrt{\rho_b}\delta)-\delta\frac{d^2}{
dt^2}(\sqrt{\rho_b})\right\rbrack\nonumber\\
-\frac{E}{a m c^2}\nabla\sigma-\frac{3\hbar^2H}{4 a
m^2 c^2}\nabla\left ( \frac{\partial\delta}{\partial t}\right )\nonumber\\
-\frac{1}{a}\left (1+\frac{2c_s^2}{c^2}\right )\vec\nabla\Phi+\frac{\hbar^2}{a
m^2 c^4}\frac{1}{\rho_b}\frac{d\rho_b}{dt}\nabla\frac{\partial\Phi}{\partial t},
\label{lp2}
\end{eqnarray}
\begin{eqnarray}
\frac{\Delta\Phi}{4\pi G \rho_b
a^2}=\left (1
+\frac{c_s^2}{c^2}\right )\delta -\frac{1}{c^2}\left (\frac{E}{m
c^2}+1\right )\sigma\nonumber\\
+\frac{E}{mc^2}\left
(\frac{E}{2mc^2}+1\right )\delta\nonumber\\
+\frac{\hbar^2}{8m^2c^4\rho_b^2}\left\lbrack
2\rho_b\frac{\partial\delta}{\partial t}+\delta\frac{d\rho_b}{dt}\right\rbrack
\frac{d\rho_b}{dt}\nonumber\\
-\frac{\Phi}{c^2}-\frac{\hbar^2}{4m^2c^4}\frac{\Phi}{c^2}\frac{1}{\rho_b^2}
\left (\frac{d\rho_b}{dt}\right )^2\nonumber\\
-\frac{E}{mc^2}\left
(\frac{E}{2mc^2}+1\right )\frac{2\Phi}{c^2}+\frac{3H}{4\pi G\rho_b
c^2}\left (\frac{\partial\Phi}{\partial t}+H\Phi\right ),\nonumber\\
\label{lp3}
\end{eqnarray}
where $\delta$, $\sigma$ and $c_s^2$ are defined as in Eqs. (\ref{j11b}) and
(\ref{j8}) where $\rho_b(t)$ is now time-dependent.

\subsection{Approximate equation for the density contrast}
\label{sec_appdelta}

A detailed study of the exact linearized equations 
(\ref{lp1})-(\ref{lp3}) derived previously  will be the
subject of a future work. Here, we derive an approximate equation for the
density contrast $\delta(\vec x,t)$ by neglecting the relativistic terms   that
involve a
time derivative, i.e. terms like
$\frac{1}{c^{\alpha}}\frac{\partial^\beta}{\partial t^\beta}$. In this way, we
simplify the equation for $\delta(\vec x,t)$
without altering the expression of the relativistic Jeans
length.

With this approximation, the linearized equations (\ref{lp1})-(\ref{lp3}) become
\begin{eqnarray}
\frac{\partial\delta}{\partial
t}+\frac{1}{a}\vec\nabla\cdot
{\vec v}=0,
\label{lp5}
\end{eqnarray}
\begin{eqnarray}
\frac{1}{c^2}\left (1+\frac{E}{mc^2}\right )\sigma=\frac{\hbar^2}{4
m^2 c^2 a^2}\Delta\delta-\left
(1+\frac{2c_s^2}{c^2}\right
)\frac{\Phi}{c^2}-\frac{c_s^2}{c^2}\delta,\nonumber\\
\label{lp8}
\end{eqnarray}
\begin{eqnarray}
\frac{\partial {\vec v}}{\partial t}+H{\vec v}=
\frac{\hbar^2}{4m^2a^3}\vec\nabla(\Delta\delta)-\frac{1}{a}\left
(1+\frac{2c_s^2}{c^2}\right )
\vec\nabla\Phi-\frac{1}{a}c_s^2 \vec\nabla \delta,\nonumber\\
\label{lp6}
\end{eqnarray}
\begin{eqnarray}
\frac{\Delta\Phi}{4\pi G \rho_b
a^2}=\left (1
+\frac{c_s^2}{c^2}\right )\delta -\frac{1}{c^2}\left (\frac{E}{m
c^2}+1\right )\sigma\nonumber\\
+\frac{E}{mc^2}\left
(\frac{E}{2mc^2}+1\right )\delta-\frac{E}{mc^2}\left
(1+\frac{E}{2mc^2}\right
)\frac{2\Phi}{c^2}\nonumber\\
-\frac{\Phi}{c^2}+\frac{3H^2}{4\pi G\rho_b
c^2}\Phi.\qquad
\label{lp7}
\end{eqnarray}
Taking the time derivative of Eq. (\ref{lp5}) and the divergence of Eq.
(\ref{lp6}), we obtain
\begin{eqnarray}
\frac{\partial^2\delta}{\partial t^2}+2H\frac{\partial\delta}{\partial
t}=\frac{c_s^2}{a^2}\Delta\delta-\frac{\hbar^2}{4
m^2a^4}\Delta^2\delta\nonumber\\
+\frac{1}{a^2}\left
(1+\frac{2c_s^2}{c^2}\right
)\Delta\Phi.
\label{lp9}
\end{eqnarray}
Substituting Eq. (\ref{lp8}) in Eq. (\ref{lp7}), we get
\begin{eqnarray}
\frac{\Delta\Phi}{4\pi G \rho_b
a^2}=\left (1
+2\frac{c_s^2}{c^2}\right )\delta-
\frac{\hbar^2}{4m^2c^2a^2}\Delta\delta\nonumber\\
+\left
(1+\frac{2c_s^2}{c^2}\right
)\frac{\Phi}{c^2}
+\frac{E}{mc^2}\left
(\frac{E}{2mc^2}+1\right )\delta\nonumber\\
-\frac{E}{mc^2}\left
(\frac{E}{2mc^2}+1\right
)\frac{2\Phi}{c^2}
-\frac{\Phi}{c^2}+\frac{3H^2}{4\pi G\rho_b
c^2}\Phi.\qquad
\label{lp10}
\end{eqnarray} 
Decomposing the perturbations in (spatial) Fourier modes, and using the
identities (\ref{j10}) and (\ref{j11}) that remain valid in the present context
where now $E(t)$ and $c_s^2(t)$ are time-dependent, we obtain a closed equation
for $\delta_k(t)$:
\begin{eqnarray}
\frac{d^2\delta_k}{dt^2}+2H\frac{d\delta_k}{dt}+\biggl\lbrack \frac{\hbar^2k^4}{
4m^2a^4 }+\frac{c_s^2}{a^2}k^2\nonumber\\
-\frac{4\pi G\rho_b}{1+\frac{3H^2a^2}{k^2c^2}}\left (1+\frac{2c_s^2}{c^2}\right
)\left (
1+\frac{3c_s^2}{c^2}+\frac{\hbar^2k^2}{4m^2c^2a^2}
\right )\biggr\rbrack\delta_k=0.\nonumber\\
\label{lp11}
\end{eqnarray} 
In a static universe ($a=1$, $H=0$), taking $\ddot\delta_k=0$ (i.e.
$\omega=0$), we can check that Eq. (\ref{lp11}) reduces to Eq. (\ref{kge27})
that determines the relativistic Jeans length. In the nonrelativistic limit
$c\rightarrow +\infty$, the equation for the
density contrast reduces to \cite{abrilMNRAS,chavaniscosmo}:
\begin{eqnarray}
\frac{d^2\delta_k}{dt^2}+2H\frac{d\delta_k}{dt}+\biggl (\frac{\hbar^2k^4}{
4m^2a^4 }+\frac{c_s^2}{a^2}k^2
-{4\pi G\rho_b}
\biggr )\delta_k=0.\nonumber\\
\label{nr21}
\end{eqnarray}

Because of the term appearing in the denominator of Eq. (\ref{lp11}), the
limits $k\rightarrow 0$ and $c\rightarrow +\infty$ do not commute. If we take
the limit $c\rightarrow +\infty$ before the limit $k\rightarrow 0$, Eq.
(\ref{lp11}) reduces to
\begin{eqnarray}
\frac{d^2\delta}{dt^2}+2H\frac{d\delta}{dt}-{4\pi
G\rho_b}\delta=0
\label{nr21b}
\end{eqnarray}
which is equivalent to the CDM model where the evolution of the perturbations is
due only to the gravitational force. By contrast, if we take the limit
$k\rightarrow 0$ before the limit $c\rightarrow +\infty$, Eq. (\ref{lp11})
reduces to
\begin{eqnarray}
\frac{d^2\delta}{dt^2}+2H\frac{d\delta}{dt}=0
\label{nr21c}
\end{eqnarray}
in which the gravitational force has ``disappeared''. This shows that the term
appearing in the denominator of Eq. (\ref{lp11}),  which is purely
relativistic, acts as an attenuation factor. Actually, this  term can be written
as $1+(\lambda/\lambda_H)^2$ where
$\lambda=2\pi/k$ is the wavelength of the perturbation and 
\begin{eqnarray}
\lambda_H={2\pi}\frac{c}{\sqrt{3}Ha}
\label{lp12}
\end{eqnarray}
is the cosmological horizon, or Hubble length.\footnote{Using the Friedmann
equation (\ref{kge12b}), the Hubble length can be written as a sort of Jeans
length $\lambda_H=2\pi c/\sqrt{8\pi G\epsilon_b/c^2}a$ in which the speed of
sound is replaced by the speed of light. It can be interpreted as a maximum
wavelength.}  This is the distance traveled
by a photon with the velocity $c$ during the Hubble time
$H^{-1}$. The cosmological horizon sets the size of the
observable universe. When $\lambda\ll\lambda_H$,
which is the most relevant case, the term $1+(\lambda/\lambda_H)^2$ can be
replaced by unity. However,
when the wavelength $\lambda$ of the perturbation approaches the cosmological
horizon, this term must be taken into account and it reduces the growth of the
perturbations (it weakens the effective gravitational constant $G_{\rm
eff}(k)=G/(1+k_H^2/k^2)$). As a result, it sets a natural upper cutoff, of the
order of the Hubble length, for the perturbations that can grow. We shall come
back to this reduction factor in Sec. \ref{sec_relat}.

\section{Evolution of the density contrast
in the matter era}
\label{sec_evolution}

In this section,  we study the
evolution of the density contrast $\delta_k$  in
the matter era. We assume that the matter era starts at 
$a_i=10^{-4}$, corresponding to the epoch of matter-radiation
equality. This is just before the recombination epoch (matter-dominated
era) starting at $a_{\rm rec}=10^{-3}$
 when the electrons recombine with the atomic nuclei, after their last
dispersion with the photons,
leaving an imprint of this
interaction in the temperature fluctuations of the CMB.

\subsection{The equation for $\delta_k(a)$ in the matter era}
\label{sec_eqdelta}

In the matter era, we can make the approximations
$\epsilon_b\sim \rho_b c^2$ and $P_b\ll \epsilon_b$ (see Sec.
\ref{sec_radmatt}). In that
limit, $\rho_b$ coincides with the rest-mass density and the Friedmann equations
(\ref{frid1}), (\ref{kge12b}) and (\ref{kge12bw}) reduce to
\begin{equation}
\frac{d\rho_b}{dt}+3H\rho_b=0,
\label{me1}
\end{equation}
\begin{eqnarray}
H^2=\left (\frac{\dot a}{a}\right )^2=\frac{8\pi G}{3}\rho_b,
\label{me2}
\end{eqnarray}
\begin{eqnarray}
\frac{\ddot a}{a}=-\frac{4\pi G}{3}\rho_b.
\label{me3}
\end{eqnarray}
They lead to the EdS solution $\rho_b\propto a^{-3}$, $a\propto t^{2/3}$,
$H=2/(3t)$ and
$\rho_b=1/(6\pi Gt^2)$. Using Eqs.
(\ref{me1})-(\ref{me3}), we can transform Eq. (\ref{lp11}) for 
$\delta_k(t)$ into
an equation for  $\delta_k(a)$ of the form
\begin{eqnarray}
\frac{d^2\delta_k}{da^2}+\frac{3}{2a}\frac{d\delta_k}{da}+\frac{3}{2a^2}\biggl
\lbrack \frac{\hbar^2k^4}{16\pi G \rho_b
m^2 a^4}+\frac{c_s^2}{4\pi G\rho_b a^2}k^2\nonumber\\
-\frac{1}{1+\frac{3H^2a^2}{k^2c^2}}\left(1+\frac{2c_s^2}{c^2}
\right)\left(1+\frac
{3c_s^2}{c^2}+\frac{\hbar^2k^2}{4m^2c^2a^2}\right)\biggr
\rbrack\delta_k=0,\nonumber\\
\label{me4}
\end{eqnarray}
where we recall that $c_s^2$ is given by Eq. (\ref{j11b}). Eq. (\ref{me4})
involves the
following characteristic wavenumbers. The quantum Jeans
wavenumber is defined by
\begin{equation}
k_Q=\left(\frac{16\pi
G\rho_b m^2a^4}{\hbar^2}\right)^{1/4}.
\label{me5}
\end{equation}
It can be written as $k_Q=\kappa_Q a^{1/4}$ with $\kappa_Q=\left({16\pi
G\rho_ba^3m^2}/{\hbar^2}\right)^{1/4}$. The self-interaction Jeans wavenumber is
defined by
\begin{equation}
k_J=\left (\frac{4\pi G\rho_b
a^2}{c_s^2}\right )^{1/2}=\left(\frac{Gm^3a^2}{a_s\hbar^2}\right)^{1/2}.
\label{me6}
\end{equation}
It can be written as $k_J=\kappa_J a$
with $\kappa_J=\left({Gm^3}/{a_s\hbar^2}\right)^{1/2}$. The Compton
wavenumber is defined by
\begin{equation}
k_C=\frac{2mca}{\hbar}.
\label{me7}
\end{equation}
It can be written as $k_C=\kappa_C a$ with
$\kappa_C={2mc}/{\hbar}$. The Hubble (or horizon) wavenumber is defined by
\begin{equation}
k_H=\frac{\sqrt{3}Ha}{c}=\left(\frac{8\pi
G\rho_b a^2}{c^2}\right)^{1/2}.
\label{me8}
\end{equation}
It can be written as $k_H={\kappa_H}/{\sqrt{a}}$ with $\kappa_H=\left({8\pi
G\rho_b a^3}/{c^2}\right)^{1/2}$. We also introduce the dimensionless parameter
\begin{equation}
\sigma=\frac{4\pi a_s \hbar^2\rho_b
a^3}{m^3 c^2}
\label{me9}
\end{equation}
so that ${c_s^2}/{c^2}={\sigma}/{a^3}$. In term of these
parameters, Eq. (\ref{me4})  can
be
rewritten as (we omit the subscript $k$ on $\delta_k$):
\begin{eqnarray}
\frac{d^2\delta}{da^2}&+&\frac{3}{2a}\frac{d\delta}{da}+\frac{3}{2a^2}
\biggl\lbrack\frac { k^4
}{\kappa^4_Qa}+\frac{k^2}{\kappa_J^2a^2}\nonumber\\
&-&\frac{1}{1+\frac{\kappa_H^2}{k^2a}}\left(1+\frac{2\sigma}{a^3}
\right)\left(1+\frac{3\sigma}{a^3}+\frac{k^2}{\kappa_C^2a^2}
\right)\biggr\rbrack\delta=0.
\nonumber\\
\label{me10}
\end{eqnarray}
It is interesting to see how the different characteristic scales
$\kappa_Q$, $\kappa_J$, $\kappa_H$ and $\kappa_C$ naturally appear in this
equation.

\subsection{Justification of the nonrelativistic limit}
\label{sec_justif}

In this section, and in Secs. \ref{sec_nid}-\ref{sec_negas}, we consider the
case
where the term associated with the Hubble
length is very small: $\kappa_H^2/k^2 a\ll 1$. For a given
wavenumber $k$, this corresponds to a scale factor $a\gg \kappa_H^2/k^2$ and,
for a given scale factor $a$, this corresponds to a wavenumber
$k\gg\kappa_H/a^{1/2}$. Therefore, this approximation is valid when  the
wavelength of the perturbation is much smaller than the cosmological
horizon on the timescale considered: $\lambda\ll \lambda_H(a)$.

A direct (non-trivial) consequence of this assumption is that  the term
$k^4/\kappa_Q^4a$
associated with the quantum Jeans length is always much larger than the term
$k^2/\kappa_C^2a^2$ associated with the Compton length. Indeed, the condition 
$k^4/\kappa_Q^4a\gg k^2/\kappa_C^2a^2$ is equivalent to $a\gg
\kappa_Q^4/k^2\kappa_C^2$. Now, from Eqs. (\ref{me5}), (\ref{me7}) and
(\ref{me8}) we can easily establish the identities
\begin{eqnarray}
(k_C k_H)^2=2k_Q^4\qquad {\rm and}\qquad  
(\kappa_C\kappa_H)^2=2\kappa_Q^4. 
\label{jn1}
\end{eqnarray}
As a result, $\kappa_Q^4/k^2\kappa_C^2=\kappa_H^2/2k^2<\kappa_H^2/k^2$.
Therefore, the
condition   $a\gg \kappa_H^2/k^2$ automatically implies  $a\gg
\kappa_Q^4/k^2\kappa_C^2$ which itself implies $k^4/\kappa_Q^4a\gg
k^2/\kappa_C^2a^2$. We stress that this inequality is independent on the mass
$m$
of the bosons.

On the other hand, the condition $\sigma/a^3\ll 1$ will be satisfied during the
whole
matter era if it is satisfied initially, i.e. if $\sigma\ll a_i^3=10^{-12}$.
This
condition corresponds to $c_s\ll c$. In Appendix \ref{sec_smallness} we show
that this condition is always satisfied in the matter era.

In conclusion, if we assume that the wavelength of the perturbation is always
much smaller than the cosmological horizon during the matter era ($\lambda\ll
\lambda_H(a)$), and if we
use the fact that $\sigma/a^3\ll 1$, Eq. (\ref{me10}) reduces to 
\begin{equation}
\frac{d^2\delta}{da^2}+\frac{3}{2a}\frac{d\delta}{da}+\frac{3}{2a^2}\left(\frac{
k^4}{\kappa_Q^4a}+\frac{k^2}{\kappa_J^2a^2}-1\right)\delta=0,
\label{jn2}
\end{equation}
which is the nonrelativistic equation for the density contrast of a
SF/BEC \cite{abrilMNRAS,chavaniscosmo}. Therefore, the nonrelativistic limit is
valid in
the matter era  when $\lambda\ll\lambda_H(a)$.

The CDM model is recovered by taking
$\kappa_Q\rightarrow +\infty$ and $\kappa_J\rightarrow +\infty$ in Eq. 
(\ref{jn2})
yielding
\begin{eqnarray}
\frac{d^2\delta_{CDM}}{da^2}+\frac{3}{2a}\frac{d\delta_{CDM}}{da}-\frac{3}{2a^2}
\delta_{CDM}=0.
\label{jn3}
\end{eqnarray}
The growing solution is $\delta_{CDM}\propto a$. It is usually
considered
 that $\delta_i\sim 10^{-5}$ at the initial
time $a_i=10^{-4}$ of matter-radiation equality \cite{dodelson}. Therefore, the
growing evolution of the density contrast in the
CDM model can be written as
\begin{equation}
\delta_{CDM}(a)=\frac{\delta_i}{a_i} a.
\label{jn4}
\end{equation}
We will take this CDM result as a reference, and compare it  with the prediction
of
the SF model. We note that Eq. (\ref{jn2})  for the density contrast of the SF 
reduces to Eq. (\ref{jn3})  when $k\rightarrow
0$  and when 
$a\rightarrow
+\infty$ because the quantum
term ${k^4}/{\kappa_Q^4a}$ and the self-interaction term
${k^2}/{\kappa_J^2a^2}$ become negligible in these limits. Therefore, the SF
is expected to behave similarly to CDM at large scales and at late times.

\subsection{Non-interacting scalar field}
\label{sec_nid}

For a noninteracting SF ($\kappa_J\rightarrow +\infty$), the equation
determining
the evolution
of the density contrast reduces to
\begin{equation}
\frac{d^2\delta}{da^2}+\frac{3}{2a}\frac{d\delta}{da}+\frac{3}{2a^2}\left(\frac{
k^4}{\kappa_Q^4a}-1\right)\delta=0.
\label{ni1}
\end{equation}
It turns out that Eq. (\ref{ni1}) can be solved analytically 
\cite{chavaniscosmo}. Its growing solution is given by
\begin{equation}
\delta(a)=\frac{A}{a^{1/4}}J_{-\frac{5}{2}}\left(\sqrt{6}\frac{k^2}{\kappa_Q^2}
\frac{1}{a^{1/2}}\right).
\label{ni2}
\end{equation}
 We shall determine
the amplitude $A$ so that the asymptotic behavior of the
density contrast of the SF at late times exactly matches the solution
(\ref{jn4}) of the
CDM
model. Using the asymptotic expansion of the Bessel function for small
arguments (see, e.g., Appendix A of \cite{chavaniscosmo}), we
obtain
\begin{equation}
A=\frac{6^{5/4}}{3}\sqrt{\frac{\pi}{2}}\frac{\delta_i}{a_i}\left (
\frac{k}{\kappa_Q} \right)^{5}.
\label{ni4}
\end{equation}
Eq. (\ref{ni2}) with $A$ given by Eq. (\ref{ni4}) determines the evolution of
the density
contrast of a non-interacting SF
during the matter era in the nonrelativistic limit $\lambda\ll \lambda_H(a)$.

We can identify two regimes. We first consider the case $k^4/\kappa_Q^4 a\gg
1$. For a given wavenumber $k$, this corresponds to a scale factor $a\ll
k^4/\kappa_Q^4$. Alternatively, for a given scale factor $a$, this corresponds
to a wavelength $\lambda\ll \lambda_Q(a)$. Since the
wavelength of the perturbation is  smaller that the quantum Jeans
length, the density contrast $\delta(a)$ displays oscillations.
Using the asymptotic expansion of the Bessel function for small arguments, we
find that 
\begin{equation}
\delta(a)\sim
-2\frac{\delta_i}{a_i}\left(\frac{k}{\kappa_Q}\right
)^4\sin\left(\sqrt{6}\frac{k^2}{\kappa_Q^2}\frac{1}{a^{1/2}}\right).
\label{ni5}
\end{equation}
This solution is valid for $a\ll k^4/\kappa_Q^4$. We see that the amplitude of
the oscillations is constant, i.e. independent on the scale factor. We now
consider the case $k^4/\kappa_Q^4 a\ll
1$. For a given wavenumber $k$, this corresponds to a scale factor $a\gg
k^4/\kappa_Q^4$. Alternatively, for a given scale factor $a$, this corresponds
to a wavelength $\lambda\gg \lambda_Q(a)$. Since the
wavelength of the perturbation is larger that the quantum Jeans
length, the density contrast $\delta(a)$ increases.  For $a\gg
k^4/\kappa_Q^4$, the perturbation behaves like in the CDM model, see Eq.
(\ref{jn4}).

In summary, when $a\ll
k^4/\kappa_Q^4$, i.e. $\lambda\ll\lambda_Q(a)$, the perturbation oscillates with
a constant amplitude and when $a\gg k^4/\kappa_Q^4$, i.e.
$\lambda\gg\lambda_Q(a)$, the perturbation grows linearly with the scale factor
like in the CDM model.

\begin{figure}[!ht]
\includegraphics[width=0.98\linewidth]{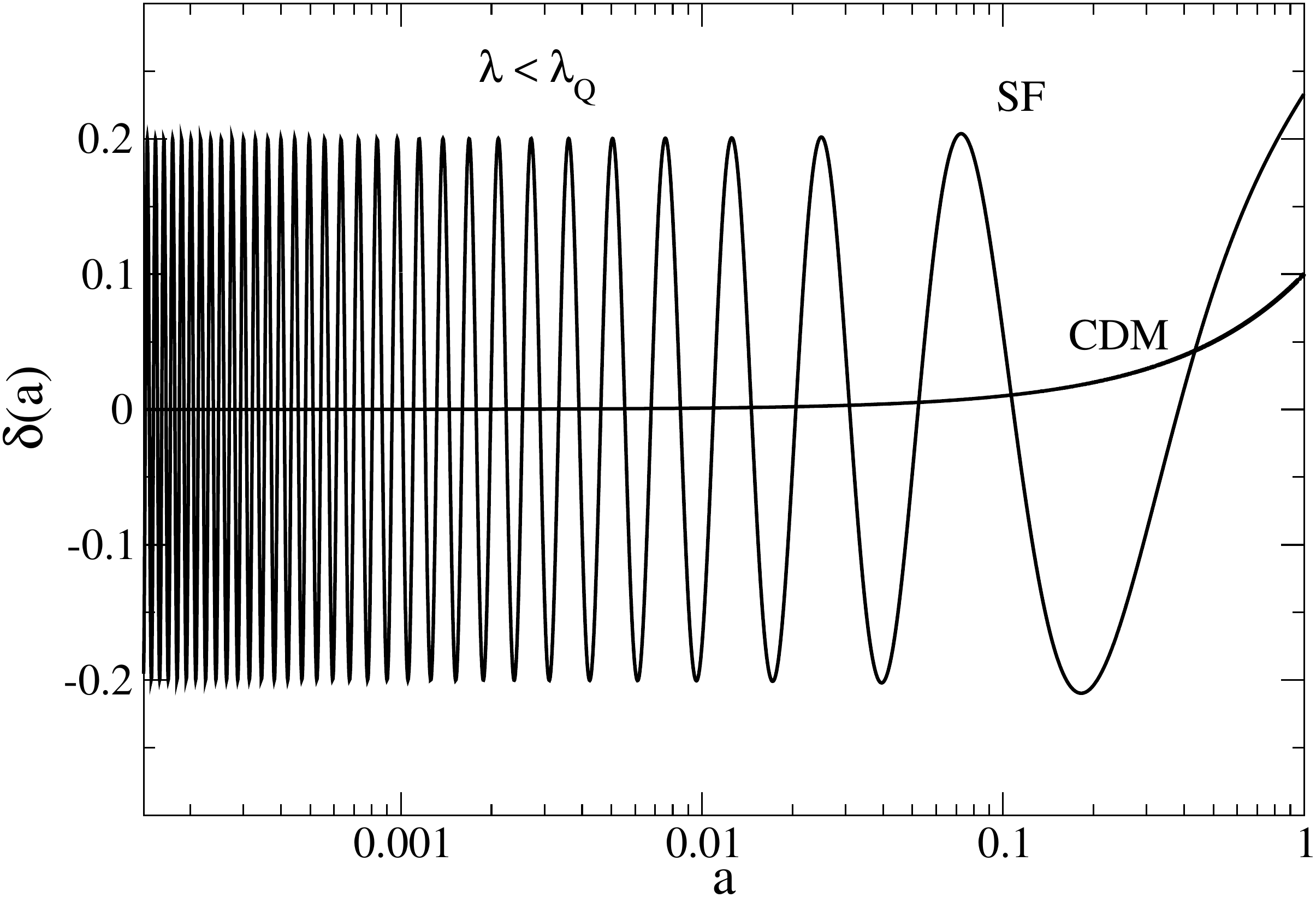}
\caption{Evolution of the density contrast $\delta(a)$ of a non-interacting SF 
for $k/\kappa_Q=1$ (semi-log plot). It is compared with the density
contrast  $\delta_{CDM}(a)$ of the CDM model.
\label{grap6}}
\end{figure}

\begin{figure}[!ht]
\includegraphics[width=0.98\linewidth]{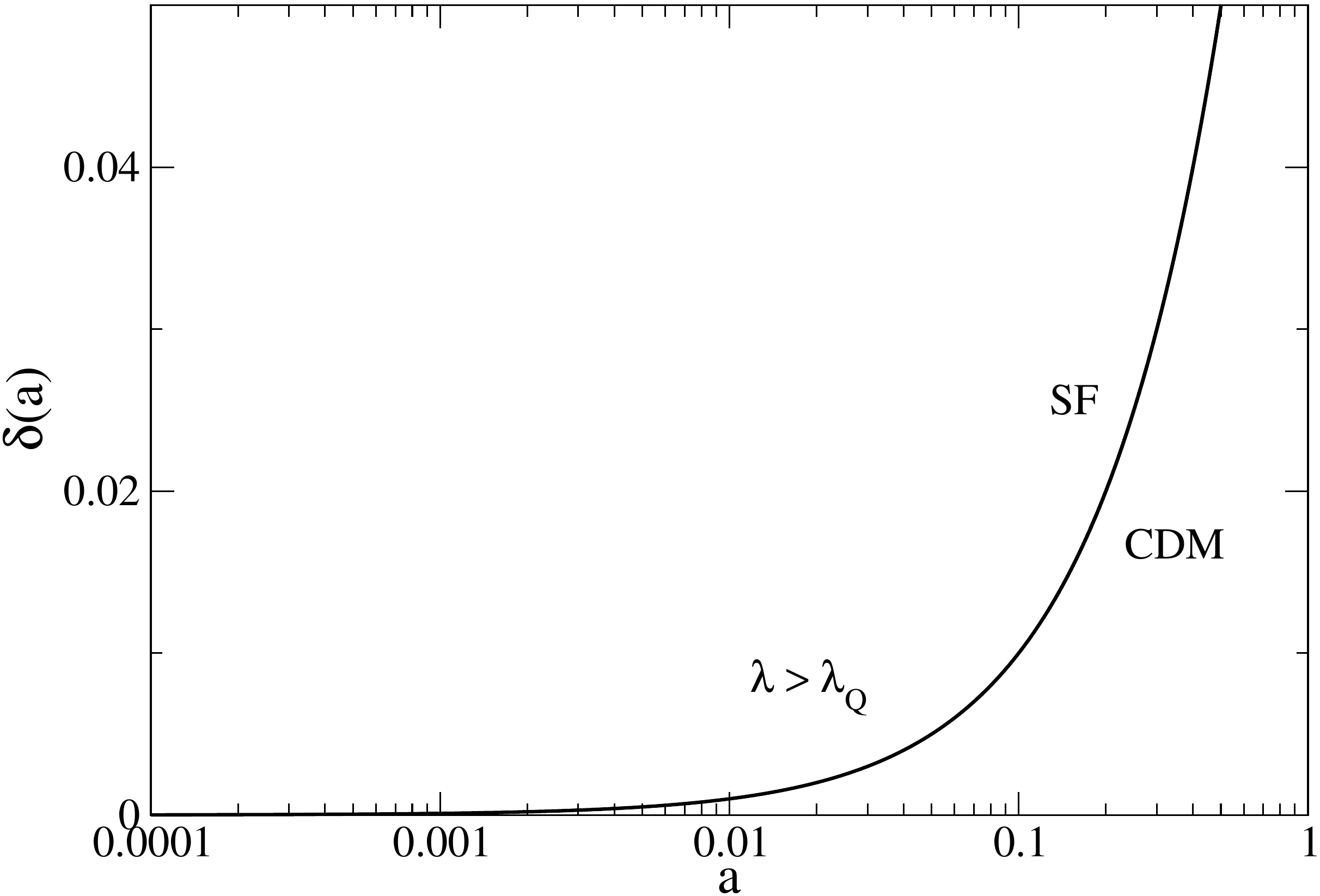}
\caption{Evolution of the density contrast $\delta(a)$ of a non-interacting SF 
for $k/\kappa_Q=0.1$ (semi-log plot).
\label{grap5}}
\end{figure}

\begin{figure}[!ht]
\includegraphics[width=0.98\linewidth]{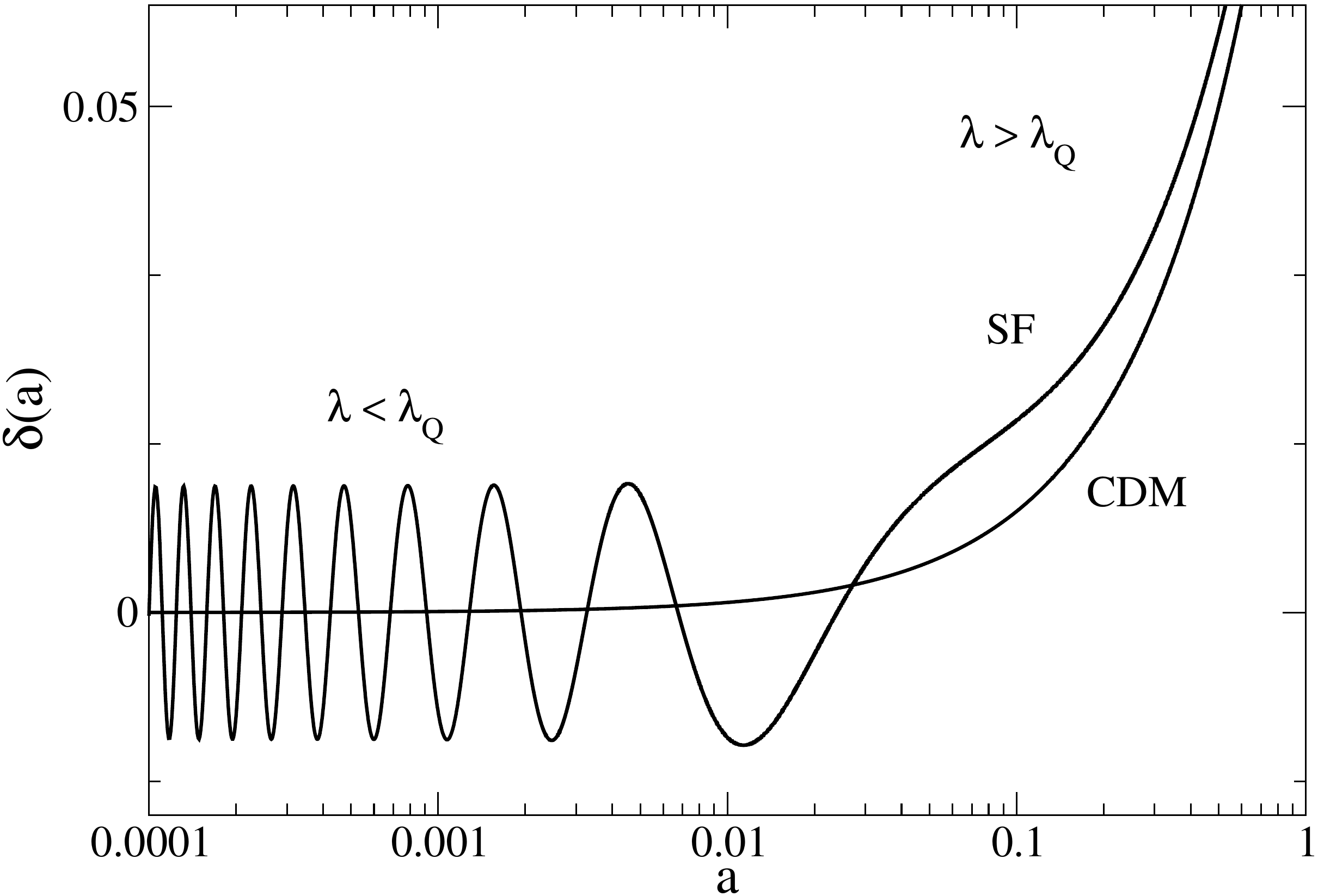}
\caption{Evolution of the density contrast $\delta(a)$ of a non-interacting SF 
for $k/\kappa_Q=0.5$ (semi-log plot).
\label{graph3}}
\end{figure}

To fix the ideas, we consider that the matter era starts at $a_i=10^{-4}$ and we
study
the evolution of the density contrast up to $a_0=1$, corresponding to
the present-day universe (of course, our linear analysis is not valid up to
that point, but our aim is just to give a simple illustration of our results).
Our nonrelativistic treatment is valid on this interval provided that $a_i\gg
\kappa_H^2/k^2$. This corresponds to wavenumbers $k\gg \kappa_H/a_i^{1/2}$.
This constraint is not restrictive since $\kappa_H/a_i^{1/2}=6.49\times
10^{-25}\, {\rm m}^{-1}$ is very small (see Table I in Appendix \ref{sec_na}).
For
$k/\kappa_Q\ge a_0^{1/4}=1$, the
evolution of the density
contrast
is purely oscillatory (see Fig. \ref{grap6}). For $k/\kappa_Q\le a_i^{1/4}=0.1$,
there is no
oscillation and
the density contrast grows from the beginning (see Fig.
\ref{grap5}).\footnote{We can
give a more general criterion for the absence of oscillations,
independent on the interval of time considered. Our nonrelativistic treatment
requires that $a\gg \kappa_H^2/k^2$ and the density contrast displays
oscillations provided that $a\ll
k^4/\kappa_Q^4$. Therefore, the density contrast will not display oscillations
if
$k\ll \kappa_Q^{2/3}\kappa_H^{1/3}$.} In
the intermediate situation
$a_i^{1/4}=0.1<k/\kappa_Q<a_0^{1/4}=1$, the density contrast first presents
oscillations
until $a\sim k^4/\kappa_Q^4$, then grows (see Fig. \ref{graph3}).  We note that
the density contrast
in the SF model grows faster than in the CDM model.

The SF has several virtues. First, it does not create an
over-abundance of small-scale  structures, contrary to the CDM model, because
the perturbations below the quantum Jeans length $\lambda_Q$ do not grow (they 
oscillate). The quantum Jeans length provides therefore a sharp small-scale
cut-off in the matter power spectrum. On the other hand, the perturbations above
the
quantum Jeans length $\lambda_Q$ grow sensibly more rapidly than in the CDM
model. This can accelerate the formation of large-scale structures in the
nonlinear regime in agreement with some observations 
\cite{nature}.

\subsection{Self-interacting scalar field in the Thomas-Fermi limit}
\label{sec_tfd}

We now consider a self-interacting SF.
We make the Thomas-Fermi (TF) approximation ($\kappa_Q\rightarrow +\infty$)
which amounts to
neglecting the quantum term in front of the self-interaction term. In that case,
the equation determining the
evolution of the density contrast reduces to
\begin{equation}
\frac{d^2\delta}{da^2}+\frac{3}{2a}\frac{d\delta}{da}+\frac{3}{2a^2}\left(\frac{
k^2}{\kappa_J^2a^2}-1\right)\delta=0.
\label{si1}
\end{equation}
This equation can be solved analytically \cite{chavaniscosmo}.  The growing
solution
is given by
\begin{equation}
\delta(a)=\frac{B}{a^{1/4}}J_{-\frac{5}{4}}\left(\sqrt{{\frac{3}{2}}}\frac{k}{
\kappa_J}\frac{1}{a}\right).
\label{si2}
\end{equation}
The amplitude $B$ is determined by requiring that the asymptotic behavior of
Eq. (\ref{si2}) for $a\rightarrow +\infty$ exactly matches the solution
(\ref{jn4}) of the CDM
model. This gives
\begin{equation}
B=\Gamma\left (-\frac{1}{4}\right )\frac{3^{5/8}}{2^{15/8}}
\left(\frac{k}{\kappa_J}\right)^{5/4}\frac{\delta_i}{a_i}.
\label{si3}
\end{equation}
Eq. (\ref{si2}) with $B$ given by Eq. (\ref{si3}) determines the evolution of
the density
contrast of a self-interacting SF  during the matter era in the
nonrelativistic limit
$\lambda\ll \lambda_H(a)$ and in the TF limit.

We can identify two regimes. We first consider the case $k/\kappa_J a\gg
1$. For a given wavenumber $k$, this corresponds to a scale factor $a\ll
k/\kappa_J$. Alternatively, for a given scale factor $a$, this corresponds
to a wavelength $\lambda\ll \lambda_J(a)$. Since the
wavelength of the perturbation is smaller that the self-interaction Jeans
length, the density contrast $\delta(a)$ displays oscillations.
Using the asymptotic expansion of the Bessel function for small arguments, we
find that  
\begin{eqnarray}
\delta(a)\sim
\Gamma\left (-\frac{1}{4}\right )\frac{3^{3/8}}{2^{9/8}}\frac{1}{\sqrt{\pi}}
\frac{\delta_i}{a_i}a^{1/4}\left (\frac{k}{\kappa_J}\right )^{3/4}\nonumber\\
\times\cos\left(\sqrt{\frac{3}{2}}\frac{k}{\kappa_J}\frac{1
}{a}+\frac{3\pi}{8}\right ).
\label{si4}
\end{eqnarray}
This solution is valid for $a\ll k/\kappa_J$. We see that the amplitude of
the oscillations grows as $a^{1/4}$. This is a difference with respect to the
non-interacting SF for which the amplitude of the oscillations is constant (see
Sec. \ref{sec_nid}). We
now
consider the case $k/\kappa_J a\ll
1$. For a given wavenumber $k$, this corresponds to a scale factor $a\gg
k/\kappa_J$. Alternatively, for a given scale factor $a$, this corresponds
to a wavelength $\lambda\gg \lambda_J(a)$. Since the
wavelength of the perturbation is larger that the self-interaction Jeans
length, the density contrast $\delta(a)$ increases. For $a\gg
k/\kappa_J$, the perturbation behaves like in the CDM model, see Eq.
(\ref{jn4}).

In summary, when $a\ll
k/\kappa_J$, i.e. $\lambda\ll\lambda_J(a)$, the perturbation oscillates with
a growing amplitude scaling as $a^{1/4}$ and when $a\gg k/\kappa_J$, i.e.
$\lambda\gg\lambda_J(a)$, the perturbation grows linearly with the scale factor
like in the CDM model.

\begin{figure}[!ht]
\includegraphics[width=0.98\linewidth]{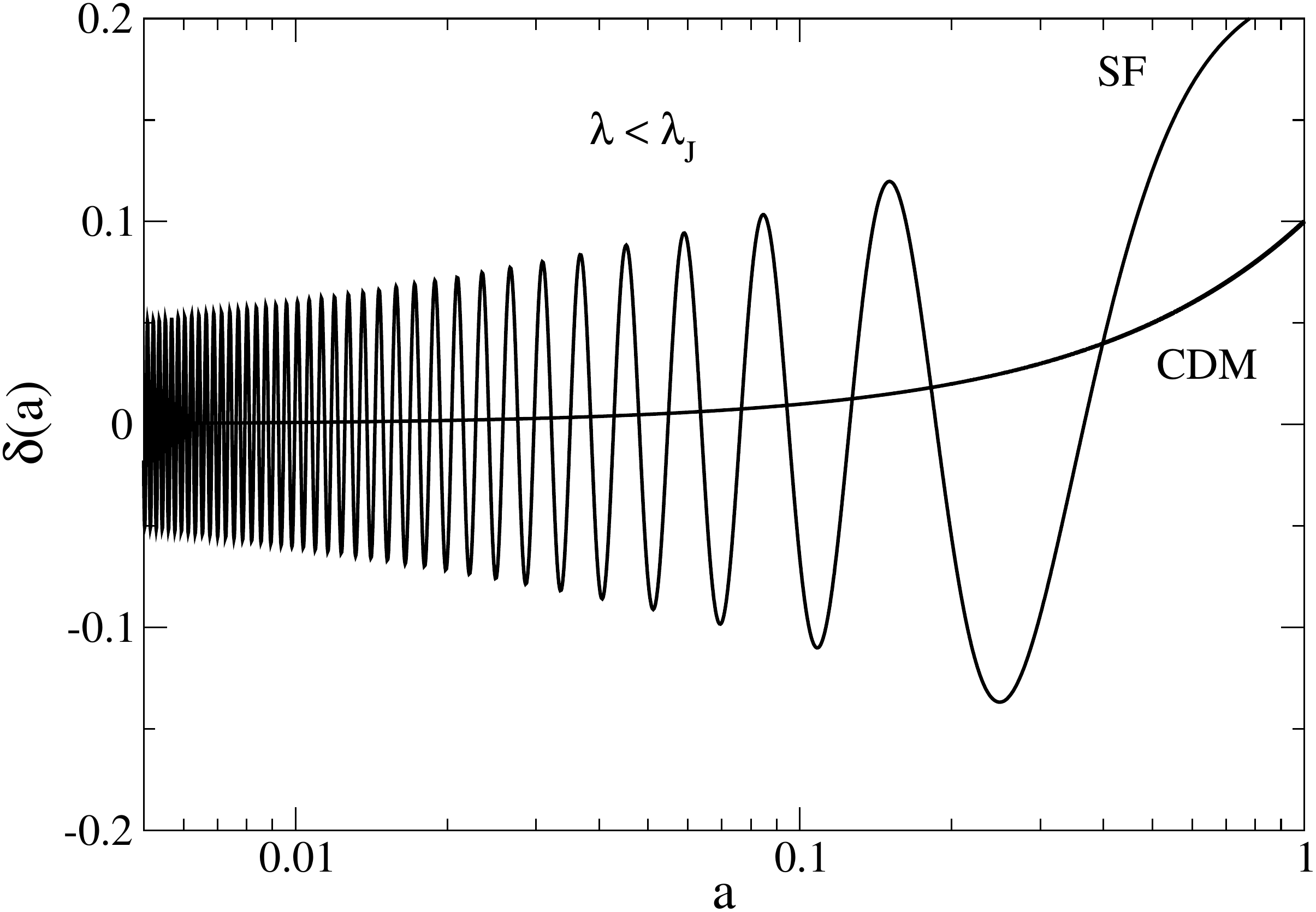}
\caption{Evolution of the density contrast $\delta(a)$ of a self-interacting SF
 in the TF limit for $k/\kappa_J=1$ (semi-log plot).
\label{grap3}}
\end{figure}

\begin{figure}[!ht]
\includegraphics[width=0.98\linewidth]{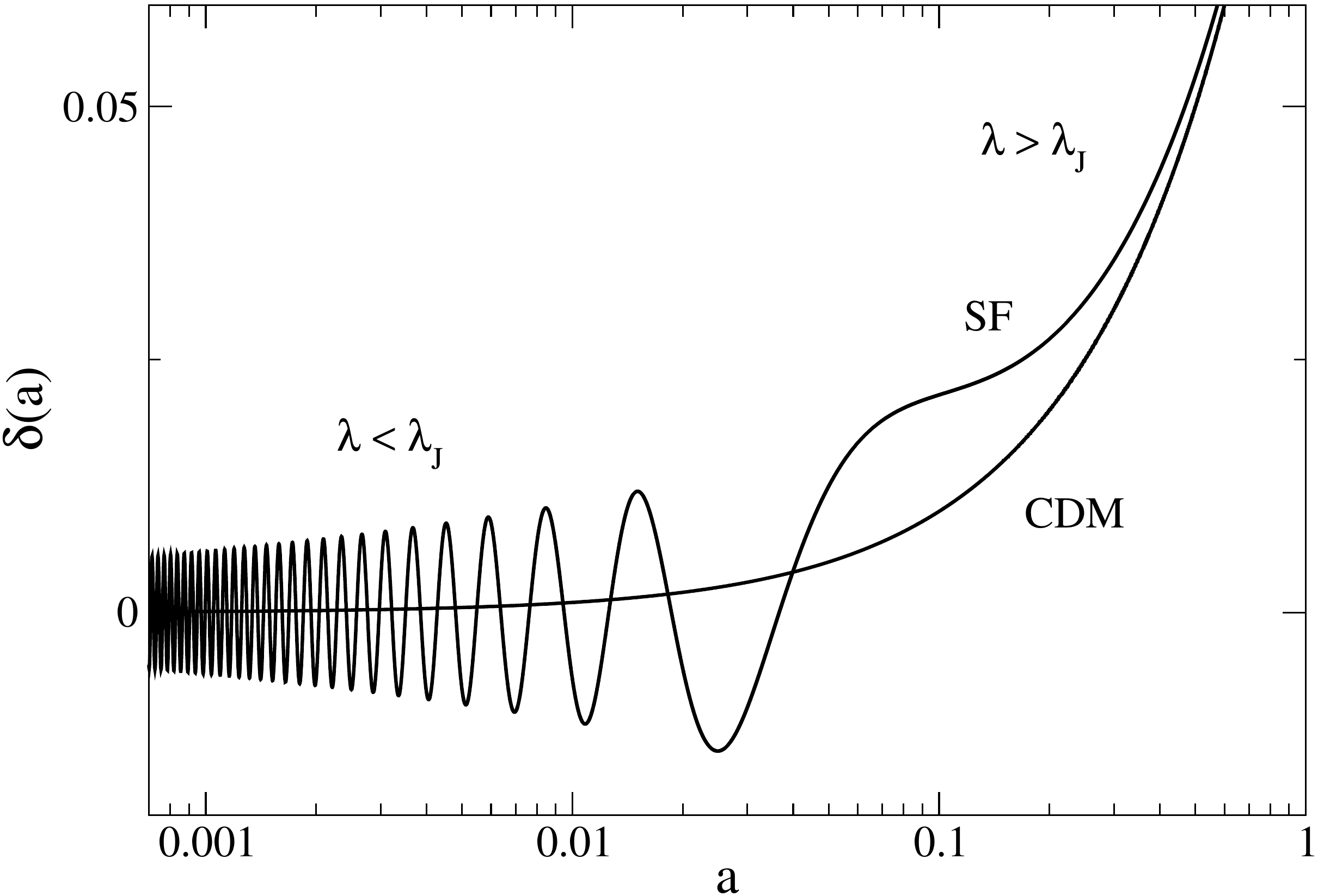}
\caption{Evolution of the density contrast $\delta(a)$ of a self-interacting SF
in the TF limit for $k/\kappa_J=0.1$ (semi-log plot).
\label{graph1}}
\end{figure}

As before, we consider that the matter era starts at $a_i=10^{-4}$ and
we study the evolution of the density contrast up to $a_0=1$. Our
nonrelativistic treatment is valid for $k\gg \kappa_H/a_i^{1/2}=6.49\times
10^{-25}\, {\rm m}^{-1}$. For $k/\kappa_J\ge a_0=1$, the
evolution of the density contrast
is purely oscillatory (see Fig. \ref{grap3}). For $k/\kappa_J\le a_i=10^{-4}$,
there
is no
oscillation and
the density contrast grows from the beginning (the evolution is similar to
the one represented in
Fig. \ref{grap5}).\footnote{The 
more general criterion for the absence of oscillations,
similar to the one obtained in footnote 11, is $k\ll
\kappa_J^{1/3}\kappa_H^{2/3}$.} In
the intermediate situation
$a_i=10^{-4}<k/\kappa_J<1$, the density contrast first presents
growing oscillations
until $a\sim k/\kappa_J$, then grows linearly with $a$ (see Fig. \ref{graph1}).
We note that
the density contrast
in the SF model grows faster than in the CDM model.

The remarks concerning the virtues of the non-interacting SF made at the end of
Sec. \ref{sec_nid} can be transposed to the self-interacting SF with the
self-interaction Jeans
length replacing the quantum Jeans length. The SF suppresses small-scale
structures below the self-interaction Jeans length $\lambda_J$ and
accelerates the
formation of
structures above the self-interaction Jeans length $\lambda_J$. This is an
advantage of the SF model with respect to the CDM model. Note that the
self-interaction
Jeans length, which provides a sharp small-scale cut-off in the matter power
spectrum,
corresponds precisely to the size of the dark matter halos (see Eq.
(\ref{smallness2})) that form in the
nonlinear regime \cite{bohmer,prd1}.

{\it Remark:} we note that the Compton term in Eq. (\ref{me10}) scales as
$-k^2/\kappa_C^2a^2$ similarly to the  self-interaction Jeans term
$k^2/\kappa_J^2a^2$.
Therefore, Eq. (\ref{si1}) with the Compton term retained can still be solved
analytically by simply replacing $1/\kappa_J^2$ by $1/\kappa_J^2-1/\kappa_C^2$.
However, we have shown in Sec. \ref{sec_justif} (see also Sec.
\ref{sec_relat}) that the  Compton term is always negligible in the matter era.

\subsection{Self-interacting scalar field beyond the TF limit}
\label{sec_both}

\begin{figure}[!ht]
\includegraphics[width=0.98\linewidth]{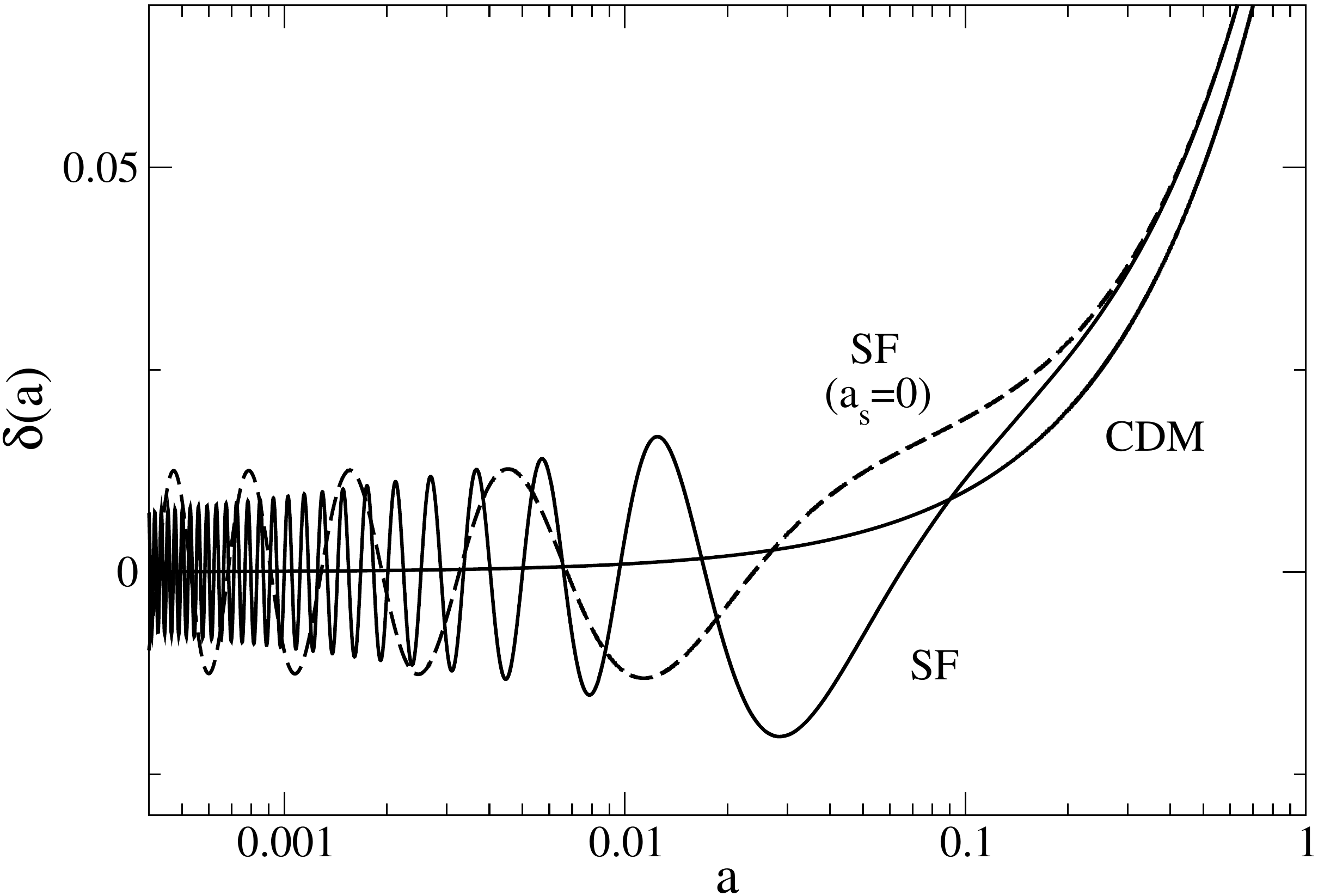}
\caption{Evolution of the density contrast $\delta(a)$ of a self-interacting SF 
for $k/\kappa_Q=0.5$ and $\kappa_Q/\kappa_J=0.1$ (semi-log plot). It is
compared with the non-interacting SF solution (dashed line).
\label{mix}}
\end{figure}

We still consider a self-interacting SF but we
now take the quantum term into account. The evolution
of the density contrast is determined by Eq. (\ref{jn2}). This equation has to
be
solved numerically. We
would like to compare the solution of Eq. (\ref{jn2}) with the solution
(\ref{jn4}) of the CDM model by imposing that they coincide at late times.
However, there is a difficulty. Since the solution of Eq. (\ref{jn2}) may be
strongly oscillating at early time, it is not convenient to specify the initial
condition at $a_i$ and solve the equation forward in time. Such a procedure
would be very sensitive to the initial condition and would not in general
reproduce the CDM result at late times.\footnote{This is a particularity of
the SF. Because of the oscillations of $\delta(a)$ at early time, the evolution
of the perturbations with small wavelengths $\lambda$ is very dependent on the
initial
values of ($\delta_i,\dot\delta_i$) prescribed at $a_i$. We need very
particular values of ($\delta_i,\dot\delta_i$) in order to
recover the results of the CDM model at later times. This implies that, in
general, the evolution of the perturbations with small
wavelengths is {\it different} in the SF and CDM models, even at late times.
This is not a drawback of the
SF model because we know that the CDM model presents discrepencies
with observations at small scales and that the SF model has been introduced
precisely to solve them (see the discussion at the end of Secs.
\ref{sec_nid} and \ref{sec_tfd}). Of course, for sufficiently large
wavelengths
$\lambda$, the fine tuning problem disappears, and the SF and CDM models give
the
same results at late time.} Instead, we can specify
the initial conditions at $a_0=1$ (today) and solve the differential equation
backwards in time until $a_i=10^{-4}$. This procedure has already been
used  in the past \cite{abrilJCAP}. Therefore, we solve Eq.
(\ref{jn2})
numerically with the initial condition ($\delta_0,\dot\delta_0$) obtained
from the analytical solution determined by Eqs. (\ref{ni2}) and (\ref{ni4}) at
$a_0=1$. For illustration, we have taken $k/\kappa_Q=0.5$ and
$\kappa_Q/\kappa_J=0.1$ in Fig. \ref{mix}.

The evolution of the perturbation can be understood qualitatively  by
using simple scaling arguments. The  self-interaction term dominates the quantum
term in Eq. (\ref{jn2}) when
${k^2}/{\kappa_J^2a^2}\gg {k^4}/{\kappa_Q^4a}$. For a given wavenumber $k$,
this corresponds to a scale factor $a\ll \kappa_Q^4/\kappa_J^2k^2$.
Alternatively, for a given scale factor $a$, this corresponds
to a wavelength $\lambda\gg\lambda_Q^2/\lambda_J$.
For $a\ll \kappa_Q^4/\kappa_J^2k^2$, the system is in the TF regime and for
$a\gg \kappa_Q^4/\kappa_J^2k^2$ it  is in the non-interacting regime.
In the TF
regime, the perturbation undergoes growing oscillations for $a\ll k/\kappa_J$
and grows linearly for $a\gg k/\kappa_J$.  In the non-interacting
regime, the perturbation undergoes oscillations with a constant amplitude for
$a\ll
k^4/\kappa_Q^4$ and grows linearly for $a\gg k^4/\kappa_Q^4$.

\begin{figure}[!ht]
\includegraphics[width=0.98\linewidth]{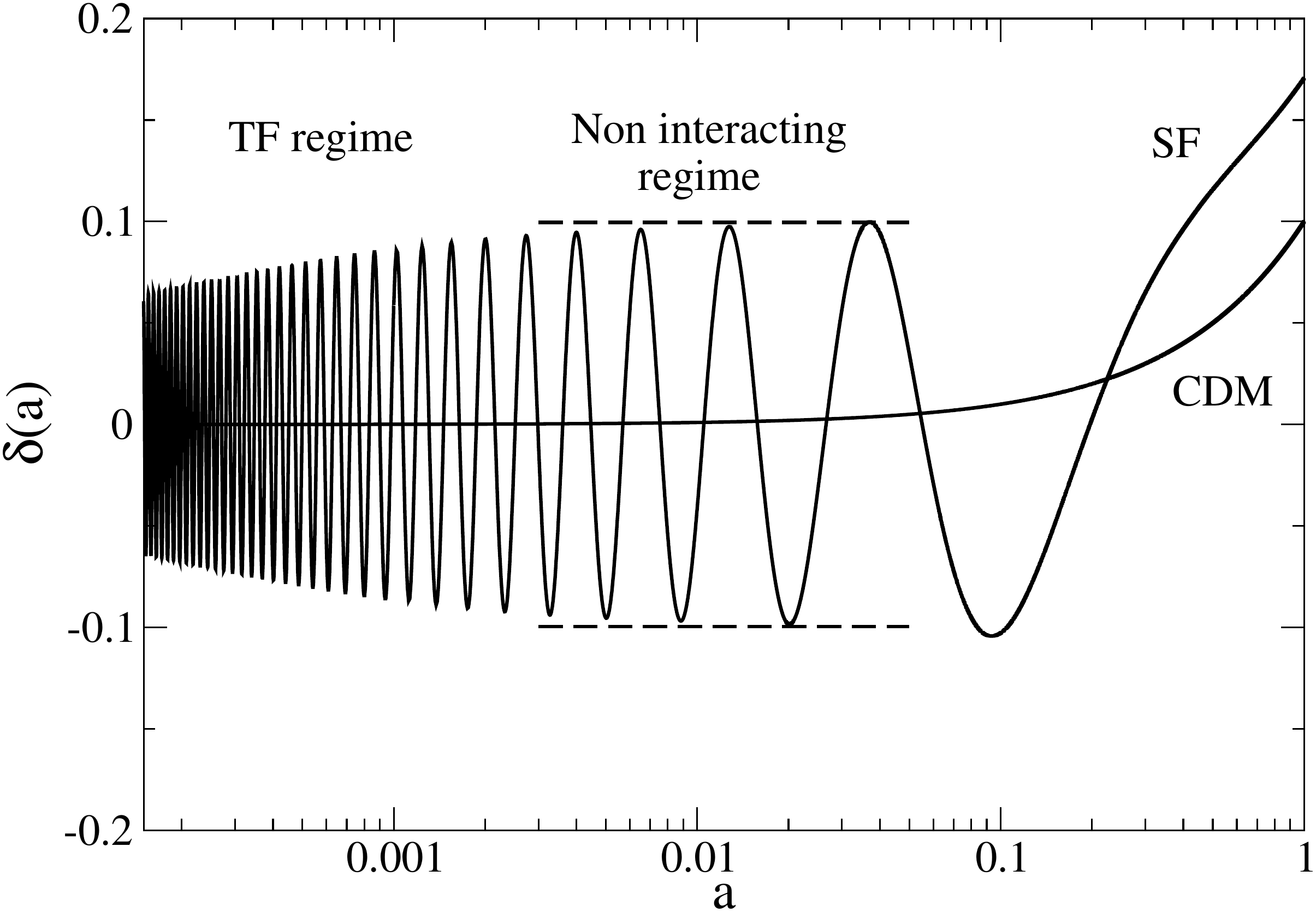}
\caption{Evolution of the density contrast $\delta(a)$ of a self-interacting SF 
for  $k/\kappa_Q=0.840$ and $\kappa_Q/\kappa_J=0.022$ (semi-log plot).
\label{plot1}}
\end{figure}

If $\kappa_Q^4/\kappa_J^2k^2\ll a_i$, the system is always in
the non-interacting regime. If $\kappa_Q^4/\kappa_J^2k^2\gg a_0=1$, the system
is
always in
the TF regime. In these cases, we are led back to the situations considered in
Secs. \ref{sec_nid} and \ref{sec_tfd}. We now assume
$a_i<\kappa_Q^4/\kappa_J^2k^2<1$. Two
situations can occur.

(i) If $k/\kappa_J<\kappa_Q^4/\kappa_J^2k^2$, we necessarily have
$k^4/\kappa_Q^4<\kappa_Q^4/\kappa_J^2k^2$. Therefore, the
perturbation undergoes growing oscillations  (provided that
$k/\kappa_J>a_i$) followed by a linear growth. The growing
oscillations take place at the begining of the TF regime. The linear growth
starts at the end of the TF regime and continues in the non-interacting
regime (there is no oscillation in the non-interacting regime because
$k^4/\kappa_Q^4<\kappa_Q^4/\kappa_J^2k^2$). It is not possible to clearly see
the
change of regime during the linear growth so the evolution of the perturbation 
looks similar to that reported in Fig. \ref{graph1}.

(ii) If $k^4/\kappa_Q^4>\kappa_Q^4/\kappa_J^2k^2$, we necessarily have 
$k/\kappa_J>\kappa_Q^4/\kappa_J^2k^2$. In that case, the perturbations
undergo growing oscillations, followed by constant-amplitude oscillations, and
finally by a linear growth (provided that $k^4/\kappa_Q^4<1$). The growing
oscillations take place in the TF
regime. The constant-amplitude oscillations and the linear growth take
place in the noninteracting regime (there is no linear growth in the TF regime
because $k/\kappa_J>\kappa_Q^4/\kappa_J^2k^2$). In that case, the change of
regime is clearly apparent since we can see the difference between growing
oscillations and constant-amplitude  oscillations. This is illustrated in Fig.
\ref{plot1}

\subsection{The case of a negative scattering length}
\label{sec_negas}

\begin{figure}[!ht]
\includegraphics[width=0.98\linewidth]{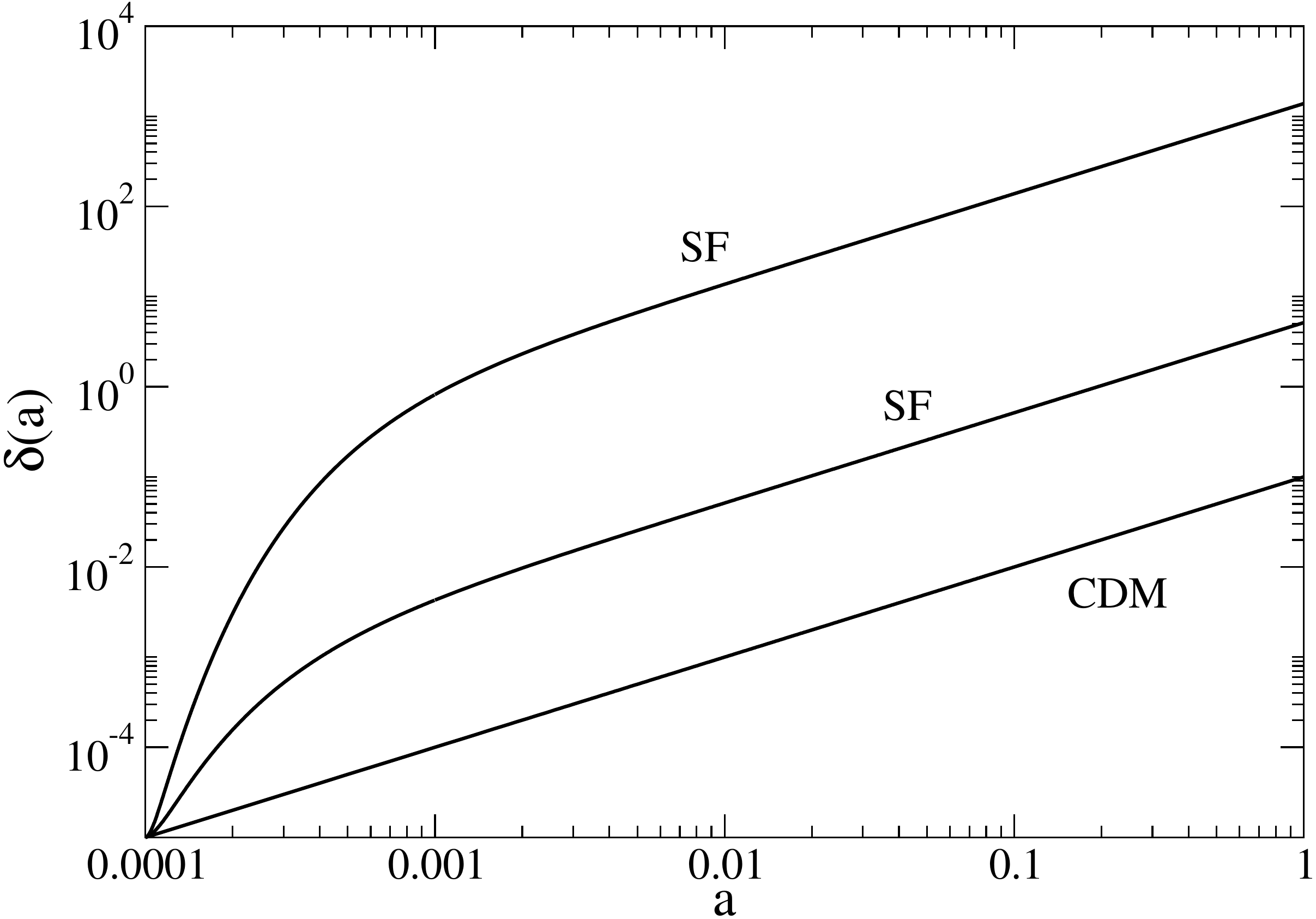}
\caption{Evolution of the density contrast $\delta(a)$ of a self-interacting
SF with $a_s<0$ in the TF limit for (bottom to top) $k/\kappa_J=0$,
$k/\kappa_J=5\times 10^{-4}$ and 
$k/\kappa_J=10^{-3}$ (log-log plot).
\label{graphatt}}
\end{figure}

The previous results implicitly assume that the scattering length is
positive ($a_s>0$), corresponding to a repulsive self-interaction. In this
section, we consider the case of a negative scattering length ($a_s<0$),
corresponding to an attractive self-interaction. The equation for the density
contrast becomes
\begin{equation}
\frac{d^2\delta}{da^2}+\frac{3}{2a}\frac{d\delta}{da}+\frac{3}{2a^2}\left(\frac{
k^4}{\kappa_Q^4a}-\frac{k^2}{\kappa_J^2a^2}-1\right)\delta=0,
\label{jn2bis}
\end{equation}
where now $\kappa_J=(Gm^3/|a_s|\hbar^2)^{1/2}$. We first neglect
the quantum
term in Eq. (\ref{jn2bis})  (TF approximation). In that case
Eq. (\ref{jn2bis}) reduces to
\begin{equation}
\frac{d^2\delta}{da^2}+\frac{3}{2a}\frac{d\delta}{da}+\frac{3}{2a^2}
\left(-\frac{k^2}{\kappa_J^2a^2}-1\right)\delta=0.
\label{jn2bisadd}
\end{equation}
The growing solution of Eq. (\ref{jn2bisadd}) is given by
\begin{equation}
\delta(a)\propto \frac{1}{a^{1/4}}I_{-\frac{5}{4}}\left(\sqrt{{
\frac { 3 } { 2 } } } \frac { k } { \kappa_J}\frac{1}{a}\right).
\label{si6}
\end{equation}
This solution diverges
exponentially rapidly towards $-\infty$ when $a\rightarrow 0$. This divergence
can be avoided by taking the decaying solution into account
\cite{chavaniscosmo}.
Therefore, we solve Eq. (\ref{jn2bisadd}) with the initial
condition $\delta_i=10^{-5}$ and
$\dot\delta_i=\delta_i/a_i=0.1$ at $a_i=10^{-4}$, obtained from Eq.
(\ref{jn4}). The result is represented in
Fig. \ref{graphatt}. For small $k$ (large scales), the density contrast
evolves as in the CDM model. For large $k$ (small scales), the density contrast
increases exponentially rapidly
with the scale factor because the attraction due to the self-interaction adds to
the gravitational  attraction \cite{chavaniscosmo}. An attractive
self-interaction
therefore favors
the growth of structures, but it does it in a rather dramatic manner. Indeed, it
leads to an over-abundance of substructures at a very early epoch. This is in
contradiction with the observations that do not show such an abundance of
substructures. The CDM model, in which only gravity is in action,
already presents too many substructures. Therefore, we need a mechanism that
erases small-scale structures, not a mechanism that enhances them. Consequently,
the SF should have in principle a positive scattering length, not a negative
scattering
length.\footnote{In addition, it is shown in \cite{prd1,prd2} that it is not
possible
to form realistic dark matter halos when $a_s<0$ because they are unstable
above a maximum mass $M_{max}=1.012\hbar/\sqrt{Gm|a_s|}$ that is usually very
small.}

\begin{figure}[!ht]
\includegraphics[width=0.98\linewidth]{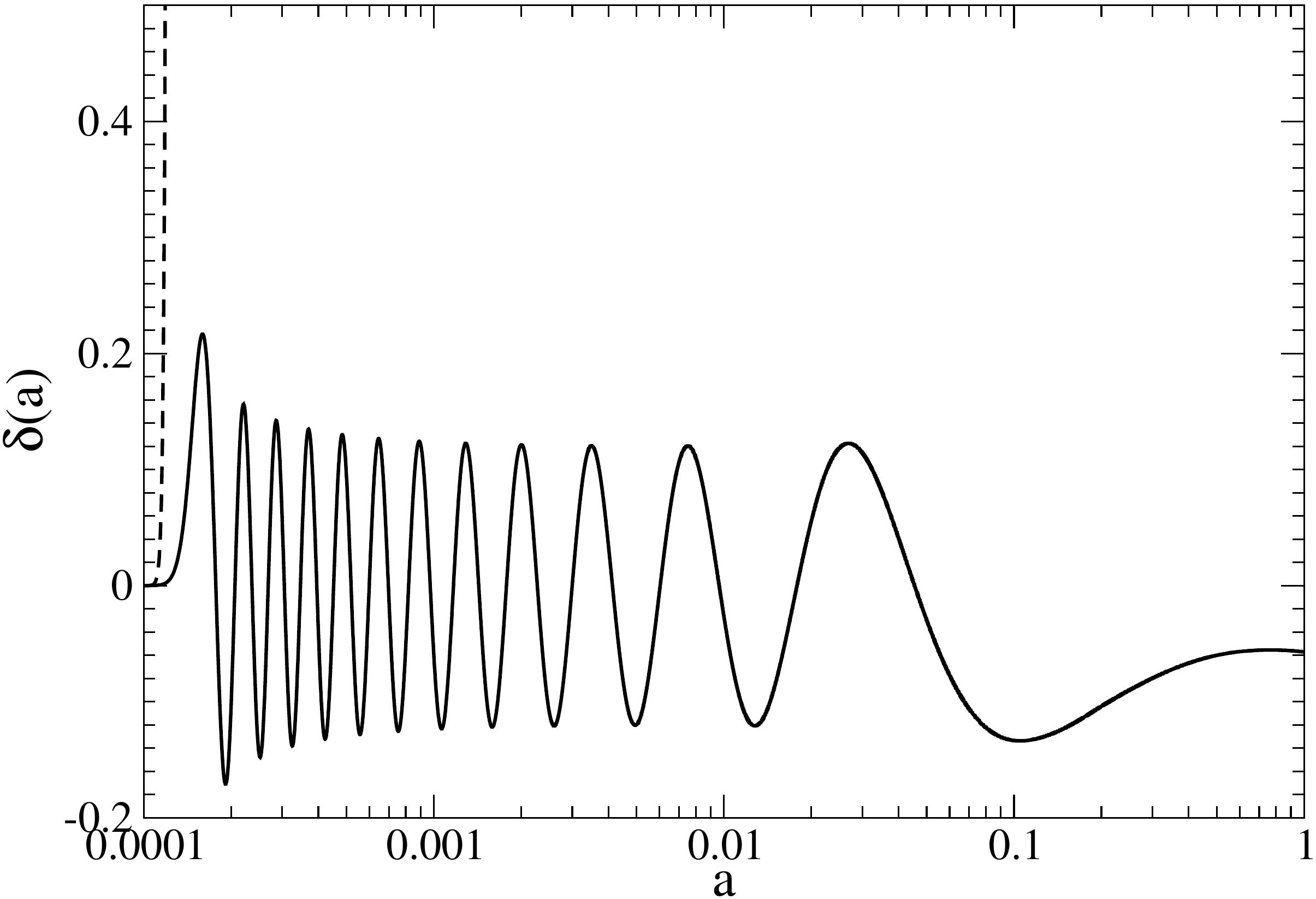}
\caption{Evolution of the density contrast $\delta(a)$ of a
self-interacting SF with $a_s<0$ for $\epsilon=\kappa_Q/\kappa_J=0.0085$ and
$k/k_Q=0.7$  (semi-log plot). The dashed curve
(exponential growth) represents the solution of Eq. (\ref{jn2bis}) without the
quantum term and the solid curve shows the stabilization due to the quantum
term.
\label{axions}}
\end{figure}

These conclusions are not true anymore if $|a_s|$ is extremely
small (typically $|a_s|\ll 10^{-60}\, {\rm fm}$) because, in that case, we need
to take into account the quantum term  in Eq. (\ref{jn2bis}) that
stabilizes the structures at small scales. The quantum term becomes efficient
when $a\gg\kappa_Q^4/\kappa_J^2k^2$. Therefore, it rapidly stabilizes the modes
with large $k$ (small scales) and stops their growth. Indeed, after an
exponential growth, the perturbation starts to oscillate (see Fig.
\ref{axions}). Modes with intermediate $k$ (intermediate scales) grow
exponentially rapidly because the range where the attractive
self-interaction dominates is longer (i.e. $\kappa_Q^4/\kappa_J^2k^2$ is
larger). Finally, modes with small $k$ (large scales) grow slower (linearly
with $a$) and behave like the CDM model.  Therefore, bosons with an extremely
small negative scattering length may be interesting DM candidates (as suggested
in \cite{chavaniscosmo}) because they
can accelerate the growth of perturbations, hence the formation of structures,
at intermediate scales without forming too many substructures at small scales.
These considerations may be relevant to QCD axions because they usually have an
extremely small negative scattering length \cite{sikivie}.

\subsection{Relativistic attenuation when $\lambda$ approaches the cosmological
horizon}
\label{sec_relat}

We now consider the case where the wavelength  $\lambda$ of the perturbation
approaches the cosmological horizon $\lambda_H$ so that relativistic effects
must be taken into account.

If we consider wavelengths $\lambda\sim
\lambda_H$, or larger, then the Compton term $k^2/\kappa_C^2a^2$ in Eq.
(\ref{me10}) is negligible in front of
unity ($k^2/\kappa_C^2a^2\ll 1$, i.e. $\lambda\gg\lambda_C$) because
$\lambda_H\gg\lambda_C$ (see Appendix \ref{sec_na}). Since we have shown in Sec.
\ref{sec_justif} that the Compton term is negligible in front of the quantum
term when $\lambda\ll \lambda_H$, we conclude that the Compton term is always
negligible in the matter era. We have also shown in Sec. \ref{sec_justif} that
$\sigma/a^3\ll 1$ in the matter era. Therefore, the evolution of the density
contrast in the matter era, valid for any perturbation, is
determined by the equation
\begin{eqnarray}
\frac{d^2\delta}{da^2}+\frac{3}{2a}\frac{d\delta}{da}+\frac{3}{2a^2}
\biggl (\frac { k^4
}{\kappa^4_Qa}+\frac{k^2}{\kappa_J^2a^2}
-\frac{1}{1+\frac{\kappa_H^2}{k^2a}}
\biggr )\delta=0.
\nonumber\\
\label{relat1}
\end{eqnarray}

\begin{figure}[!ht]
\includegraphics[width=0.98\linewidth]{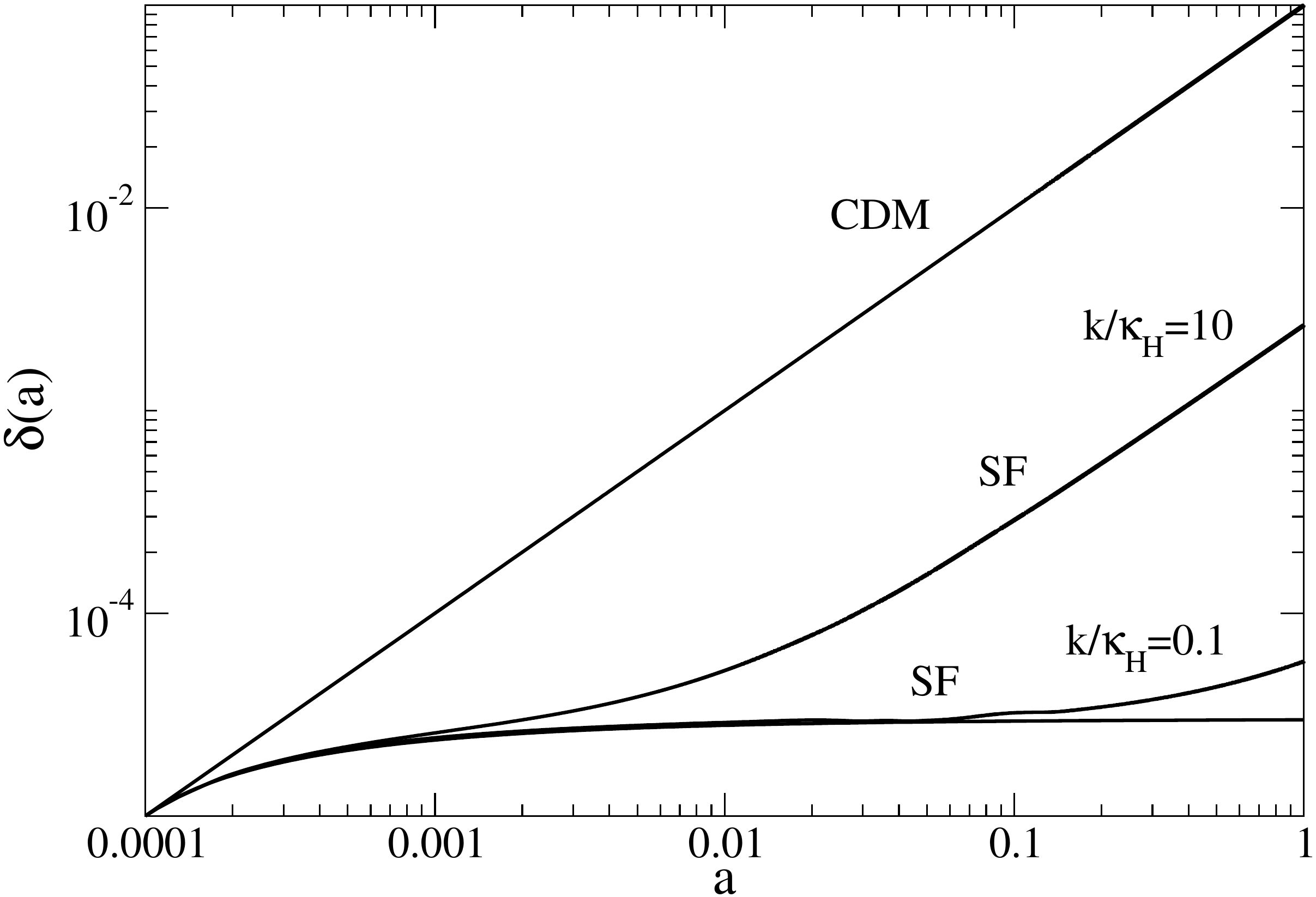}
\caption{Evolution of the density contrast $\delta(a)$ of a
self-interacting
SF for $k/\kappa_H=10$ and  $k/\kappa_H=0.1$ (log-log plot). We have also
plotted
the CDM solution (recovered for $k\gg
\kappa_H/\sqrt{a}$) and the asymptotic solution (\ref{relat4}) valid for
$k\rightarrow 0$.
\label{relat}}
\end{figure}

In the matter era, we have $\lambda_C\ll (\lambda_Q,\lambda_J)\ll\lambda_H$ (see
Appendix
\ref{sec_na}). For $\lambda\ll (\lambda_Q,\lambda_J)$, we can use the
nonrelativistic results of Secs. \ref{sec_nid}-\ref{sec_both} since
$\lambda\ll\lambda_H$. In that case, we have seen that the perturbation
oscillates because the wavelength is smaller than the Jeans length. 
We now assume $\lambda\gg (\lambda_Q,\lambda_J)$. In that case, Eq.
(\ref{relat1}) can be approximated by 
\begin{eqnarray}
\frac{d^2\delta}{da^2}+\frac{3}{2a}\frac{d\delta}{da}-\frac{3}{2a^2}\frac{1}{
1+\frac{\kappa_H^2}{k^2a}}\delta=0.
\label{relat2}
\end{eqnarray}
We solve Eq. (\ref{relat2}) with the initial condition $\delta_i=10^{-5}$ and
$\dot\delta_i=\delta_i/a_i=0.1$ at $a_i=10^{-4}$, obtained from Eq.
(\ref{jn4}). The last term in Eq. (\ref{relat2}) corresponds to the
gravitational attraction. For $\lambda\ll\lambda_H$, we can use the
results of Secs. \ref{sec_nid}-\ref{sec_both} and we find that the perturbation
grows linearly with $a$, exactly like in the CDM model. Indeed, the term
${\kappa_H^2}/{k^2a}$ can be neglected and Eq. (\ref{relat2}) reduces to Eq.
(\ref{jn3}) which has the growing solution of Eq. (\ref{jn4}).
We now consider the situation where $\lambda$ approaches the Hubble length 
$\lambda_H$.
In that case, the weight of the gravitational term is reduced because of
relativistic effects and, consequently, the
perturbations grow slower. This is illustrated in Fig. \ref{relat}. When
$\lambda\rightarrow +\infty$, the gravitational term becomes negligible and Eq.
(\ref{relat2}) reduces to
\begin{eqnarray}
\frac{d^2\delta}{da^2}+\frac{3}{2a}\frac{d\delta}{da}=0.
\label{relat3}
\end{eqnarray}
It has the solution 
\begin{eqnarray}
\delta(a)=\delta_i\left (3-2\sqrt{\frac{a_i}{a}}\right )
\label{relat4}
\end{eqnarray}
which  tends to a constant $3\delta_i$ when $a\rightarrow +\infty$. This shows
that when the wavelength approaches or overcomes the cosmological horizon, the
perturbations do not grow anymore. There is a attenuation of relativistic
origin.

Coming back to the general equation (\ref{relat1}) and summarizing our results,
we have shown that the relativistic term $1+k_H^2/k^2$ sets a natural upper
cutoff, of the
order of the Hubble length $\lambda_H$, above which  the perturbations are
attenuated. On the other hand, the terms $k^4/k_Q^4$ and 
$k^2/k_Q^2$ set natural lower cutoffs, of the order of the Jeans length
$\lambda_Q$ or $\lambda_J$, below which the perturbations oscillate. In
between, the perturbations grow linearly with $a$ as in the CDM model.
Therefore, the effect of the relativistic SF is to introduce sharp cut-offs in
the matter power spectrum at small scales $\lambda_Q$ or $\lambda_J$ because of
quantum mechanics and at large scales  $\lambda_H$ because of general
relativity.

\section{About the importance of relativistic effects}

Before concluding this paper, we would like to discuss the importance, or
non-importance, of relativistic effects in the SFDM model.

In the matter era, we have shown that relativistic effects can be neglected
as long as the wavelength of the perturbations is much smaller than the
cosmological horizon. This is the case for most perturbations of physical
interest. Since there is a wide range of scales satisfying
$(\lambda_Q,\lambda_J)\ll\lambda\ll\lambda_H$, the perturbations can grow during
the matter era and lead
to the formation of structures in the nonlinear regime. However, in the stiff
matter era and in the
radiation era, relativistic effects are important because the
wavelength of
the perturbations may more easily reach the horizon (which is smaller). Since
general relativity tends to prevent the growth of perturbations when their
wavelength
approaches the horizon, this may explain why the
perturbations do not grow in the
radiation era (the Jeans length may be larger than the horizon). This
interesting
effect will be studied in detail in future works.

On the other hand, even if relativistic corrections are weak in the matter era,
we may wonder whether their effect could be detected in an era of ``precision
cosmology" \cite{precision}. In particular, it would be interesting to see if
one
can observe differences between the KGE equations considered in this paper and
the heuristic KGP equations studied  in the past, in which gravity is introduced
by hand in the KG
equations (see Appendix \ref{sec_kgp}).

Relativistic effects are important for other astrophysical systems
described by SFs or BECs. We can mention, for example, the case of boson
stars
\cite{kaup,rb,thirring,breit,takasugi,colpi,bij,gleiserseul,
ferrellgleiser,gleiser,
seidel90,
kusmartsev,kusmartsevbis,leepang,jetzer,seidel94,balakrishna,schunckliddle,
mielkeschunck,torres2000,wang,mielke,guzmanbh,chavharko} and the
case of
microscopic quantum black holes made of BECs of gravitons stuck at a
quantum critical point \cite{dvali,casadio,bookspringer}. It has
also
been proposed \cite{chavharko} that, because of their superfluid core, neutron
stars could be BEC
stars. Indeed, the neutrons (fermions) could form Cooper pairs and behave as
bosons of mass $2m_n$. This idea may solve certain issues regarding the maximum
mass of neutron stars. To be complete, we should also mention analog 
models of gravity in which condensed matter systems described by the GP
equation or by the KG equation are used to simulate
results of classical and quantum field theory in curved space-time
\cite{unruh,volovik,novello}.
One of the first
models for experimentally simulating black hole (Hawking) evaporation was
suggested by Unruh
in 1981 \cite{unruh}. Fifteen years later, a sonic analog of black holes 
specifically using BECs was developed \cite{garay}. Since then, various
configurations of relativistic BECs have been proposed to simulate different
scenarios of gravity \cite{bl1,ff,bl2}.

\section{Conclusion}
\label{sec_conclusion}

In this paper, we have developed  a formalism based on a relativistic
SF described by the KGE equations in the weak field limit. We have transformed
these equations into equivalent hydrodynamic
equations. These equations are arguably more tractable than the KGE equations
themselves. In the nonrelativistic limit, they reduce to the  hydrodynamic
equations directly
obtained from the GPP equations \cite{chavaniscosmo}. Therefore, our
formalism clarifies the
connection between the relativistic and nonrelativistic treatments. We note
that, in the relativistic regime, the hydrodynamic variables $\psi$, $\rho$,
$S$, $\vec v$, $p$,... that we have introduced do not have a direct physical
interpretation. It is only in the nonrelativistic limit $c\rightarrow
+\infty$ that they coincide with
the wavefunction, rest-mass density, action, velocity, and pressure. However,
these variables are perfectly well-defined mathematically from the SF 
$\varphi$ in any regime, and they are totally legitimate. Furthermore, in terms
of these variables, the relativistic hydrodynamic equations take a relatively
simple form that provides a natural generalization of the nonrelativistic
hydrodynamic equations. The main advantage in using
the hydrodynamic representation of the SF is the facility of interpreting the
results since the hydrodynamic equations are expressed in terms of
familiar physical variables (density, velocity, pressure...). In addition, the
use of these variables avoids the need of making averages over the oscillations
of the SF as discussed in Sec. \ref{sec_b}. This is valuable because it is not
easy in practice to deal with the oscillations of the SF when directly solving
the KGE equations numerically.

The complete study of the relativistic hydrodynamic equations is of
considerable interest but it is, of course, of great complexity. In this paper,
we 
have started their study in simple cases and we have obtained explicit results.
Our paper shows therefore that these equations can be useful in practice. 

First, we have considered the relativistic Jeans
instability problem in a static universe. We have derived the relativistic
dispersion relation and the relativistic Jeans length. Physical  applications of
these results  will be developed in our companion papers. 

Secondly, we have checked that the hydrodynamic equations of the SFDM
model reproduce the
evolution of the homogeneous background obtained previously by Li {\it et al.}
\cite{shapiro} directly from the KGE equations: a stiff matter era, followed by
a radiation era (for a self-interacting SF), and a matter era.

Finally, we have started to analyze the evolution of the perturbations in an
expanding universe dominated by a SF. We have considered the linear regime and
we have focused on
the matter era during which the growth of structures is expected to take place.
In this regime, we can use the nonrelativistic limit of our formalism to treat
the  perturbations that are well inside the cosmological
horizon ($\lambda\ll\lambda_H$). We have evidenced  analytically and
numerically different regimes. Perturbations with $\lambda\ll\lambda_J$
oscillate with an
amplitude growing as $a^{1/4}$, perturbations with $\lambda\ll\lambda_Q$
oscillate with a constant amplitude,
perturbations with $(\lambda_Q,\lambda_J)\ll\lambda\ll\lambda_H$
grow linearly with $a$ like in the CDM model, and perturbations with
$\lambda\sim\lambda_H$ are
attenuated by general relativity and do not grow anymore. Therefore,
the
relativistic SF introduces sharp cut-offs in
the matter power spectrum at small scales ($\lambda_Q$ or $\lambda_J$) because
of
quantum mechanics and at large scales  ($\lambda_H$) because of general
relativity.

Generically, we can use the TF approximation at early times, then the
non-interacting approximation, and finally the classical (non quantum)
approximation. As a result, the
perturbations first display growing oscillations, followed by constant amplitude
oscillations, and finally grow linearly with $a$ similarly to the CDM model (see
Fig. \ref{plot1}). The growth of
perturbations in the SF model is substantially faster than in the CDM model
in agreement with certain cosmological observations
where large-scale structures are observed at high redshifts \cite{nature}.
Depending on the wavelength of the perturbations, the duration of these
different regimes in the matter-dominated era can change. We have
provided all the scaling laws necessary to have a complete picture of the
evolution of the perturbations in any case.  In future works, we will extend our
study of perturbations in the radiative era
and in the stiff matter era where relativistic effects are expected to play an 
important role. 

In this paper, we have studied the development of perturbations of the SFDM
model in the linear regime.
In future works, it will be important to consider
the nonlinear
regime where structure formation actually
occurs. In
general, this problem must
be addressed numerically. The hydrodynamic equations
derived in this paper may be very helpful because they may be easier to solve
than the KGE equations.\footnote{Since our equations
are based on the Newtonian metric assuming that $\Phi/c^2\ll 1$, they cannot be
extrapolated to the nonlinear relativistic regime where terms like $\Phi^2/c^4$
must be taken into account. However, our equations are exact in the
linear relativistic regime which is relevant to describe the
radiation era, and in 
the nonlinear nonrelativistic regime ($c\rightarrow +\infty$) which is relevant
to study the formation of structures in the matter era.} Therefore, numerical
simulations
using fluid dynamics should be developed in the future. As a first step,
relativistic effects could be neglected and the nonrelativistic equations of
Appendix \ref{sec_nr} could be used.\footnote{It may be mentioned that very
nice numerical simulations have been made recently for nonrelativistic
non-interacting SFDM/BECDM by directly solving the SP equations \cite{ch2,ch3}.}
These equations are similar to the
hydrodynamic equations of CDM
except that they include a quantum potential (Heisenberg) and a pressure term
(scattering) that avoid singularities at small scales and may solve the missing
satellite
problem \cite{chavaniscosmo}. In
this
respect, it may be recalled that the SP equations
were introduced early by Widrow and Kaiser \cite{widrow} as a
procedure of small-scale regularization (a sort of mathematical trick) to
prevent singularities in collisionless simulations of {\it
classical} particles. In their approach, $\hbar$ is not the Planck
constant, but rather an ajustable parameter that controls the spatial
resolution. Their procedure finds a
physical justification if DM is made of self-gravitating BECs \cite{chavkpz}.

In this paper, we have studied the evolution of a universe
containing exclusively a SF representing dark matter. If we want to construct a
more general cosmological
model, we must take into account the contribution of other components such as
normal\footnote{We call it ``normal'' in order to distinguish it
from the radiation produced by a relativistic self-interacting SF, which has a
different
physical origin (see \cite{shapiro} and Sec. \ref{sec_radmatt}).} radiation
(due to relativistic particles: photons, neutrinos...), baryonic matter,
dark energy (e.g. the cosmological constant), and possibly non-gravitational
couplings (e.g. electromagnetism). Normal radiation, baryons, and the
cosmological constant can be introduced straightforwardly in the
energy-momentum tensor or in the Friedmann
equations. Since
our perturbation analysis is carried out during the
matter-dominated era (dark matter), corresponding to $10^{-4}<a<1$, the
contribution  to the density contrast of the components different from the SF
is subdominant, so these components are not expected to change the evolution of
the
perturbations significantly. In the radiation-dominated era, corresponding to
$a<10^{-4}$, the growth of structures is inhibited because the velocity
dispersion of the particles dominates their gravitational attraction (free
streaming) and the wavelength of the perturbations that could trigger
gravitational collapse lies outside the horizon making these perturbations
ineffective. Non-trivial
gravitational couplings must be treated specifically. For example,
electromagnetism can be taken into account by adding the contribution of the
Maxwell tensor in the Lagrangian. The hydrodynamic equations can then
be generalized. In the case of a complex SF (charged),
electromagnetic forces can have a significant influence on the formation of
structures due to vortical motion.

We would like to conclude this paper by recalling
the interest of the SFDM model with respect to the CDM model for what
concerns astrophysical and cosmological observations. One of the
main observations that can be related to our study is
the recently observed large-scale structures at high redshifts \cite{nature}.
Contrary to the CDM model which does not seem to give sufficient time for such
structures to form, SFDM perturbations seem to grow somewhat faster,
thereby facilitating the formation of structures at earlier times, or more
rapidly. Another observation that can be related to our work is the ``missing
satellite problem'' \cite{satellites}.  
Numerical simulations of CDM lead to an over-abundance of small structures
because the Jeans length is equal to zero (or is
very small). Such a large number of satellite galaxies is not
observed in the Universe. By contrast, the  SFDM model provides  a natural
cutoff
for the formation of small structures, due to quantum mechanics, giving a
possible solution to the missing satellite problem because the Jeans
length is finite. Another problem of the CDM model is the ``cusp-core problem''.
The CDM model predicts that DM halos should be cuspy (the central density should
diverge as $r^{-1}$) \cite{nfw} while observations reveal that they have a flat
core \cite{observations}. Once again, the quantum properties of the SFDM model
may solve this problem. Indeed, the quantum potential arising from the
Heisenberg uncertainty principle (for a noninteracting SF) or the pressure due
to the
scattering (for a self-interacting SF) prevent gravitational collapse at
small scales and lead to central density cores instead of cusps. Other
astrophysical and cosmological applications of our SFDM formalism will
be
given in future papers.

\appendix

\section{The value of $A$}
\label{sec_cons}

The GP equation is obtained from the KG equation by means of the transformation
\begin{eqnarray}
\varphi=A e^{-i m c^2
t/\hbar}\psi.
\label{k1}
\end{eqnarray}
The constant $A$ can be computed as follows. Substituting Eq. (\ref{k1})
in Eq. (\ref{kge5}), we find that the energy density of the SF is
given by
\begin{eqnarray}
\frac{\epsilon}{c^2}=\frac{T_0^0}{c^2}=\frac{1}{2}\left
(1-\frac{2\Phi}{c^2}\right
)\frac{m^2}{\hbar^2}A^2|\psi|^2+\frac{m^2}{2\hbar^2}A^2
|\psi|^2\nonumber\\
+\frac{A^2}{2c^4}\left (1-\frac{2\Phi}{c^2}\right )\left
|\frac{\partial\psi}{\partial t}\right |^2+\frac{A^2}{2a^2c^2}\left
(1+\frac{2\Phi}{c^2}\right )|\nabla\psi|^2\nonumber\\
+\frac{2\pi a_s m A^4}{\hbar^2c^2}|\psi|^4-\frac{m A^2}{\hbar
c^2}\left (1-\frac{2\Phi}{c^2}\right ){\rm Im} \left
(\frac{\partial\psi}{\partial t}\psi^*\right ).  \nonumber\\
\label{k2}
\end{eqnarray}
Taking the nonrelativistic limit $c\rightarrow +\infty$ of this equation, we
obtain
\begin{eqnarray}
\frac{\epsilon}{c^2}\rightarrow
\frac{m^2A^2}{\hbar^2}|\psi|^2=\frac{m^2 A^2}{\hbar^2}\rho,
\label{k3}
\end{eqnarray}
where $\rho=|\psi|^2$ is the rest-mass density. Since $\epsilon\sim\rho c^2$ in
the nonrelativistic limit $c\rightarrow +\infty$, we find
\begin{eqnarray}
A=\frac{\hbar}{m}.
\label{k5}
\end{eqnarray}

\section{Spatially inhomogeneous clusters}
\label{sec_inh}

We consider a spatially inhomogeneous SF/BEC cluster at equilibrium,
representing for example a dark matter star, a boson star, or a dark
matter halo. This cluster results from the nonlinear development of the Jeans
instability. It corresponds to a steady state of the
hydrodynamic equations (\ref{kge15})-(\ref{kge17}) with $a=1$, $H=0$, $\rho(\vec
x,t)=\rho(\vec x)$,  $\Phi(\vec
x,t)=\Phi(\vec x)$, $\vec v(\vec x,t)=\vec 0$, and $S(\vec
x,t)=S(t)$.

The equation of continuity (\ref{kge15}) implies that $S=-Et$ where $E$ is a
constant. Then, the Hamilton-Jacobi equation (\ref{kge18}) and the Einstein
equation (\ref{kge17}) reduce to 
\begin{eqnarray}
\left (1+\frac{4\Phi}{c^2}\right )\frac{\hbar^2}{2
m}\frac{\Delta\sqrt{\rho}}{\sqrt{\rho}}
-m\Phi-\frac{4\pi a_s \hbar^2\rho}{m^2}\left
(1+\frac{2\Phi}{c^2}\right )\nonumber\\
+E\left (1+\frac{E}{2mc^2}\right )=0,\qquad
\label{he2}
\end{eqnarray}
\begin{eqnarray}
\frac{\Delta\Phi}{4\pi G}=\left (1-\frac{\Phi}{c^2}\right )\rho
+\frac{E}{mc^2}\left (1+\frac{E}{2mc^2}\right )\left (1-\frac{2\Phi}{c^2}\right
)\rho\nonumber\\
+\frac{\hbar^2}{8m^2c^2}\left (1+\frac{2\Phi}{c^2}\right
)\frac{(\vec\nabla\rho)^2}{\rho}
+\frac{2\pi a_s \hbar^2}{m^3c^2}\rho^2.\qquad
\label{he1}
\end{eqnarray}

These two coupled equations determine the equilibrium state of a spatially
inhomogeneous relativistic self-gravitating SF/BEC.  If we neglect the terms in
$\Phi/c^2$ (see the simplified model of Appendix \ref{sec_kgp}), they reduce to 
\begin{eqnarray}
\frac{\hbar^2}{2
m}\frac{\Delta\sqrt{\rho}}{\sqrt{\rho}}
-m\Phi-\frac{4\pi a_s \hbar^2\rho}{m^2}
+E\left (1+\frac{E}{2mc^2}\right )=0,\qquad
\label{he4}
\end{eqnarray}
\begin{eqnarray}
\frac{\Delta\Phi}{4\pi G}=\rho
+\frac{E}{mc^2}\left (1+\frac{E}{2mc^2}\right )\rho\nonumber\\
+\frac{\hbar^2}{8m^2c^2}\frac{(\vec\nabla\rho)^2}{\rho}
+\frac{2\pi a_s \hbar^2}{m^3c^2}\rho^2.
\label{he3}
\end{eqnarray}
Taking the Laplacian of Eq. (\ref{he4}) and substituting the result in Eq.
(\ref{he3}), we obtain a single differential equation for the density
\begin{eqnarray}
\frac{\hbar^2}{8\pi G m^2}\Delta\left
(\frac{\Delta\sqrt{\rho}}{\sqrt{\rho}}\right
)-\frac{a_s\hbar^2}{Gm^3}\Delta\rho=\rho+\frac{\hbar^2}{8m^2c^2}\frac{
(\vec\nabla\rho)^2}{\rho}\nonumber\\
+\frac{2\pi
a_s\hbar^2}{m^3c^2}\rho^2+\frac{E}{mc^2}\left (1+\frac{E}{2mc^2}\right
)\rho.\quad
\label{he5}
\end{eqnarray}
In the nonrelativistic limit $c\rightarrow +\infty$, it reduces to
\begin{eqnarray}
\frac{\hbar^2}{2m^2}\Delta\left
(\frac{\Delta\sqrt{\rho}}{\sqrt{\rho}}\right
)-\frac{4\pi a_s\hbar^2}{m^3}\Delta\rho=4\pi G\rho.
\label{he6}
\end{eqnarray}
This returns the differential equation  describing
the equilibrium state of a
spatially inhomogeneous nonrelativistic self-gravitating BEC studied in
\cite{prd1,prd2}.

\section{The nonrelativistic limit}
\label{sec_nr}

In the nonrelativistic limit $c\rightarrow +\infty$, the evolution of the wave
function $\psi(\vec x,t)$ in an expanding universe is determined by the GPP
equations \cite{chavaniscosmo}:
\begin{eqnarray}
i\hbar\frac{\partial\psi}{\partial
t}+\frac{3}{2}i\hbar
H\psi=-\frac{\hbar^2}{2 m a^2}\Delta\psi+m\Phi \psi
+\frac{4\pi a_s \hbar^2}{m^2}|\psi|^2\psi,\nonumber\\
\label{nr1}
\end{eqnarray}
\begin{eqnarray}
\frac{\Delta\Phi}{4\pi G a^2}=|\psi|^2 -\frac{3H^2}{8\pi G}.
\label{nr2}
\end{eqnarray}
These equations are equivalent to the quantum Euler-Poisson equations
\cite{chavaniscosmo}:
\begin{eqnarray}
\frac{\partial\rho}{\partial t}+3H\rho+\frac{1}{a}\vec\nabla\cdot (\rho
{\vec v})=0,
\label{nr3}
\end{eqnarray}
\begin{eqnarray}
\frac{\partial S}{\partial t}+\frac{(\nabla S)^2}{2 m
a^2}=\frac{\hbar^2}{2
m a^2}\frac{\Delta\sqrt{\rho}}{\sqrt{\rho}}
-m\Phi-\frac{4\pi a_s \hbar^2\rho}{m^2},
\label{nr6}
\end{eqnarray}
\begin{eqnarray}
\frac{\partial {\vec v}}{\partial t}+H{\vec v}+\frac{1}{a}({\vec v}\cdot
\vec\nabla){\vec v}=\nonumber\\
\frac{\hbar^2}{2m^2a^3}\vec\nabla\left( \frac{\Delta\sqrt{\rho}}{\sqrt{
\rho}} \right )-\frac{1}{a}
\vec\nabla\Phi-\frac{
1}{\rho a}\vec\nabla p,
\label{nr4}
\end{eqnarray}
\begin{eqnarray}
\frac{\Delta\Phi}{4\pi G a^2}=\rho-\frac{3H^2}{8\pi
G},
\label{nr5}
\end{eqnarray}
where the pressure $p$ is given by Eq. (\ref{kgp20}). The
Euler equation (\ref{nr4}) includes a quantum potential
$Q=-\hbar^2/(2ma^3)\Delta(\sqrt{\rho})/\sqrt{\rho}$ arising from the
Heisenberg uncertainty principle. In the classical limit
$\hbar\rightarrow 0$ (or in the TF limit), we recover the classical 
Euler-Poisson equations for a barotropic fluid \cite{bt}.

For a spatially homogeneous SF/BEC, the hydrodynamic equations
(\ref{nr3})-(\ref{nr5}) reduce to
\begin{eqnarray}
\frac{d\rho_b}{dt}+3H\rho_b=0,
\label{nr15}
\end{eqnarray}
\begin{eqnarray}
E=-\frac{dS_b}{dt}=\frac{4\pi a_s \hbar^2\rho_b}{m^2},
\label{nr16}
\end{eqnarray}
\begin{eqnarray}
\frac{3H^2}{8\pi G}=\rho_b.
\label{nr17}
\end{eqnarray}
From Eq. (\ref{nr15}), we find that $\rho_b\propto a^{-3}$, so the
homogeneous SF/BEC behaves as CDM. Then, Eq.
(\ref{nr17}) implies $a\propto t^{2/3}$ and $\rho_b=1/(6\pi G t^2)$, which is
the EdS solution. Combining these results with Eq.
(\ref{nr16}), we find that the energy of the homogeneous SF decreases
as
\begin{eqnarray}
E(t)=\frac{2a_s \hbar^2}{3Gm^2t^2}.
\label{nr19}
\end{eqnarray}
Accordingly, its phase decreases as $S_b(t)={2a_s
\hbar^2}/{3Gm^2t}+C$. To our
knowledge, these results have not been given before.
The energy density
and the pressure of the homogeneous SF are given by 
\begin{eqnarray}
\epsilon_b=\rho_b c^2,\qquad P_b=\frac{2\pi a_s
\hbar^2}{m^3}\rho_b^2.
\label{nr18}
\end{eqnarray}

The case of a static Universe is recovered by taking $a=1$ and $H=0$ in the
foregoing equations. For a static homogeneous SF, we find that
$S_b(t)=-Et$ with
\begin{eqnarray}
E=\frac{4\pi
a_s \hbar^2\rho_b}{m^2}=mc_s^2.
\label{nr12}
\end{eqnarray}
The energy of the SF is given by a sort of Einstein equation
``$E=mc^2$'' where the speed of light is replaced by the speed of sound.

\section{Generalized Klein-Gordon-Poisson equations}
\label{sec_kgp}

In this Appendix, we consider a simplified model in which we introduce the
gravitational potential $\Phi(\vec x,t)$ in the ordinary KG
equation by hand, as an external potential, and assume that this potential
is produced
by the SF itself via a generalized Poisson equation in which the source is the
energy density $\epsilon$. This leads to
the generalized KGP equations. We then show that these equations can be
rigorously justified from the KGE
equations in the limit $\Phi/c^2\rightarrow 0$. Finally, we argue that this
simplified model is not sufficient to study the evolution of the perturbations
in the linear relativistic regime.

\subsection{The FLRW metric and the phenomenological interaction Lagrangian}

We consider the FLRW metric that describes an isotropic and
homogeneous expanding background. The line element in the comoving frame is
\begin{equation}
ds^2=g_{\mu\nu}dx^{\mu}dx^{\nu}=c^2dt^2-a(t)^2\delta_{ij}dx^idx^j.
\label{kgp1}
\end{equation}
For this metric, the d'Alembertian operator
(\ref{tb7}) writes
\begin{equation}
\Box=\frac{1}{c^2}\frac{\partial^2}{\partial
t^2}+\frac{3H}{c^2}\frac{\partial}{\partial t}-\frac{1}{a^2}\Delta.
\label{kgp2}
\end{equation}

In order to take the self-gravity of the SF into account, we
introduce a Lagrangian of interaction that couples the gravitational potential
$\Phi(\vec x,t)$ to the scalar
field $\varphi(\vec x,t)$ according to
\begin{equation}
\mathcal{L}_{\rm int}=-\frac{m^2}{\hbar^2}\Phi|\varphi|^2.
\label{kgp3}
\end{equation} 
The total Lagrangian of the system (SF $+$ gravity) is
given by $\mathcal{L}=\mathcal{L}_\varphi+\mathcal{L}_{\rm int}$.

\subsection{The Klein-Gordon equation}

The equation of motion resulting from the stationarity of the
total action $S=S_{\varphi}+S_{\rm int}$, obtained by writing $\delta
S=0$, is the KG equation
\begin{equation}
\Box\varphi+2V(|\varphi|^2),_{\varphi^*}+\frac{2m^2}{\hbar^2}\Phi\varphi=0,
\label{kgp5}
\end{equation}
where the d'Alembertian operator is given by Eq. (\ref{kgp2}) and the
gravitational
potential  $\Phi(\vec x,t)$ acts here as an external potential.
For the specific SF potential  (\ref{tb3}), we obtain
\begin{eqnarray}
\frac{1}{c^2}\frac{\partial^2\varphi}{\partial
t^2}+\frac{3H}{c^2}\frac{\partial\varphi}{\partial
t}-\frac{1}{a^2}\Delta\varphi\nonumber\\
+\left (1+\frac{2\Phi}{c^2}\right ) \frac{m^2
c^2}{\hbar^2}\varphi
+\frac{8\pi a_s m}{\hbar^2}|\varphi|^2\varphi =0.\label{kgp6}
\end{eqnarray}
The energy density
and the pressure, defined from the diagonal part of the
energy-momentum tensor (\ref{tb12}), are given by
\begin{equation}
\epsilon=\frac{1}{2c^2}\left |\frac{\partial\varphi}{\partial
t}\right|^2+\frac{1}{2a^2}|\vec\nabla\varphi|^2+V(|\varphi|^2),
\label{kgp7}
\end{equation}
\begin{equation}
P=\frac{1}{2c^2}\left |\frac{\partial\varphi}{\partial
t}\right|^2-\frac{1}{6a^2}|\vec\nabla\varphi|^2-V(|\varphi|^2).
\label{kgp8}
\end{equation}

\subsection{The generalized Poisson equation}

Eq. (\ref{kgp6}) is the ordinary KG equation for a SF in an
external potential $\Phi(\vec x,t)$ in an expanding background. We now state
that $\Phi(\vec x,t)$ is actually the gravitational potential produced by the
SF itself. We phenomenologically assume that the gravitational
potential is determined by a generalized Poisson equation of the
form
\begin{eqnarray}
\frac{\Delta\Phi}{4\pi
Ga^2}=\frac{1}{c^2}(\epsilon-\epsilon_b)
\label{kgp9}
\end{eqnarray}
in which the source of the gravitational potential is the energy density
$\epsilon$ of the SF (more precisely, its deviation from the
homogeneous background density $\epsilon_b(t)$). Using Eq. (\ref{kgp7}) for the
energy
density of a SF, and recalling the Friedmann equation (\ref{kge12b}),
the generalized Poisson equation can be written
as
\begin{eqnarray}
\frac{\Delta\Phi}{4\pi
Ga^2}=\frac{1}{2c^4}\left |\frac{\partial\varphi}{\partial
t}\right |^2&+&\frac{1}{2a^2c^2}|\vec\nabla\varphi|^2+\frac{m^2}{2\hbar^2}
|\varphi|^2\nonumber\\
&+&\frac{2\pi a_s m}{\hbar^2c^2}|\varphi|^4-\frac{3H^2}{8\pi
G}.
\label{kgp11}
\end{eqnarray}
Eqs. (\ref{kgp6}) and (\ref{kgp11}) form the generalized KGP equations. They
have been introduced in an {\it ad hoc} manner but they
can be rigorously justified from the  KGE equations 
(\ref{kge4}) and (\ref{kge12}) in the limit $\Phi/c^2\rightarrow 0$
(which, of course, is different from the nonrelativistic limit $c\rightarrow
+\infty$). We see that the gravitational potential $\Phi$ appears in the KG
equation (\ref{kgp6}) due to the cancelation of $c^2$ in the product
$\Phi/c^2\times c^2$ in Eq. (\ref{kge4}). Therefore, we do not have to
introduce $\Phi$ by hand: the generalized KGP equations can be
obtained from the KGE equations  by simply neglecting terms of order $\Phi/c^2$
in these equations.
Similarly, the equations related to the generalized KGP equations (e.g. the
generalized GPP equations, the corresponding hydrodynamic equations...) can be
obtained from the ones related to the KGE equations by neglecting terms of
order $\Phi/c^2$. Therefore, we do not write them explicitly (they are given in
\cite{proceedings}). Now, a warning
is required. This model
correctly describes the homogeneous background for which $\Phi=0$ but it  is not
sufficient to describe the evolution of the perturbations in the linear
relativistic regime
because we must precisely take into account the terms of order $\Phi/c^2$ in
this regime (except, of course, in the nonrelativistic limit $c\rightarrow
+\infty$). Therefore, the
use of the KGE equations is mandatory to study the evolution of the
perturbations in the relativistic regime. To show that
the generalized KGP equations may give wrong results, we derive in the next two
subsections  the dispersion
relation of the perturbations in a static universe (Jeans problem) and the
equation for the density contrast in an
expanding universe based on the  KGP equations and compare their expressions 
with those obtained from the KGE equations.

{\it Remark:} We could also assume that the
gravitational potential is determined by a Poisson equation 
of the form
\begin{eqnarray}
\frac{\Delta\Phi}{4\pi
Ga^2}=\rho
\label{kgp9w}
\end{eqnarray}
in which the source of the gravitational potential is the pseudo rest-mass
density $\rho=|\psi|^2$ of the SF.  Eqs. (\ref{kgp6}) and (\ref{kgp9w}) form the
KGP equations. This approximation has been considered in \cite{abrilMNRAS}.
However, there is an inconsistency in coupling the relativistic KG equation
(\ref{kgp6}) to the classical Poisson equation (\ref{kgp9w}).

\subsection{Dispersion relation in a static universe}

We consider the linear evolution of small perturbations in a static universe.
For the generalized KGP equations, Eqs.  (\ref{kge21})-(\ref{kge24}) can be
reduced to a system of two coupled wave equations
\begin{eqnarray}
\frac{1}{c^2}\frac{\partial^2\sigma}{\partial
t^2}&-&\Delta\sigma=\left
(1+\frac{E}{mc^2}\right )\frac{\partial^2\delta}{\partial
t^2},\label{j13}
\end{eqnarray}
\begin{eqnarray}
\left
(1+\frac{E}{mc^2}\right )\left (\Delta\sigma-\frac{4\pi
G\rho_b}{c^2}\sigma\right )=\frac{\hbar^2}{4m^2}
\Delta\left(\Delta\delta-\frac{1}{c^2} \frac{
\partial^2\delta}{\partial t^2}\right)\nonumber\\
-c_s^2\Delta\delta
-4\pi G\rho_b
\left(1+\frac{c_s^2}{c^2}\right)\delta
-4\pi G\rho_b \frac{E}{mc^2}\left (1+\frac{E}{2mc^2}\right )\delta.\nonumber\\
\label{j14}
\end{eqnarray}
Decomposing the perturbations in Fourier modes, we obtain the dispersion
relation
\begin{eqnarray}
\frac{\hbar^2}{4m^2c^4}\omega^4-\biggl (
1+\frac{3c_s^2}{c^2}+\frac{\hbar^2k^2}{2m^2c^2}\biggr )\omega^2\nonumber\\
+\left\lbrack \frac{\hbar^2k^4}{4m^2}+c_s^2k^2-4\pi
G\rho_b\left(1+\frac{2c_s^2}{c^2}\right)\right\rbrack=0.
\label{j15}
\end{eqnarray}
Taking $\omega=0$ in Eq. (\ref{j15}), we find that the Jeans length is
determined by the equation
\begin{eqnarray}
\frac{\hbar^2k_J^4}{4m^2}+c_s^2k_J^2-4\pi
G\rho_b\left(1+\frac{2c_s^2}{c^2}\right)=0.
\label{j16}
\end{eqnarray}
These equations differ from Eqs. (\ref{kge25}) and (\ref{kge27}) obtained in
Sec. \ref{sec_di} starting from the KGE equations.

\subsection{Approximate equation for the density contrast in an expanding
universe}

We consider the evolution of small perturbations in an expanding universe.
For the generalized KGP equations, Eqs. (\ref{lp5})-(\ref{lp7}) can be reduced
to a system of two coupled wave equations
\begin{eqnarray}
\frac{\partial^2\delta}{\partial t^2}+2H\frac{\partial\delta}{\partial
t}=\frac{c_s^2}{a^2}\Delta\delta-\frac{\hbar^2}{4
m^2a^4}\Delta^2\delta+\frac{1}{a^2}\Delta\Phi,
\label{lx9}
\end{eqnarray}
\begin{eqnarray}
\frac{\Delta\Phi}{4\pi G \rho_b
a^2}=\left (1
+2\frac{c_s^2}{c^2}\right )\delta-
\frac{\hbar^2}{4m^2c^2a^2}\Delta\delta+\frac{\Phi}{c^2}\nonumber\\
+\frac{E}{mc^2}\left
(\frac{E}{2mc^2}+1\right )\delta.\qquad
\label{lx10}
\end{eqnarray} 
Decomposing the perturbations in (spatial) Fourier modes, we obtain a closed
equation for the density contrast
\begin{eqnarray}
\frac{d^2\delta_k}{dt^2}+2H\frac{d\delta_k}{dt}+\biggl\lbrack \frac{\hbar^2k^4}{
4m^2a^4 }+\frac{c_s^2}{a^2}k^2\nonumber\\
-\frac{4\pi G\rho_b}{1+\frac{4\pi G\rho_b
a^2}{k^2c^2}}\left (
1+\frac{3c_s^2}{c^2}+\frac{\hbar^2k^2}{4m^2c^2a^2}
\right )\biggr\rbrack\delta_k=0.\nonumber\\
\label{lx11}
\end{eqnarray} 
In a static universe ($a=1$, $H=0$), taking $\ddot\delta_k=0$ (i.e.
$\omega=0$), we can check that Eq. (\ref{lx11}) reduces to Eq. (\ref{j16})
that determines the relativistic Jeans length derived from the generalized KGP
equations. However, Eq. (\ref{lx11}) differs from Eq. (\ref{lp11}) obtained in
Sec.
\ref{sec_appdelta} starting from the KGE equations.

\section{Typical mass and scattering length of bosonic particles}
\label{sec_typ}

\subsection{Dark matter halos}

We assume that dark matter halos are made of bosons in the form
of BECs. We determine the mass of the bosons that compose dark matter halos
according to whether they are noninteracting or self-interacting. For dark
matter halos, we can use Newtonian gravity. The smallest known dark
matter halo is Willman 1. It has a radius $R=33\, {\rm pc}$ and a mass 
$M=0.39\, 10^6\, M_{\odot}$ \cite{vega2,vega3}. We assume that this most
compact halo is completely condensed, i.e. that it corresponds to the
ground state ($T=0$) of a self-gravitating Bose gas without radiative
halo.\footnote{For a value of the boson mass in the range
$2.57\times
10^{-20}\, {\rm eV}/c^2<m<1.69\times 10^{-2}\, {\rm eV}/c^2$ (see below), we
have $T\ll
T_c$ (where $T_c$ is the condensation temperature) for all
the dark matter halos of the Universe (with mass $M\sim 10^6-10^{11}\,
M_{\odot}$), so they can be
considered to be at $T=0$
\cite{clm}. They have a core-halo structure with a solitonic core (BEC), which
is a stationary solution of the GP equation, surrounded by a halo of scalar
radiation in which the density decreases as $r^{-3}$
similarly to the Navarro-Frank-White (NFW) \cite{nfw} and Burkert
\cite{observations} profiles. This core-halo structure results from a
process of gravitational cooling \cite{seidel94}. Dwarf dark matter halos are
compact objects that have just a solitonic core (BEC) without atmosphere.
Therefore, their size is equal to the size of the soliton. By contrast, large
dark matter halos are extended objects with a core-halo structure. It is the
radiative atmosphere that fixes the size of large dark matter halos. The
atmosphere can be much larger than the size of the soliton (core). The presence
of the
radiative atmosphere solves the apparent paradox that BEC halos at $T=0$
should all have the same radius (in the self-interacting case) or that their
radius should decrease with their mass (in the
non-interacting case), in
contradiction with the observations \cite{prd1,prd2}.}

A completely condensed system of self-gravitating bosons (BEC) without
self-interaction at $T=0$  has the mass-radius relation $MR=9.95\, \hbar^2/(G
m^2)$ \cite{rb,membrado,prd2}. This gives
\begin{equation}
\frac{m}{{\rm eV}/c^2}=9.22\, 10^{-17} \left (\frac{\rm pc}{R}\right
)^{1/2}\left (\frac{M_{\odot}}{M}\right )^{1/2}.
\label{fb2}
\end{equation}
Using the values of $M$ and $R$ corresponding to Willman 1, we obtain a boson
mass $m=2.57\, 10^{-20}\, {\rm eV}/c^2$.\footnote{This result
assumes that (i) Willman 1 is completely condensed without radiative halo,
and that (ii) the observational values of $r_h$ and $M_h$ are accurate. More
precise observational data may change the value of the boson mass $m$ but its
order of magnitude should remain the same.}

A completely condensed system of self-gravitating bosons (BEC)  with
self-interaction at $T=0$ in the TF limit has a unique radius
$R=\pi(a_s\hbar^2/Gm^3)^{1/2}$ 
\cite{goodman,arbey,bohmer,prd1}. This gives
\begin{equation}
\left (\frac{{\rm fm}}{a_s}\right )^{1/3} \left (\frac{m}{{\rm eV}/c^2}\right
)=6.73 \left (\frac{\rm pc}{R}\right )^{2/3}.
\label{fb3}
\end{equation}
Using the value of $R$ corresponding to Willman 1, we obtain $({\rm
fm}/a_s)^{1/3}(mc^2/{\rm eV})=0.654$.  In order to
determine the mass of the bosons, we need another relation (see footnote 1 in
\cite{becstiff}). This relation is
provided by the constraint $\sigma/m<1.25\, {\rm cm}^2/{\rm g}$ set by the
Bullet Cluster \cite{bullet}, where $\sigma=4\pi a_s^2$ is the self-interaction
cross section. Assuming that the bound is reached (this gives a maximum bound on
the mass and on the scattering length of the bosons) we get $(a_s/{\rm
fm})^2({\rm
eV}/mc^2)=1.77\, 10^{-8}$. From these two constraints, we obtain $m=1.69\,
10^{-2}\, {\rm eV}/c^2$ and $a_s=1.73\, 10^{-5}\, {\rm fm}$. This boson mass is
in
agreement with the limit $m<1.87\, {\rm eV}/c^2$ obtained from cosmological
considerations \cite{limjap}. 

The mass $m=2.57\, 10^{-20}\, {\rm eV}/c^2$ obtained for bosons without
self-interaction gives a lower bound on the mass of the bosonic dark matter
particle. Inversely, the mass $m=1.69\, 10^{-2}\, {\rm eV}/c^2$ obtained for
self-interacting bosons in the TF limit gives an upper bound on the
mass of the bosonic dark matter particle. Therefore, the typical mass of
the bosonic particle lies in the range $2.57\, 10^{-20}\, {\rm eV}/c^2<m<1.69\,
10^{-2}\, {\rm eV}/c^2$. The TF limit is valid for sufficiently large scattering
lengths. An estimate of the critical scattering length can be obtained by
substituting $m=2.57\, 10^{-20}\, {\rm eV}/c^2$ in the relation $({\rm
fm}/a_s)^{1/3}(mc^2/{\rm eV})=0.654$. This gives $a_c=6.07\, 10^{-59}\, {\rm
fm}$. For $a_s<a_c$, the mass of the bosonic particle is $m=2.57\, 10^{-20}\,
{\rm eV}/c^2$ and, for $a_c<a_s<1.73\, 10^{-5}\, {\rm fm}$, the mass of the
bosonic particle is $mc^2/{\rm eV}=0.654(a_s/{\rm fm})^{1/3}<1.69\, 10^{-2}\,
{\rm eV}/c^2$.

\subsection{Boson stars}

General relativity is required to describe boson stars.
The maximum mass of boson stars made of non-interacting bosons is
$M_{max}=0.633\, \hbar
c/Gm$ and their minimum radius is $R_{min}=9.53 \, GM_{max}/c^2$
\cite{kaup}. Introducing scaled variables, we get
\begin{equation}
\frac{M_{max}}{M_{\odot}}=8.48\, 10^{-11}\frac{{\rm
eV}/c^2}{m},\quad \frac{R_{min}}{{\rm km}}=14.1 \frac{M_{max}}{M_{\odot}}.
\label{brelat2}
\end{equation}
For $m=1\, {\rm GeV}/c^2$, we
obtain $M_{max}=8.48\,
10^{-20}\,
M_{\odot}$
and $R_{min}=1.20\, 10^{-18}\, {\rm km}$. For $m=10^{-10} \, {\rm eV}/c^2$, we
obtain $M_{max}=0.848\,
M_{\odot}$
and $R_{min}=12.0\, {\rm km}$. For $m=10^{-17} \, {\rm eV}/c^2$, we
obtain $M_{max}=8.48\times 10^{6}\,
M_{\odot}$
and $R_{min}=1.20\times 10^8\, {\rm km}$. For $m=2.57\times
10^{-20}\, {\rm eV}/c^2$, we obtain $M_{max}=3.30\times 10^{9}\, M_{\odot}$
and $R_{min}=1.51\times 10^{-3}\, {\rm pc}$. Some applications of
these numerical results are given in the Introduction.

The maximum mass of boson stars made of self-interacting bosons is
$M_{max}=0.307\, \hbar c^2\sqrt{a_s}/(Gm)^{3/2}$ and their minimum radius is
$R_{min}=6.25 \, GM_{max}/c^2$
\cite{chavharko}. Introducing scaled variables, we get
\begin{equation}
\frac{M_{max}}{M_{\odot}}=1.12\, \left (\frac{a_s}{{\rm fm}}\right
)^{1/2}\left (\frac{{\rm
GeV}/c^2}{m}\right )^{3/2},
\label{brelat3}
\end{equation}
\begin{equation}
\frac{R_{min}}{{\rm km}}=9.27
\frac{M_{max}}{M_{\odot}}.
\label{brelat4}
\end{equation}
We note that these results do not depend 
on the specific mass $m$ and scattering length $a_s$ of the bosons, but only on
the ratio $m^3/a_s$. For $m=1\, {\rm GeV}/c^2$ and $a_s=1\, {\rm fm}$, we
obtain $M_{max}=1.12\,
M_{\odot}$
and $R_{min}=10.4\, {\rm km}$. For $({\rm
fm}/a_s)^{1/3}(mc^2/{\rm
eV})=0.654$, we obtain $M_{max}=6.70\times
10^{13}\, M_{\odot}$ and $R_{min}=20.2\, {\rm pc}$.

\section{Numerical applications}
\label{sec_na}

The complete equation governing the evolution of the density
contrast in the matter era, Eq. (\ref{me10}), can be written in dimensionless
form as 
\begin{eqnarray}
\frac{d^2\delta}{da^2}&+&\frac{3}{2a}\frac{d\delta}{da}+\frac{3}{2a^2}\biggl [
\frac { \kappa^4}{a}+\frac{\epsilon^2\kappa^2}{a^2}\nonumber\\
&-&\frac{1}{1+\frac{\eta^2}{\kappa^2 a}}\left(1+\frac{2\sigma}{a^3}
\right)\left(1+\frac{3\sigma}{a^3}+\frac{\nu^2\kappa^2}{a^2}\right)\biggr ]
\delta=0,\nonumber\\
\label{na1}
\end{eqnarray}
where we have defined $\kappa=k/\kappa_Q$,
$\epsilon={\kappa_Q}/{\kappa_J}=(16\pi\rho_b a^3
a_s^2\hbar^2/Gm^4)^{1/4}$, $\eta=\kappa_H/\kappa_Q=(4\pi G\rho_b
a^3\hbar^2/m^2c^4)^{1/4}$ and $\nu=\kappa_Q/\kappa_C=(\pi G\hbar^2\rho_b
a^3/m^2c^4)^{1/4}$. We note that $\eta=\sqrt{2}\nu$ as a consequence of
Eq. (\ref{jn1}).

\begin{table*}[t]
\centering
\begin{tabular}{|c|c|c|c|c|c|c|}
\hline 
Physical system & $\kappa_Q ({\rm m}^{-1})$ & $\kappa_J ({\rm m}^{-1})$ &
$\kappa_C ({\rm m}^{-1})$ & $\kappa_H ({\rm m}^{-1})$ \\
\hline
DM halos & & & & \\  
\hline
$a_s=0$, $m=2.57\times10^{-20}$eV/c$^2$
& $3.46\times10^{-20}$  & $+\infty$ & $2.60\times10^{-13}$ &
$6.49\times10^{-27}$ \\ 
\hline
$a_s=1.73\times10^{-5}$fm, $m=1.69\times10^{-2}$eV/c$^2$
& $2.81\times10^{-11}$ & $3.08\times10^{-18}$ & $1.71\times10^{5}$ &
$6.49\times10^{-27}$ \\
\hline\
$a_s=6.07\times10^{-59}$fm, $m=2.57\times10^{-20}$eV/c$^2$
& $3.46\times10^{-20}$ & $3.08\times10^{-18}$ & $2.60\times10^{-13}$ &
$6.49\times10^{-27}$ \\
\hline\
Boson stars & & & & \\
\hline 
$a_s=0$, $m=1\times10^{9}$eV/c$^2$
& $6.83\times10^{-6}$  & $+\infty$ & $1.01\times10^{16}$ &
$6.49\times10^{-27}$ \\
\hline
$a_s=0$, $m=1\times10^{-10}$eV/c$^2$  
& $2.16\times10^{-15}$ & $+\infty$ & $1.01\times10^{-3}$ &
$6.49\times10^{-27}$
\\ 
\hline
$a_s=0$, $m=1\times10^{-17}$eV/c$^2$
& $6.83\times10^{-19}$ & $+\infty$ & $1.01\times10^{-10}$ &
$6.49\times10^{-27}$  \\
\hline
$a_s=1$fm, $m=1\times10^{9}$eV/c$^2$
& $6.83\times10^{-6}$ & $1.84\times10^{-4}$ & $1.01\times10^{16}$
& $6.49\times10^{-27}$  \\
\hline
\end{tabular}
\caption{Values of the parameters that appear in the
equation
for the density contrast for different
astrophysical systems. }
\label{table1}
\end{table*}

\begin{table*}[t]
\centering
\begin{tabular}{|c|c|c|c|c|}
\hline 
Physical system & $\epsilon$ & $\sigma$ & $\eta$ & $\nu$ \\
\hline
DM halos & & & & \\
\hline  
$a_s=0$, $m=2.57\times10^{-20}$eV/c$^2$
& $0$ & 0 & $1.87\times10^{-7}$ & $1.33\times10^{-7}$  \\ 
\hline
$a_s=1.73\times10^{-5}$fm, $m=1.69\times10^{-2}$eV/c$^2$
& $9.11\times10^{6}$ & $2.22\times10^{-18}$ & $2.31\times10^{-16}$ &
$1.65\times10^{-16}$ \\
\hline\
$a_s=6.07\times10^{-59}$fm, $m=2.57\times10^{-20}$eV/c$^2$
& $1.12\times10^{-2}$ & $2.21\times10^{-18}$ & $1.87\times10^{-7}$ &
$1.33\times10^{-7}$ \\
\hline\
Boson stars & & & & \\
\hline 
$a_s=0$, $m=1\times10^{9}$eV/c$^2$
& 0 & 0 & $9.49\times10^{-22}$ & $6.77\times10^{-22}$   \\
\hline
$a_s=0$, $m=1\times10^{-10}$eV/c$^2$  
& 0 & 0 & $3.00\times10^{-12}$ & $2.14\times10^{-12}$   \\ 
\hline
$a_s=0$, $m=1\times10^{-17}$eV/c$^2$
& $0$ & $0$ & $9.49\times10^{-9}$ &
$6.77\times10^{-9}$ \\
\hline
$a_s=1$fm, $m=1\times10^{9}$eV/c$^2$
& $3.70\times10^{-2}$ & $6.19\times10^{-46}$ & $9.49\times10^{-22}$ &
$6.77\times10^{-22}$ \\
\hline
\end{tabular}
\caption{Values of the parameters that appear in the equation
for the density contrast for different
astrophysical systems. }
\label{table2}
\end{table*}

It is instructive to evaluate the different quantities appearing in Eq.
(\ref{na1}) by
specifying the mass $m$ and the scattering length $a_s$ of the
bosons. To that purpose, we rewrite them in the convenient form
\begin{eqnarray}
\frac{\kappa_Q}{{\rm m}^{-1}}=2.16\times10^{-10}\left(\frac{m}{{\rm
eV}/c^2}\right)^{1/2},
\label{na2}
\end{eqnarray}
\begin{eqnarray}
\frac{\kappa_J}{{\rm m}^{-1}}=5.83\times10^{-18}\left(\frac{m}{{\rm
eV}/c^2}\right)^{3/2}\left(\frac{\rm fm}{a_s}\right)^{1/2},
\label{na3}
\end{eqnarray}
\begin{eqnarray}
\frac{\kappa_C}{{\rm m}^{-1}}=1.01\times10^{7}\left(\frac{m}{{\rm
eV}/c^2}\right),
\label{na4}
\end{eqnarray}
\begin{eqnarray}
\frac{\kappa_H}{{\rm m}^{-1}}=6.49\times10^{-27},
\label{na5}
\end{eqnarray}
\begin{eqnarray}
\sigma=6.19\times10^{-19}\left(\frac{{\rm
eV}/c^2}{m}\right)^3\left(\frac{a_s}{\rm fm}
\right),
\label{na6}
\end{eqnarray}
\begin{eqnarray} 
\epsilon=3.70\times
10^{7}\left(\frac{{\rm
eV}/c^2}{m}\right)\left(\frac{a_s}{\rm fm}\right)^{1/2},
\label{na7}
\end{eqnarray}
\begin{eqnarray} 
\eta=\sqrt{2}\nu=3.00\times
10^{-17}\left(\frac{{\rm
eV}/c^2}{m}\right)^{1/2}.
\label{na8}
\end{eqnarray}
To obtain  these results we have used the relation $\rho_b
a^3=\Omega_{DM,0}\rho_0$, where
$\Omega_{DM,0}=0.268$ is the present fraction of dark matter and
$\rho_0={3H_0^2}/{8\pi G}=8.42\times 10^{-27}\, {\rm kg}\, {\rm
m}^{-3}$ is the present density of the Universe calculated from the Hubble
constant
$H_0=67.11\, {\rm km}\,  {\rm s}^{-1}\, {\rm Mpc}^{-1}$ obtained from the
2014 Planck survey. 

The numerical values of these parameters are given in Tables I
and II for
typical values of $m$ and $a_s$ appropriate to dark matter halos and boson
stars (see Appendix \ref{sec_typ}).
In the matter era ($a>a_i=10^{-4}$), we have
\begin{eqnarray} 
\lambda_C\ll (\lambda_Q,\lambda_J)\ll\lambda_H.
\label{na9}
\end{eqnarray}
Indeed,
$\lambda_C/\lambda_Q=\kappa_Q/\kappa_Ca^{3/4}=\nu/a^{3/4}<10^{-3}$,
$\lambda_C/\lambda_J=\kappa_J/\kappa_C=\nu/\epsilon<10^{-3}$,
$\lambda_Q/\lambda_H=\kappa_H/\kappa_Qa^{3/4}=\eta/a^{3/4}<10^{-3}$,
$\lambda_J/\lambda_H=\kappa_H/\kappa_J a^{3/2}=\eta\epsilon/a^{3/2}<10^{-3}$,
and
$\lambda_C/\lambda_H=\kappa_H/\kappa_Ca^{3/2}=\eta\nu/a^{3/2}<10^{-6}$.
The Tables make clear that relativistic effects are very weak in the matter era
(except for perturbations with very large wavelengths)
in agreement with the discussion given in the paper.

\section{The smallness of the speed of sound in the matter era}
\label{sec_smallness}

The speed of sound is defined by $c_s^2=P'(\epsilon)c^2$, where $\epsilon$
is the energy density. In the matter era, $\epsilon\sim\rho c^2$, where $\rho$
is the rest-mass density, so that $c_s^2=P'(\rho)$. For a self-interacting SF,
the speed of sound is given by Eq. (\ref{j11b}). It may be considered as
obvious that $c_s\ll c$ during the matter era. Actually, there is a nice way to
prove this statement. Using $\rho_b a^3=\Omega_{DM,0}\rho_0$, and Eq.
(\ref{me2}),
we obtain
\begin{equation}
\frac{c_s^2}{c^2}=\frac{3 a_s\hbar^2\Omega_{DM,0}H_0^2}{2Gm^3a^3c^2}.
\label{smallness1}
\end{equation}
This quantity depends on the mass $m$ and scattering length $a_s$ of
the bosons only through the ratio $a_s/m^3$. It turns out that this ratio also
determines the size of dwarf dark matter halos through the relation
(see \cite{goodman,arbey,bohmer,prd1} and Appendix \ref{sec_typ}):
\begin{equation}
R=\pi\left (\frac{a_s\hbar^2}{Gm^3}\right )^{1/2}.
\label{smallness2}
\end{equation}
As a result, Eq. (\ref{smallness1}) can be rewritten as
\begin{equation}
\frac{c_s}{c}=\left (\frac{3\Omega_{DM,0}}{2\pi^2}\right
)^{1/2}\frac{H_0R}{a^{3/2}c}.
\label{smallness3}
\end{equation}
Since $a>a_i=10^{-4}$ in the matter era, and since the size of the DM halos is
obviously much smaller than the present horizon ($R\ll c/H_0$), we conclude that
$c_s\ll c$ during the matter era. We have reached this conclusion without
having to specify the mass and scattering length of the bosons. We can
explicitly check on Table II that the condition $\sigma/a^3\ll 1$, equivalent to
$c_s\ll c$, is indeed satisfied for the
typical values of the mass $m$ and scattering length $a_s$ of the
bosons considered in the literature.

\acknowledgments{A. S. acknowledges CONACyT for the postdoctoral grant
received.}

\end{document}